\documentclass[12pt,letter]{iopart}

\usepackage{verbatim}
\usepackage{float}
\usepackage{makeidx}
\usepackage[dvips]{graphicx}
\usepackage{amssymb}
\usepackage{subfigure}
\usepackage{booktabs}
\usepackage[draft]{hyperref}
\usepackage[english]{babel}
\usepackage{color}
\usepackage[latin1]{inputenc}
\usepackage[top=1.5in,bottom=1in]{geometry}
\usepackage{setspace}

\usepackage{longtable}
\usepackage{multirow}

\usepackage{relsize}

\begin{document}
\title[Identifying the community structure of the food-trade international multi network]{Identifying the community structure of the international food-trade multi network }
\author{S Torreggiani$^1$, G Mangioni$^2$,
M J Puma$^3$\footnote{Corresponding author. NASA Goddard Institute for Space Studies, 2880 Broadway, New York, NY 10025 USA. Email:  \texttt{mjp38@columbia.edu}} and G Fagiolo$^4$}
\address{$1$ SOAS University of London, U.K., and Istituto di Economia, Scuola Superiore Sant'Anna, Pisa, Italy.}
\address{$2$ Dipartimento di Ingegneria Elettrica, Elettronica e Informatica, University of Catania, Catania, Italy}
\address{$3$ Center for Climate Systems Research and Center for Climate and Life, Columbia University; NASA Goddard Institute for Space Studies, New York, USA}
\address{$4$ Istituto di Economia, Scuola Superiore Sant'Anna, Pisa, Italy.}
\begin{abstract}
Achieving international food security requires improved understanding of how international trade networks connect countries around the world through the import-export flows of food commodities. The properties of food trade networks are still poorly documented, especially from a multi-network perspective. In particular, nothing is known about the community structure of food networks, which is key to understanding how major disruptions or ``shocks'' would impact the global food system. Here we find that the individual layers of this network have densely connected trading groups, a consistent characteristic over the period 2001 to 2011. We also fit econometric models to identify social, economic and geographic factors explaining the probability that any two countries are co-present in the same community. Our estimates indicate that the probability of country pairs belonging to the same food trade community depends more on geopolitical and economic factors -- such as geographical proximity and trade agreements co-membership -- than on country economic size and/or income. This is in sharp contrast with what we know about bilateral-trade determinants and suggests that food country communities behave in ways that can be very different from their non-food counterparts.
\end{abstract}
\submitto{\ERL}
\maketitle
\ioptwocol

\section{Introduction\label{sec:intro}}

Achieving international food security \cite{Porkka_etal_2013} is undoubtedly one of the major challenges of the forthcoming decades and a globally recognized priority \cite{UN_2015}. However, understanding how and why the availability of \textit{and access to} food commodities change across time and space is a dauntingly difficult task, due to its inherent multidimensional nature \cite{DOdorico_etal_2014}. International food security may indeed depend on many intertwined phenomena \cite{Godfray_etal_2010}, including population growth \cite{UN_2015_pop}; agricultural productivity and (over) exploitation of natural resources \cite{Hazell_Wood_2008,Hanjra_Qureshi_2010,Woods_etal_2010}; climate change \cite{Coumou_Rahmstorf_2012,Battisti_etal_2009,Gornall_2010}; regional conflicts and epidemics \cite{McCloskey_etal_2014}; and the evolution of consumption habits \cite{Nonhebel_Kastner_2011,Tilman_etsl_2011,Cassidy_etal_2013}. 

The resulting impact of these interacting factors may generate unexpected volatility and substantial shocks in the supply and availability of food commodities, possibly leading to global crises \cite{Clapp_2009_book}. International trade, in this respect, may act both as a dampening force and as an amplifying device to regional shocks \cite{Clapp_2015}. On the one hand, international trade may provide new channels to meet increasing food demand through the transfer of food commodities and resources to food-scarce regions. Empirical evidence indeed shows that the amount of traded food has more than doubled in the last 30 years, and it now accounts for 23\% of global production \cite{DOdorico_etal_2014}. Furthermore, whereas in the past insufficient domestic production generally implied scarcity in food supplies, production shortfalls in more recent years have been increasingly dealt with by increasing food imports \cite{Porkka_etal_2013,Puma_etal_2015}.

On the other hand, import-export linkages across countries can boost shock diffusion: increased connectivity in the international trade network (ITN, cf. \cite{Fagiolo2009pre}) can lead to a growing fragility \cite{Lee_etal_plosone_2011,Amour_etal_2016,Puma_etal_2015}. This parallels what happened during the 2007-2008 global financial crisis (GFC henceforth), when seemingly minor shocks spread quickly in a complex, networked world, with disastrous effects \cite{Haldane_May_2011}.     

To better understand how shocks can spread beyond a regional scope, it is therefore important to shed light on the structure of the networks connecting countries through import-export flows of food commodities. Despite advances in understanding the ITN at the aggregate level \cite{Fagiolo_survey_2017} and for a set of highly-traded commodities (not necessarily food related) \cite{Barigozzi_etal_2010,Barigozzi_etal_2011}, the properties of food trade networks are still poorly documented  \cite{Brooks_etal_2013,Fracasso_etal_2016,Gephart_Pace_2015,Wu_Guclu_2013,Puma_etal_2015}, especially from a multi-network perspective \cite{Battiston_etal_2014,Kivela_etal_2014,Boccaletti_etal_2014}. In particular, nothing is known about the community structure (CS) of food networks \cite{Fortunato_2010}, that is the organization of network nodes in clusters, where nodes within a cluster are comparatively more intensively connected than they are with nodes belonging to different clusters. Documenting the CS of the international food trade multi-network (IFTMN) may help us better understand how food crises propagate. Indeed, if trade across countries is organized into well-defined clusters, shocks originating within a cluster would likely spread more readily within that group than across groups.  

Here we start to fill this gap using data on international trade flows taken from FAOSTAT, with a focus on the 16 most internationally traded staple food commodities for the period 1992-2011. We document the evolution of CSs in the IFTMN both across layers (i.e., when the IFTMN is analyzed as a collection of separate layers, each one representing bilateral trade for a specific food commodity, e.g. wheat) and in the multi-layer graph (i.e., when the IFTMN is conceived as a single multi-layer network where countries are connected by multiple import-export relationships, e.g. for maize, wheat, rice, etc.).

We then fit econometric models to identify social, economic and geographic factors explaining the probability that any two country are co-present in the same community. Results reveal that countries in the IFMN tend to organize into densely connected trading groups that remain sufficiently stable over time. Overall, our estimates indicate that the probability for country pairs to belong to the same food trade community depends more on geopolitical and economic factors ---such as geographical proximity and trade agreements co-membership--- than country economic sizes and/or incomes. This is in sharp contrast with what we know about bilateral-trade determinants and suggest that food country communities behave in ways that can be very different from their non-food counterparts.      
 
\section{Materials and Methods\label{sec:methods}}

\subsection{Data and Definitions}
We use FAOSTAT data on international trade flows, which contain bilateral export-import yearly figures for food and agricultural products in the period 1986-2013\footnote{Data are available at \texttt{fao.org/faostat}.} 

We select the 16 most-traded commodities in 2013, ranked according to the total kilocalories (kcal henceforth) embodied, so as to account for about 90\% of the total kcal trade for food-related goods\footnote{To compute total kcal embodied we explicitly consider caloric values of secondary and derivative products, see table \ref{tab:products} in \ref{app:products} for details. Primary and secondary products are aggregated after converting them to kcal.}.

Table \ref{tab:top16} lists the top 16 commodities according to kcal embodied (in 2013) and their trade value (in current USD). As expected, the two rankings are not correlated. For example, there are traded commodities with an extremely high economic value that contribute much less in terms of kcal (e.g., meat and animal products). Notice also that the distribution of kcal is extremely skewed: more than 55\% of total kcal are accounted for by wheat, soybean, maize and rice (which together form just 23\% of total value in USD).

\begin{table*}[h!]
\centering{\caption{\label{tab:top16}Top world 16 import commodities in 2013 according to kcal embodied.}
\begin{indented}
\item[]\begin{tabular}{@{}lcccc}
\mr
  Code & Commodity & kcal$^a$ & USD & \% kcal \\ 
\br	
  1 & Wheat & 6.45 $\times 10^{14}$ & 9.71 $\times 10^{10}$ & 21.11 \\ 
  2 & Soybeans & 5.93 $\times 10^{14}$ & 1.07 $\times 10^{11}$ & 19.43 \\
  3 & Maize & 4.44 $\times 10^{14}$ & 4.22 $\times 10^{10}$ & 14.54 \\ 
  4 & Sugar & 2.25 $\times 10^{14}$ & 3.31 $\times  10^{10}$ & 7.38 \\ 
  5 & Rice & 1.36 $\times 10^{14}$ & 2.61 $\times  10^{10}$ & 4.47 \\ 
  6 & Barley & 1.32 $\times 10^{14}$ & 2.74 $\times  10^{10}$ & 4.33 \\ 
  7 & Oil, Palm & 9.74 $\times 10^{13}$ & 4.20 $\times  10^{10}$ & 3.18 \\ 
  8 & Oil, Sunflower & 7.22 $\times 10^{13}$ & 1.01 $\times 10^{10}$ & 2.37 \\ 
  9 & Milk & 6.81 $\times 10^{13}$ & 8.23 $\times 10^{10}$ & 2.21 \\ 
  10 & Cassava & 5.33 $\times 10^{13}$ & 4.07 $\times  10^{9}$ & 1.75 \\ 
  11 & Pulses & 4.64 $\times 10^{13}$ & 1.02 $\times 10^{10}$ & 1.49 \\ 
  12 & Cocoa & 4.51 $\times 10^{13}$ & 4.22 $\times  10^{10}$ & 1.46 \\ 
  13 & Pig Meat & 4.47 $\times 10^{13}$ & 4.21 $\times  10^{10}$ & 1.43 \\
  14 & Poultry Meat & 2.82 $\times 10^{13}$ & 3.45 $\times 10^{10}$ & 0.92\\
  15 & Nuts & 2.61 $\times 10^{13}$ & 2.03 $\times 10^{10}$ & 0.86 \\
  16 & Sorghum & 2.40 $\times 10^{13}$ & 2.01 $\times 10^{9}$ & 0.78 \\
\br
\end{tabular}
\item[] Source: Our computation on FAOSTAT data (see \texttt{fao.org/faostat}).
\end{indented}
}
\end{table*}

Selecting commodities according to a mass-to-kcal conversion ---rather than value or volume--- allows us to better address the role of trade in the nutritional security of countries\footnote{Other factors such as water \cite{Konar_etal_2011,Sartori_Schiavo_2014,Tamea_etal_2014,Fracasso_etal_2016} or nutritional \cite{billen2014biogeochemical} content of the food may be included in future studies.}. Furthermore, the 16 commodities selected also have a substantial environmental footprint, as they typically use the most cropland and strongly influence irrigation water consumption \cite{MacDonald_etal_2015}.

In order not to bias our analysis with issues related to the collapse of the USSR and of the former Yugoslavia, we do not include the years 1986-1991. We also remove the two most recent years (2012-2013) from the sample, as bilateral updated data are still not available for some products and/or countries\footnote{Note that our selected commodities are still the top-16 most-traded agricultural products in terms of kcal also in 2011.}. We eventually end up with $N=178$ countries (see table \ref{tab:countries} in \ref{app:countries} for a complete list)\footnote{A country is inserted in our sample if it is involved in a positive bilateral flow for at least one year or one commodity.}, whose bilateral trade flows for the 16 selected commodities are observed from 1992 to 2011 ($T=20$).    

We define the IFTMN as the sequence of $T$ multi-layer networks, where each layer represents bilateral trade among our $N$ countries for a specific commodity $c=1,\dots,C$ ($C=16$) in a given year. More formally, in each year $t=1992,\dots,2011$, let $\mathbf{X}^t$ be the 3-dimensional weight matrix whose generic entry $x_{ij,c}^t\geq 0$ represents exports (in kcal) from country $i$ to country $j$ for commodity $c$ in year $t$. As usual, we posit that $x_{ii,c}^t=0$ for all $i$, $c$ and $t$. We define the IFTMN as the time sequence of multi-layer networks characterized by the time sequence of weighted-directed matrices $\{\mathbf{X}^t, t=1,\dots,T\}$. In other words, each snapshot (year) of the IFTMN is a multi-layer network, where the nodes are the 178 countries connected by multiple directed links, each of which represents an exporter-importer flow for a particular commodity, weighted by its correspondent intensity in terms of kcal traded. A directed link $(i\rightarrow j)_{c}^t$ is therefore present for a given commodity-year combination $(c,t)$ if $i$ exports to $j$ a non-zero volume for commodity $c$ in year $t$. All zero off-diagonal entries therefore represent either a missing value or a sheer zero-trade flow.\footnote{In the IFTMN, links between any two commodity layers $c_1$ and $c_2$, $c_1\neq c_2$ are present only between \textit{copies} of the same country, i.e. any country $i$ is connected to herself in all the layers. Two different countries are not linked across different layers. In this respect, the IFTMN can be defined as a \textit{multiplex} or \textit{colored network}.}           

\subsection{Network Structure}
Prior to performing community detection, we explore the properties of the time sequence of multi-networks $\mathbf{X}^t$ using a principal component analysis in the space of network statistics computed over each single layer. More precisely, given link weights $x_{ij,c}^t$ of layer $(c,t)$, let $\mathbf{W}^t_c$ be the associated log-transformed weight matrix\footnote{As it is customary in this literature \cite{Fagiolo2009pre}, positive trade levels are log-transformed in order to reduce the skewness of their distribution.} and $\mathbf{A}^t_c$ the correspondent adjacency matrix. In each year $t$, we compute a number of network statistics over $\mathbf{W}^t_c$ and $\mathbf{A}^t_c$ to fully characterize the weighted and binary topological properties of the layer (see \ref{app:netprop} for details). We include measures of binary and weighted connectivity (e.g., network density, size of largest connected component, average and standard deviation of link weights), assortativity, node clustering and network centrality, in order to provide a full topological characterization of each layer. After removing the statistics that turn out to be redundant (i.e., too highly correlated with the most basic statistics like density), we perform a principal-component analysis to reduce the dimensionality of the space of remaining statistics, and we then interpret the results. This allows us both to identify network measures that better characterize the topological structure IFTMN in each year and to explore similarities and differences among commodity networks. 

\subsection{Community Structure Detection}
Identifying communities in a network is fundamental for gaining insights about its fundamental structure, its robustness, and the ways in which shocks percolate through it \cite{Porter_etal_2009}. Essentially, communities are clusters of vertices characterized by a higher ``within'' connectivity, but a much sparser connectivity ``between'' nodes belonging to different clusters.  Community detection is a very difficult task and a host of different techniques and definitions have been recently proposed in the literature for the case of simple or multi-graphs  \cite{Fortunato_2010,Malliaros_etal_2013,Kim_Lee_2015}. 

Here, we tackle the problem of community detection in the IFTMN using two complementary approaches. 

First, in any given $t$, we treat the IFTMN as a collection of $C$ different commodity-specific weighted-directed simple graphs, and we analyze the CS of each layer separately. To identify communities, we employ the modularity optimization approach originally introduced by \cite{Newman_Girvan_2004} and subsequently extended to the case of weighted directed graphs by \cite{Arenas_etal_2007}. In this case, the modularity function to be maximized is:

\begin{equation}
	Q_c^t=\frac{1}{X_c^t}\sum_{ij}{( x_{ij,c}^t - \mathbf{E} [x_{ij,c}^t])}\delta(\xi_{i,c}^t, \xi_{j,c}^t),
\end{equation}    
where $X_c^t$ is the volume of the layer $(c,t)$ and $\delta$ is a Kronecker delta function equal to 1 if nodes $i$ and $j$ are in the same community and 0 otherwise. $\mathbf{E}$ is the expected value of the link weight $x_{ij,c}^t$, which following \cite{Arenas_etal_2007} reads:
\begin{equation}
	\mathbf{E} [x_{ij,c}^t]= \frac{s_{i,c}^t(out) \cdot s_{j,c}^t(in)}{X_c^t},
\end{equation}   	
where $s_{i,c}^t(out)$ and $s_{j,c}^t(in)$ are respectively out-strength of node $i$ and in-strength of node $j$ \cite{Barrat_etal_2004}. To optimize $Q_c^t$, we employ the modularity-clustering heuristic developed by \cite{Rotta_Noack_2011}, which extends and improves the well-known ``Louvain'' algorithm pionereed by \cite{Blondel_etal_2008_LouvainMeth} (see \ref{app:communities} for more details). This procedure ends up, for any given year $t$ and commodity-layer $c$, with a univocal assignment of countries into clusters, the number of which is not fixed ex-ante, in such a way that each country belongs to a single cluster (i.e., communities are not overlapping). Clusters can also contain a single country, e.g., if that country is an isolated node in the network.

\setcounter{footnote}{0}

Second, we check the results of the former procedure when the IFTMN is described, for any $t$ as a single multi-layer network. More precisely, following \cite{Mucha_etal_2010,Carchiolo2011}, we consider the $C$ layers making up a time snapshot of the IFTMN as being connected through weighted, non-directed links that join the same node across all the layers. The weight of such links ($\theta$) is homogeneous across time, nodes and layers, and is treated as a system parameter. In such a multi-layer perspective, communities are formed by country-commodity pairs. So, for example, the same country can end up in different clusters in association with different commodities; or different countries can belong to the same cluster in association with the same commodity. Here, we perform a multilayer community-detection analysis as in \cite{Mucha_etal_2010}, who extend modularity to multi-layer graphs on the base of generalized null models obtained by considering a Laplacian dynamics on the multi-layer. More specifically, we use the implementation of the algorithm in \cite{Mucha_etal_2010} available in MuxViz \cite{muxviz}, which is based on a generalization of the ``Louvain'' algorithm \cite{Blondel_etal_2008_LouvainMeth} (see \ref{app:communities} for further details).

\subsection{Econometric Models}
As mentioned, identifying communities in the IFTMN treated as a collection of $C$ separate layers, results in a univocal assignment of countries to clusters for any given choice of $t$ and $c$. Clusters are multilateral entities, as they emerge whenever a group of countries trades comparatively more among them than they do with countries outside the cluster. But what are the factors underlying the emergence of such clusters? Here, we address this issue fitting probit and logit models \cite{Winkelmann_2008} that explain the probability that any two countries belong to the same cluster (for a given $(c,t)$ slice of the IFTMN) as a function of economic, socio-political and geographical, bilateral relationships. More precisely, we perform two sets of exercises. 

First, for all $c=1,\dots,16$ and two selected years ($t_0=2001$ and $t_1=2011$)\footnote{These two years have been chosen in order to focus on two time periods sufficiently far from the GFC.}, we fit to the data the following probit model using a maximum-likelihood procedure:

\begin{equation}
	Prob\{\gamma_{ij,c}^t=1\}=\Phi(\alpha+\mathbf{\beta}\mathbf{Z}_{ij}^t),\label{eq:probit}
\end{equation}   
where $\gamma_{ij,c}^t$ is a binary indicator for the event that countries $i$ and $j$ belong to the same community for product $c$ and year $t\in \{t_0,t_1\}$, $\Phi$ is the cumulative distribution function for the standard normal variate\footnote{All our econometric results are robust when we employ a logit specification instead of a probit, i.e. when we let $\Phi$ be the cumulative distribution of a logistic random variate.}, $\alpha$ is a constant, $\beta$ is a vector of slopes and $\mathbf{Z}_{ij}^t$ is a set of bilateral covariates (more on that below). 

\setcounter{footnote}{0}

Second, we run a panel-data estimation of the probit model in Eq. (\ref{eq:probit}) on the pooled dataset containing all the years in our sample, for some selected commodities (i.e., wheat, maize and rice). We choose wheat, maize, and rice (and their associated commodities) as they are among the most important internationally traded grains and are fundamental to staple food supplies around the world. Panel estimations feature the same covariates of the cross-section setup, but they now become time-varying. Furthermore, as it is customary in this approach \cite{Baldwin_Taglioni_2006}, we control for unobserved heterogeneity and common trend effects including in panel regressions time-invariant country fixed-effects and time dummies. 

To choose the covariates, we rely on the literature on the empirical trade-gravity model \cite{Anderson_2011}, see \ref{app:covariates} and Table \ref{tab:covars} for details. We employ five classes of covariates: economic variables (i.e., combined measures of economic country size and income); trade policy variables (e.g., whether the two countries belong to the same preferential trade agreements); geographical variables (e.g., distance between countries and whether they share a border); historical/political variables (e.g., former colonial relationships); and cultural variables (i.e., whether countries share the same language).  

Despite the fact that our probit specification has an obvious gravity flavor, it departs from traditional trade-gravity models in the way we treat directionality of relationships. Indeed, since the co-presence relations are symmetric by definition, the binary response model in Eq. \ref{eq:probit} does not distinguish between importer and exporter, as, on the contrary, gravity models with trade flows as dependent variable often do. Therefore, sign and intensity of the impact of covariates cannot differ between origin and destination markets.

\section{Results\label{sec:results}}
We now turn to a description of our main results. First, we describe some basic network properties of the IFTMN, both across commodity-layers and time. Second, we discuss the CS of ITMN considered as a collection of $C$ separate layers. Third, we explain co-presence in clusters using probit models. Finally, we check what happens when CS detection is performed over the IFTMN described as a multi-layer network.  
  
\subsection{Overview of network properties\label{subsec:net_props}}
The IFTMN is characterized by low variability over the time interval under observation but substantial heterogeneity across layers in each year. A comparison of results in Tables \ref{tab:stats_2001}-\ref{tab:stats_2011} in \ref{app:netprop}, which report network statistics in 2001 and 2011, suggests that network structure did not go through dramatic changes before and after the GFC. 

\begin{figure}[h!]
 \centering
    {\includegraphics[width=8cm]{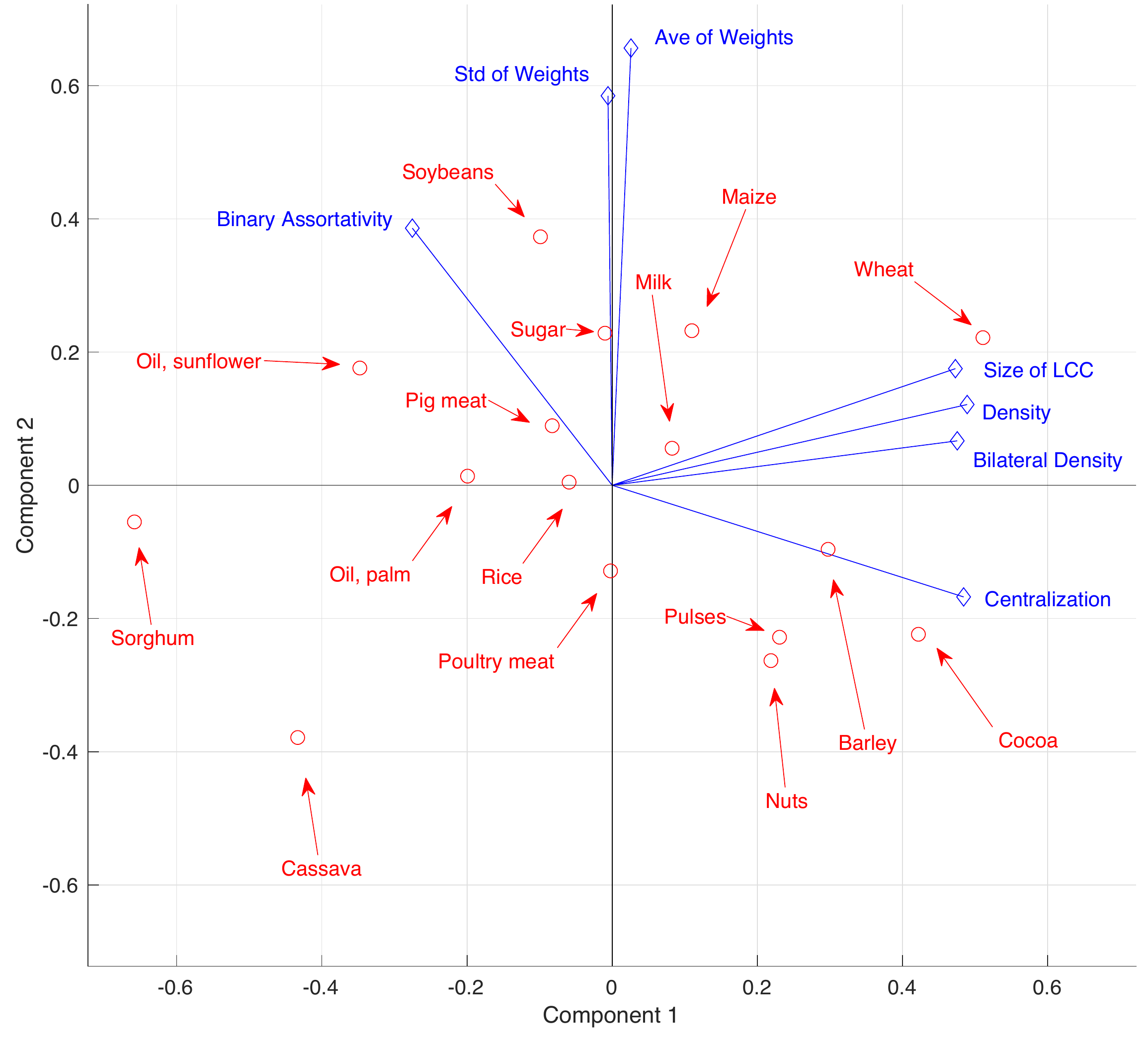}}
 \caption{\label{fig:pca_2011}The IFTMN in year 2011. Principal component (PC) analysis in the space of network statistics. First two PCs explain 83\% of total variance. }
\end{figure}

However, our analysis indicates considerable variation in the topological properties across commodity layers. For example, the IFTMN is composed of small-density layers (as compared to the aggregate ITN), whose link probabilities range from 0.01 to 0.16. Substantial variation is also detected in the size of the largest connected component (LCC) -- from 87 to 171 -- and many other statistics. Therefore, a principal component (PC) analysis can help in summarizing the most important dimensions of variability. Results for year 2011 are reported in Figure \ref{fig:pca_2011}. We use a bi-plot to represent both the units (commodities) in the space of the first two PCs (which together explain 83\% of total variance) and network statistics as vectors (whose direction and length indicate how each variable contributes to the two principal components in the plot). The first PC is positively correlated with connectivity measures (density and size of LCC), network symmetry and centralization, and negatively correlated with binary assortativity (i.e., the larger the x-axis coordinate, the smaller the assortativity coefficient). The second PC is instead positively correlated with average and standard deviation of link weights (in addition to assortativity). This means that, overall, commodity layers tend to display a higher density and size of LCC, and to be more centralized and symmetric, but less assortative. Moreover, more intense  bilateral connections are gained, on average, at the expense of a larger standard deviation thereof. 

Zooming inside commodities, the position of layers in the bi-plot suggests the existence of two paradigmatic cases. The first one is represented by layers such as wheat, cocoa and barley, which are characterized by a relatively high connectivity, centralization and symmetry, but a relatively smaller assortativity, and a lower intensity and variability of import-export relationships. To the second one belong layers such as sorghum and cassava, who are much less connected and symmetric, and they are structured over more intense and less variable trade relationships. Other important layers like maize, rice and soybeans play instead an intermediate role, being less internally connected than wheat but displaying stronger and more variable bilateral connections.

Network statistics in Tables \ref{tab:stats_2001}-\ref{tab:stats_2011} and their correlations (see Figure \ref{fig:corr_stats}) reveal two important additional facts. First, the layers of the IFTMN are mostly assortative: more-intensively connected countries tend to import and export to countries which are themselves more connected. This conflicts with widespread evidence observed both in the aggregate ITN and across commodity-specific trade layers, not necessarily related to food, representing import-export relationships for specific product classes at a two-digit breakdown (e.g., cereals, pharmaceutical products, iron and steel, etc.), see Ref. \cite{Fagiolo2009pre,Barigozzi_etal_2010}. 

Second, the weighted version of statistics such as asymmetry, clustering and assortativity are almost linearly correlated with their binary counterpart, suggesting that in the IFTMN, unlike in the aggregate ITN, the creation of new trade channels are more important than increases in trade flows of already existing connections (i.e., in economics jargon, extensive trade margins are more important than intensive ones). 

\begin{figure}[h!]
 \centering
    {\includegraphics[width=8cm]{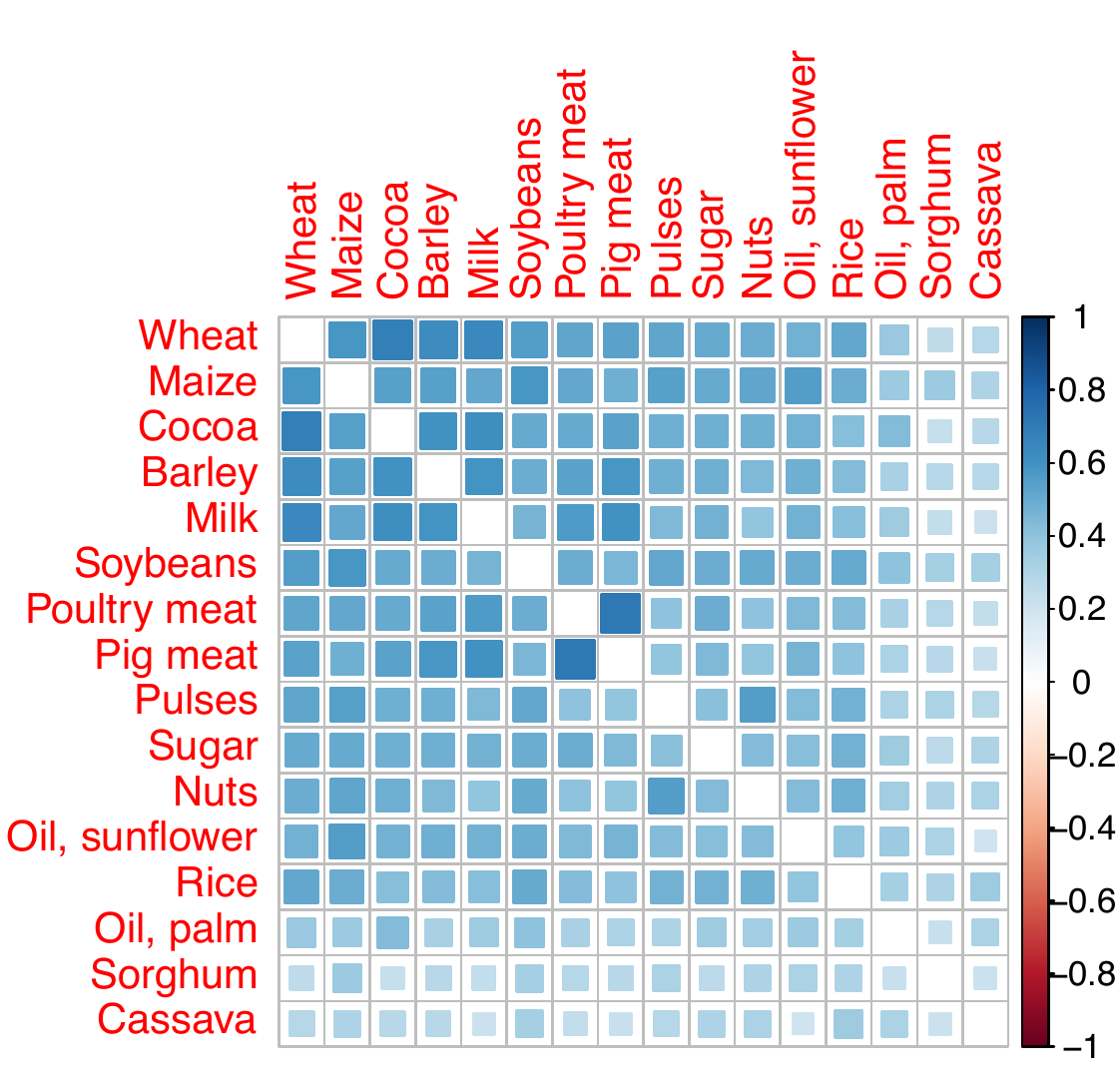}}
 \caption{\label{fig:lw_corr_2011}Correlation between logged link weights of commodity layers. Year=2011. Commodities have been ordered using a (Ward) hierarchical clustering.}
\end{figure}

We now explore across-layer correlation in (logs of) link-weight distributions $w_{ij,c}^t=\log(x_{ij,c}^t)$, cf. Figure \ref{fig:lw_corr_2011} for year 2011 and Figure \ref{fig:lw_corr_2001} in \ref{app:netprop} for year 2001. We notice that almost all commodities are traded as complements (i.e., all correlations are positive and significant). The only exceptions are palm oil, sorghum and cassava, which are traded in an almost uncorrelated way with all the others. This may probably be due to the fact that these are either markets extremely concentrated around a handful of producers (i.e., palm oil) or extremely agglomerated geographically (i.e., cassava and sorghum).

\begin{figure}[h!]
 \centering
    {\includegraphics[width=8cm]{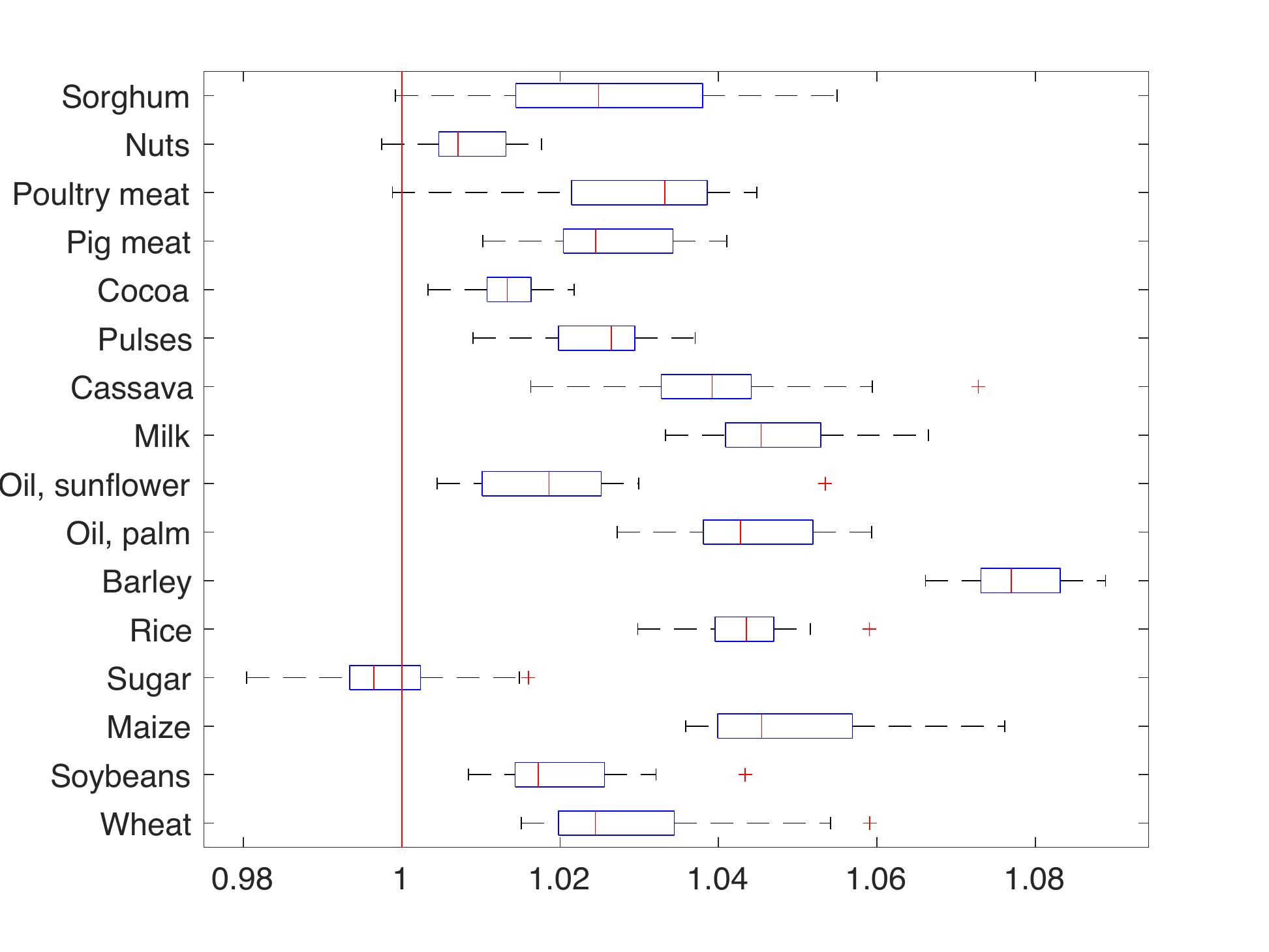}}
 \caption{Time-series distributions for the average import/export intensity ratio. The central red mark of each box is the median, the edges of the box are the 25th and 75th percentiles, the whiskers extend to the most extreme not-outlier observations, and the outliers are plotted individually (red plus). \label{fig:nsin_nsout_boxplot}}
\end{figure}

Finally, we investigate the extent to which export per outward link is associated with imports per inward link, across years and layers. Figure \ref{fig:nsin_nsout_boxplot} depicts time-series distributions for the ratio between layer-average import intensity vs. average export intensity (i.e., the import/export intensity ratio). Import (resp. export) intensity is defined as total country import (resp. export) per importing (resp. exporting) partner, that is, in network jargon, the ratio between node in (resp. out) strength and node in (resp. out) degree. Note how almost all layers have been characterized by ratios always larger than one across the years. This means that, on average, countries tend to have, irrespective of the commodity traded and its share on the world market, more intensive import relations than export ones. This result is consistent with the evidence shown by Ref. \cite{Barigozzi_etal_2010} for a more aggregated set of commodity-specific -- not necessarily food-related -- networks (and it is, in particular, true for coarse cereals). This evidence could be a symptom of the high dependency of several countries on few relevant import channels for their staple-food supply.  

\subsection{Layer-by-layer community structure\label{subsec:comm_descr}}
We now discuss community-detection findings when the IFTMN is treated, in each year, as a collection of independent food-staple trade layers. We begin with results related to two temporal cross sections -- for the individual years 2001 and 2011 -- across all layers. Then, for three selected commodities (wheat, maize and rice), we document the evidence on community-detection for the 2001-2011 panel.

As Table \ref{tab:comm_stats_2001_2011} shows, the first general observation is that the IFTMN exhibits a very high level of (maximum) modularity in almost all layers and years. This suggests that the IFTMN is characterized throughout by a strong community structure, with countries that organize into densely linked groups. Indeed, maximum modularity levels typically fall in the range [0.2,0.5], which, as suggested in Ref. \cite{Newman_Girvan_2004}, is strong evidence for the existence of well-defined clusters. The only exception to this general rule is cassava, which displays an almost negligible level of modularity. In each layer, we identify on average 6 clusters (or communities) with number ranging from 3 (for poultry meat in 2011, the least dispersed layer on average) to 10 (for sorghum in 2001, the most dispersed layer on average). 

\begin{figure*}[h!]
 \centering
 \subfigure
   {\includegraphics[width=10cm]{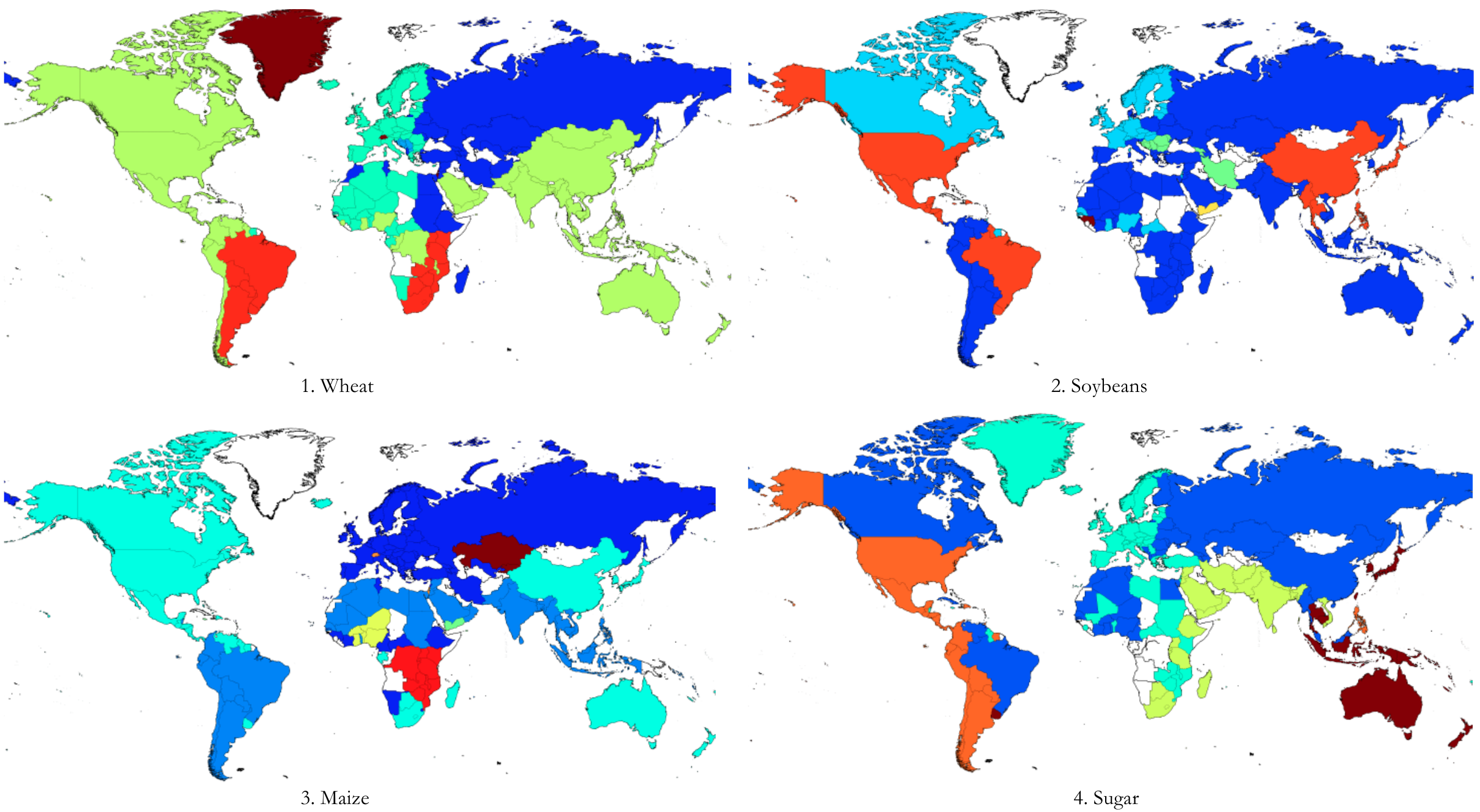}}
 \hspace{5mm}
 \subfigure
   {\includegraphics[width=10cm]{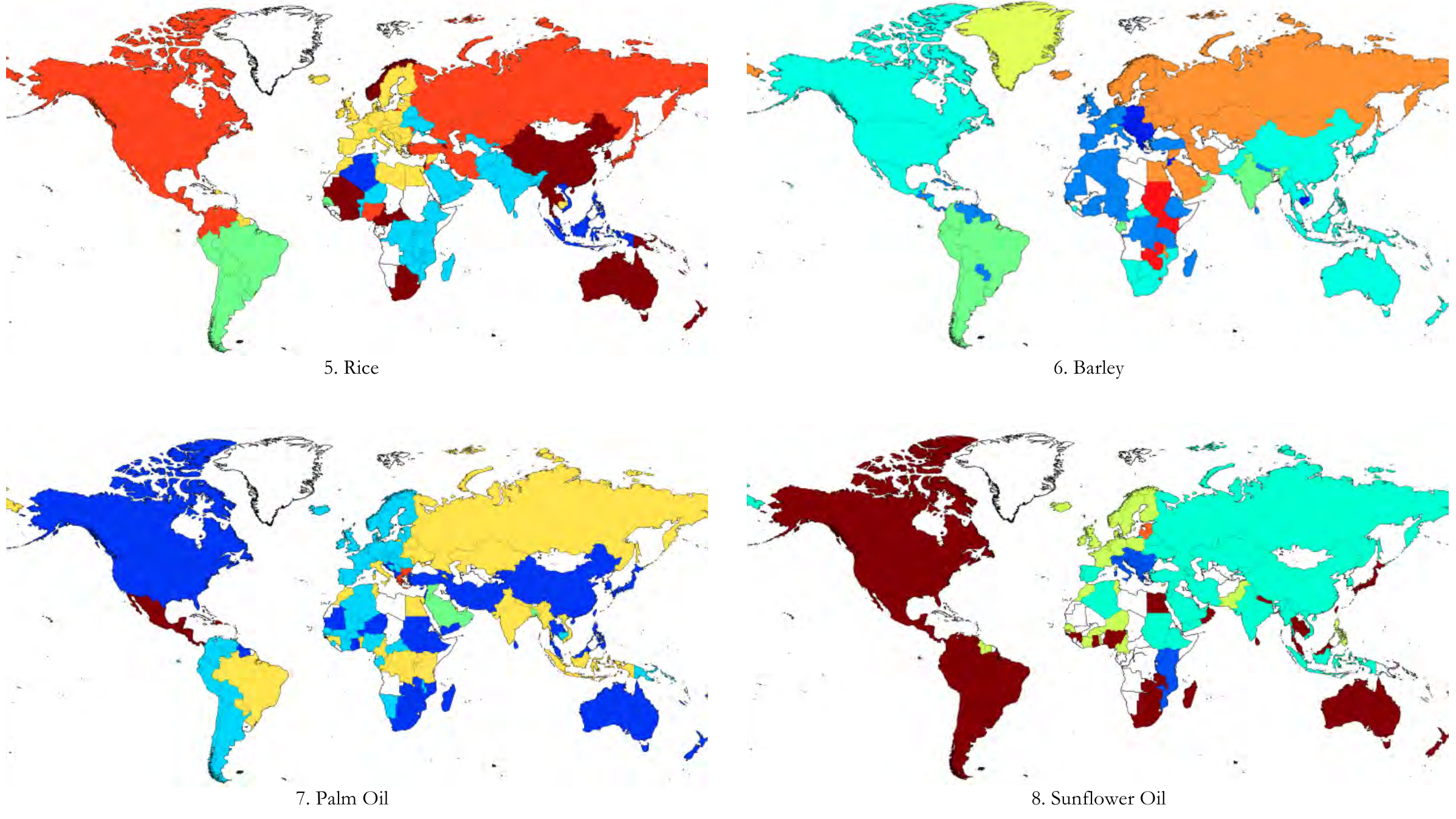}}
 \protect\\
 \subfigure
   {\includegraphics[width=5cm]{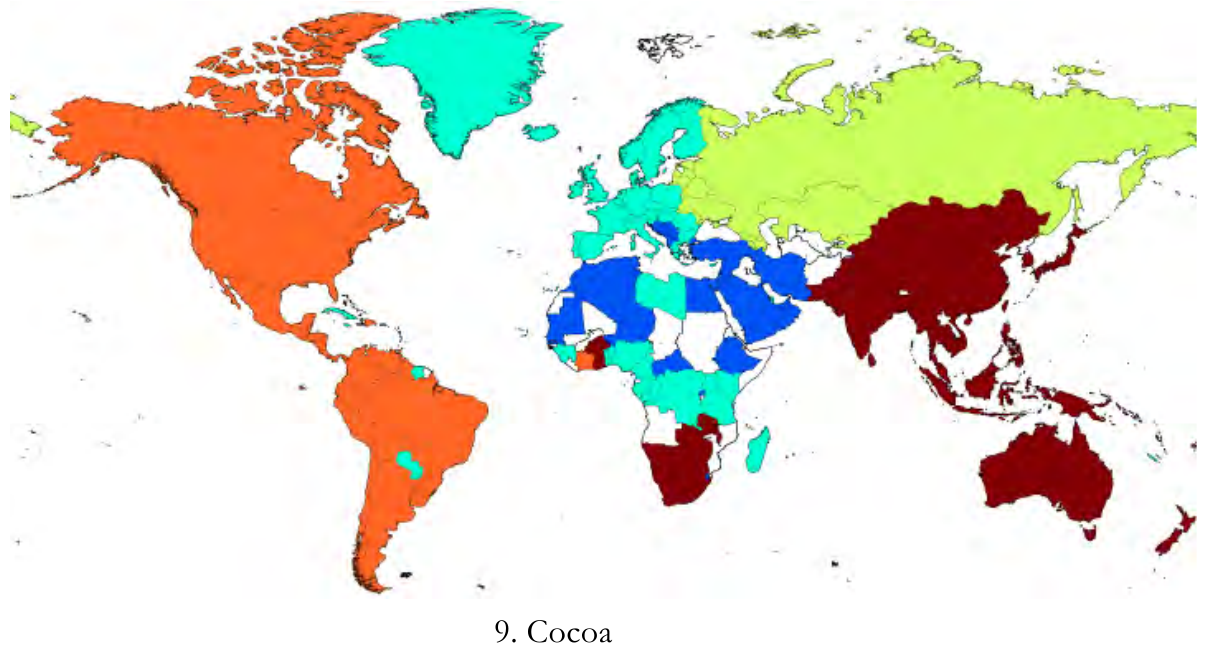}}
 \subfigure
   {\includegraphics[width=5cm]{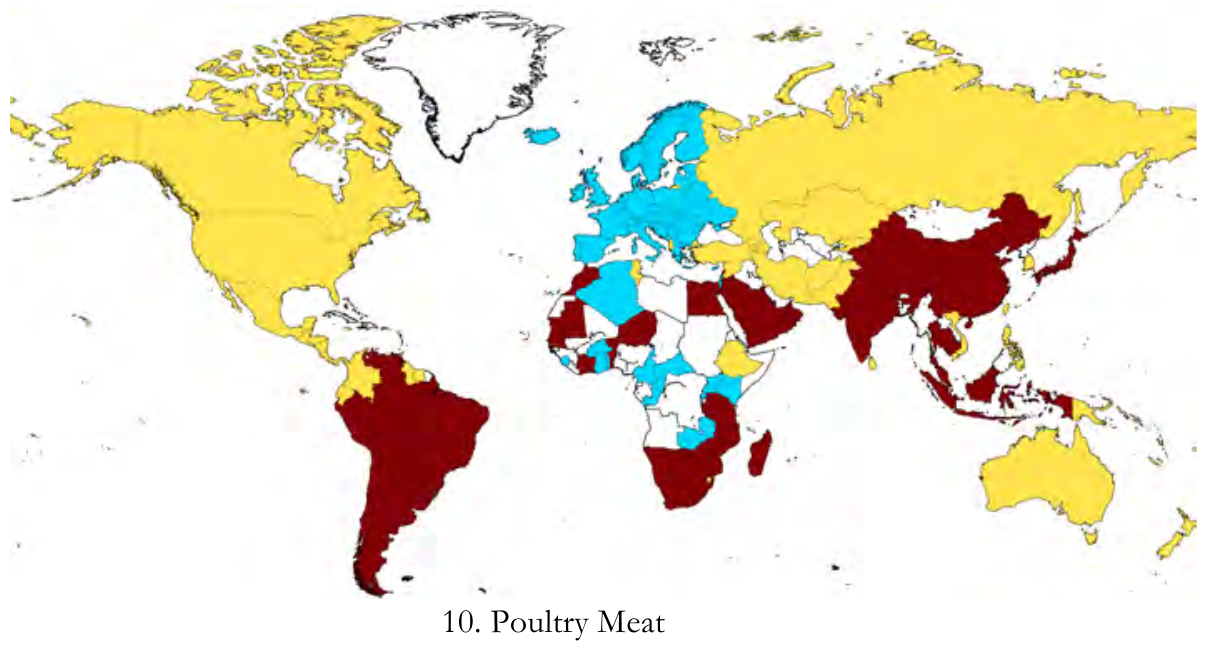}}
 \caption{Community detection in year 2011. Choropleth maps display country membership to communities for selected commodities. In white, countries not belonging to any community or for which no data are available.\label{fig:choro_2011}}
\end{figure*}

More importantly, our community detection exercises indicate that countries in the IFTMN tend to cluster into trading blocs that display relevant geopolitical and socioeconomic patterns. This can be seen in Figure \ref{fig:choro_2011}, where we plot choropleth maps with countries colored according to their community membership in 2011 for selected commodities.
     
Choropleth maps for year 2011 reveal interesting across-layer regularities. First, there often exists a North American cluster (with the US and Canada often linked to Central and Latin America countries), whereas relevant breadbaskets such as Brazil and Argentina often set up alternative communities independently. Second, Russia generally forms a cluster together with Central, Caucasian and East- European (non EU-members) states, often absorbing some MENA region countries (especially Egypt). A unified European cluster often emerges, sometimes linked with the Russian cluster and rarely linked with the US, confirming that Europe is not such an open market for many agricultural products. Furthermore, a consolidated and independent Asian cluster seems to exist only in the case the region is a net importer for that commodity (i.e., wheat, milk and diary products, and cocoa). East Asian (e.g., China, India and Japan) and Southeast Asian (e.g.,  Vietnam, the Philippines, and Thailand) countries instead typically belong to different communities, orbiting around other clusters such as the North American and South American ones. Finally, Africa and the Middle East are often divided -- independently of the commodity examined -- and only in a few cases we can observe a small independent Eastern Sub-Saharan cluster. 

Apart from these macro regularities, several cross-sectional differences also emerge among commodity-specific community structures\footnote{In \ref{app:remarks_2011} we discuss in details economic factors that can explain the pattern of each commodity-specific community structure in 2011}, the most striking of which concerns concentration in their size distributions (see Figure \ref{fig:sizedistr_2011} in \ref{app:add_tabs_figs} for the case of year 2011). The most concentrated community structures are those of soybeans, palm oil, poultry meat and nuts, whereas rice exhibits the most homogeneous size distribution.\footnote{This result is confirmed when one computes the Herfindahl concentration index (see description that follows).}
 
\begin{figure}[h!]
 \centering
    {\includegraphics[width=8cm]{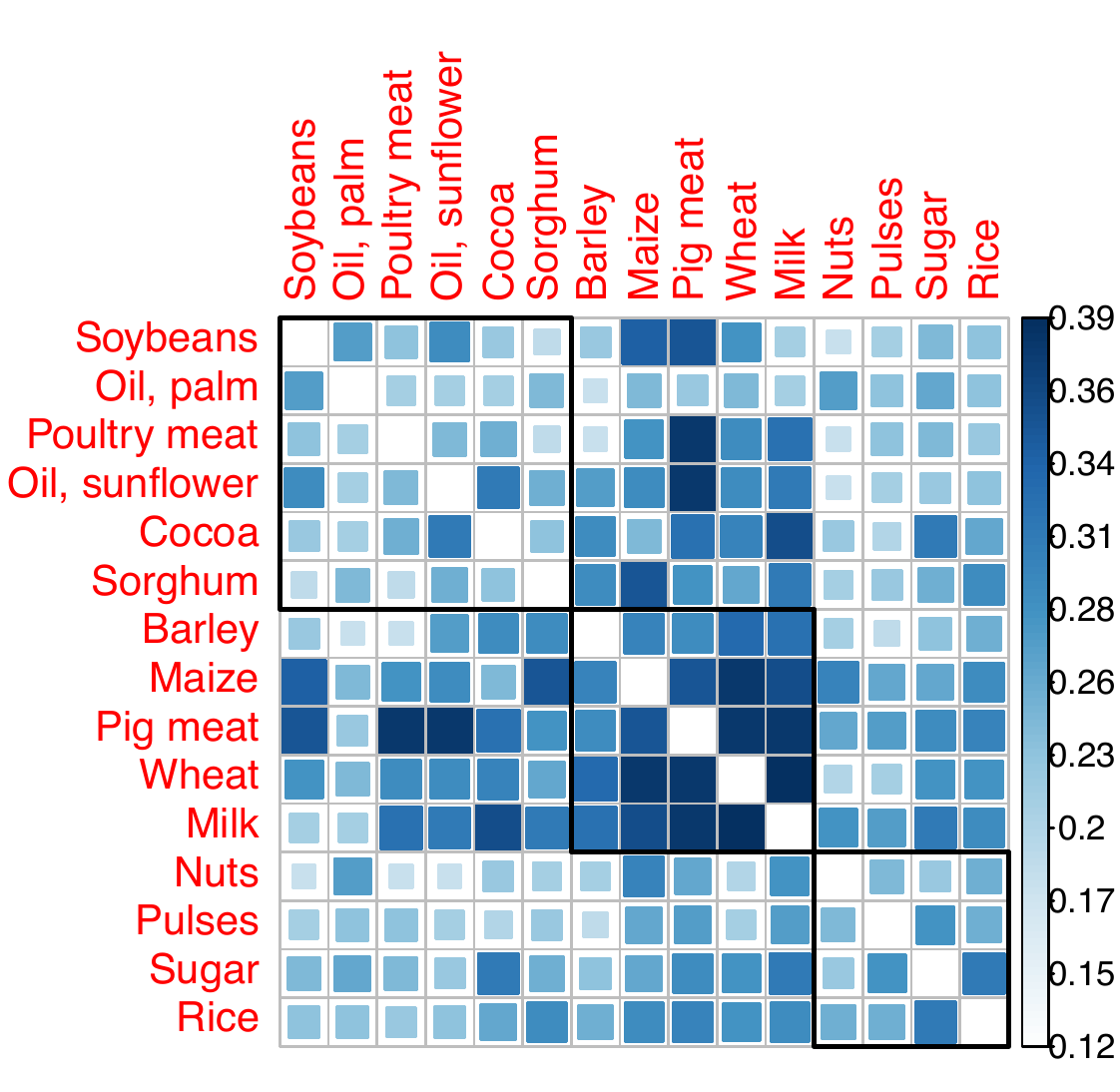}}
 \caption{\label{fig:nmi_2011}Normalized mutual information (NMI) index in year=2011. Higher values of the index suggest that the two community structures are similar. Commodities have been ordered using a (Ward) hierarchical clustering. Squares identify clusters.}
\end{figure}

Similarities and differences among community structures can be better appreciated computing the normalized mutual information (NMI) index between pairs of community structures (see Figure \ref{fig:nmi_2011} and \ref{app:nmi} for details). The NMI index ranges between 0 and 1 and increases the more the two community structures are similar. Three groups of commodities can be identified (outlined by the three squares in the figure). The first one comprises the most similar structures, i.e. coarse grains (barley, maize, wheat), pig meat and milk. The other two consist of commodities that exhibit quite different trading blocs, and differ from the other groups. These are (i) nuts, pulses, sugar and rice; (ii) soybeans, poultry meat, oil, cocoa and sorghum. Note that pig and poultry meat are very similar in terms of their community structures but belong to different groups.

We now explore whether community structures have changed from 2001 to 2011. Figure \ref{fig:choro_2001} in \ref{app:add_tabs_figs} shows, for a few commodities, country community membership in 2001. A qualitative comparison with Figure \ref{fig:choro_2011} shows that in 2011 the European trading bloc became larger, possibly due the Eastern enlargement of the Union (from 15 to 27 members). This evidence is particularly strong in the case of wheat, maize, sugar, rice, palm oil and cocoa, whereas holds to a lesser extent for barley, milk, pulses and poultry meat. Overall, this may be interpret as a first evidence of the effectiveness of the Common Agricultural Policy (CAP) of the European Union. Furthermore, comparing 2001 and 2011 maps reveals an increasing influence of  Brazil, Russia, India and China (i.e., the BRIC countries) in the African continent. This evidence may be partly explained by the increasing hegemony of Russia and India  in Eastern Africa, which has gradually undermined that of Australia in wheat and rice trade. Similarly, maps seem to be coherent with the increasing importance that Brazil gained as maize supplier in African and Middle Eastern countries, at the expense of the Northern American and the European clusters.

More generally, community structures in 2001 differ from those in 2011 because the size distributions of the latter are typically more concentrated. Figure \ref{fig:h_index_2001_2011} in \ref{app:add_tabs_figs} plots the normalized Herfindahl concentration index computed in 2001 and 2011 for all commodity networks (expect cassava) and shows that the lion's share of layers lie above the main diagonal. Rice, soybeans, poultry meat and sunflower oil display the largest increase in concentration. A more concentrated community structure implies that a larger share of countries belong to existing trading groups. Therefore, increases in H index can be interpreted as a tendency to a more globalized trade network. Notice that increasing concentration levels are not necessary associated with a decrease in the number of detected communities (cf Table \ref{tab:comm_stats_2001_2011}). This suggests that, when detected, increasing concentration levels in community size distributions are attained through country switching among clusters and not due to a reduction in the number of trading blocs.

To delve further into the time dynamics of community structures, we focus on three selected commodities, i.e. wheat, maize and rice. We document how community structure for these three products evolve across the whole time sample (1992-2011). Figure \ref{fig:maize_wheat_rice_1992_2011} plot the time series of community number (left) and maximum modularity (right). Note that in general modularity has been increasing over time, suggesting that the IFTMN, at least in the three layers considered in the figure, has exhibited a stronger and stronger tendency to clusterize into well-defined trading blocs. Furthermore, the three commodities considered have followed quite distinct time patterns as far as the number of detected communities is concerned. Maize trade network has been organizing itself into an increasing number of clusters, whereas the number of trading blocs in the wheat network has decreased and stabilized around four. Finally, the rice network has experiencing a lot of turbulence, oscillating between 6 and 9 trading groups over time.

\subsection{Econometric models\label{subsec:comm_econometrics}}
As visual inspection of Figures \ref{fig:choro_2001} and \ref{fig:choro_2011} shows, community structures in the IFTMN exhibits evident geopolitical and socioeconomic regularities. In order to quantitatively explore this issue, we run a set of probit-regression exercises where we explain the probability that any two countries belong to the same trade bloc as a function of a host of covariates (see Section \ref{subsec:comm_econometrics} and Table \ref{tab:covars}), capturing country-pair (dis)similarity along geographical, economic, social, and political dimensions.    

Covariates employed in the analysis are borrowed from the trade-gravity literature \cite{Anderson_2011}, which suggests that bilateral trade flows typically increase in the importer and exporter market size and income (proxied by country total and per-capita GDP) and decrease the stronger trade frictions. The latter are usually proxied by geographical distance and a number of bilateral indicators (e.g., dummy variables) that control ---among other things--- for whether the importer and the exporter share a border, a common language, a trade agreement, any colonial relationship, and whether they belong to the same geographical macro-area.      

\begin{figure*}[h!]
 \centering
    {\includegraphics[width=\textwidth]{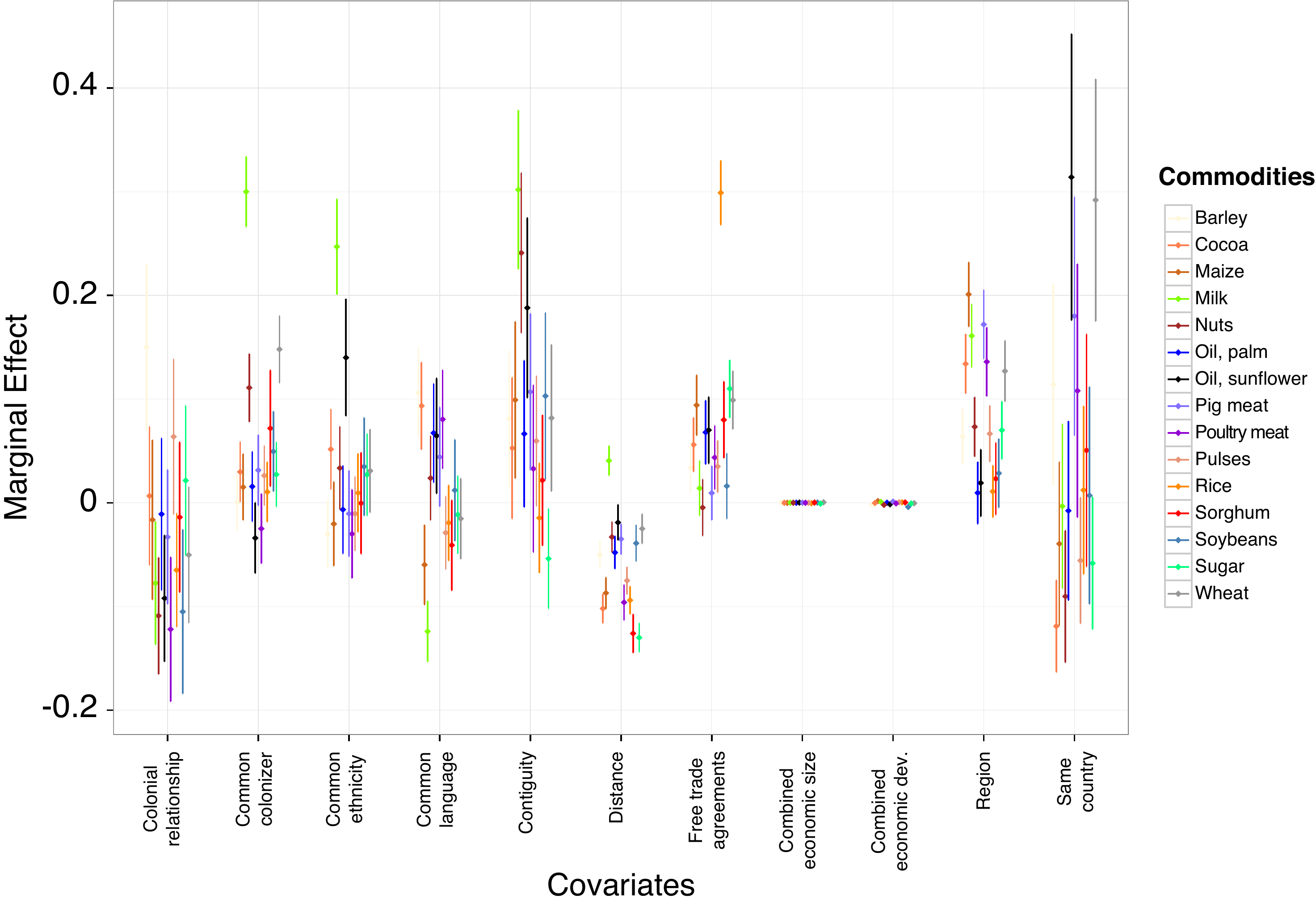}}
 \caption{\label{fig:cs_regs_2011} Probit estimation for year 2011. Marginal effects obtained fitting Eq. (\ref{eq:probit}) to each commodity layer separately using maximum-likelihood. X-axis: covariates used in the model. Y-axis: marginal effect of the covariate on the probability that two countries belong to the same community. Dots represent the point estimate of marginal effects and bars are 95\% confidence intervals.}
\end{figure*}

We begin by fitting Eq. (\ref{eq:probit}) cross-sectionally to year 2001 and year 2011, for all commodity layers. Results for year 2011 are visually presented in Figure \ref{fig:cs_regs_2011}, where point estimates of marginal effects of covariates are plotted together with their 95\% confidence intervals for all commodities (see Figure \ref{fig:cs_regs_2001} in \ref{app:add_tabs_figs} for year 2001)\footnote{All models turn out to be nicely specified according to standard goodness-of-fit tests, e.g., the Akaike information criterion (AIC).}.   
  
Our findings indicate that distance has a negative and statistically significant impact on the probability that two countries belong to the same trade community, for all products considered (but milk). Other geographically-related covariates such as contiguity and regional membership have a product-specific effect, both in terms of significance and sign, notwithstanding they generally boost the co-presence of country pairs in the same trade bloc. Furthermore, free-trade agreements almost always promote co-presence, and their importance has become higher in 2011 as compared to 2001. The role of past colonial relationships and common language is instead less relevant in explaining joint membership. Most importantly, regressions suggest that economic indicators, i.e. absolute and per-capita GDP, are not significant either in statistical and in economic terms, because of too high standard errors and too small marginal effects. 

These results are confirmed by panel-data exercises run for the cases of wheat, rice and maize. We regress co-presence probabilities against the same set of covariates used in the cross-section setup, but now employing the entire time sample in a dynamic fashion, and controlling for common trends and country-specific unobserved heterogeneity with an appropriate use of dummy variables. Again, as Figure \ref{fig:panel_regs} shows, distance and free trade agreements\footnote{More precisely, the EU27 trade agreement and NAFTA seem to strongly affect co-presence probabilities, as well as AFTA for maize and EFTA for wheat.} are two important determinants of the co-presence of country pairs in the same trade community, whereas economic factors are almost not significant ---and their impact is very weak if they are.

Overall, our econometric estimates are in line with the trade-gravity literature, as they show that distance, trade frictions and trade agreements are important determinants of country co-presence in trade communities as they are for bilateral trade flows. However, they strongly depart from traditional gravity exercises as they indicate a very weak impact of country economic size and income in shaping food-trade blocs, whereas it is well known that these two covariates explain to a great extent the intensive margins of aggregate trade \cite{Anderson_2011}\footnote{Country GDP and, in particular, country per capita GDP are not only significant determinants of aggregate bilateral trade in general, but also of staple-food specific bilateral trade flows. To double check that this is the case, we have run a set of standard gravity models where the dependent variable is bilateral trade for our set of staple-food commodities and covariates are as in all our exercises above, finding that country economic size and income are in general much more statistically and economically significant than they are when the dependent variable is country co-membership in food-trade communities.}.

We suggest that such a mismatch with trade gravity results may partly depend on the fundamental difference existing between the dependent variable in gravity exercises and in those explaining country co-membership in trade communities. Whereas in the former the dependent variable mostly concerns a bilateral relationship, in the latter the dependent variable refers to co-presence in a group of countries, and therefore is mostly about a multi-lateral relationship. Therefore, regional and trade-policy variables that describe bilateral relationship in a multi-lateral setup (e.g. regional trade agreements or geographic positioning) may better explain co-presence of countries in trading blocs. At the same time, the differences between our exercises and traditional gravity models suggest that community detection techniques are really able to statistically elicit multi-lateral relationship among countries, even they start from fundamentally bilateral trade relationships among pairs of countries.         

\subsection{Multi-layer community detection\label{subsec:comm_multi}}
In the last subsection, we have performed a community-detection analysis assuming that the IFTMN consists of independent layers in each time period. Here, we ask what communities look like if they can span across layers. More precisely, we suppose that each country is coupled with itself across commodity slices. Therefore, in each year, the IFTMN becomes a multi-layer network, where nodes are country-commodity pairs. Identifying communities in such an object means finding clusters where countries and commodities can possibly repeat themselves many times: the same country (respectively, commodity) may belong to different clusters as it can appear coupled with different commodities (respectively, countries). 

A first question that naturally arises is whether projecting communities into the space of commodities results in country clusters that are similar to those obtained assuming that the IFTMN consists of independent layers. Of course, communities now span over commodity layers. Therefore, this exercise must be just intended as a robustness check as it entails loosing a lot of information. Figure \ref{app:add_tabs_figs} in \ref{app:add_tabs_figs} shows NMI values when comparing community structures in the multi-layer and in the independent-layer cases, for year 2001 and year 2011. NMI values appear to be quite high, especially in year 2011, where for most products communities in the multi-layer become more similar to the independent-layer case. The fact that results previously obtained in the independent-layer case are in general robust to a multi-layer representation can be visually appreciated looking at choropleth maps of projections of multi-layer communities into the space of commodities, see Figure \ref{fig:choro_ml} for the cases of wheat, rice and maize (and the correspondent maps in Figure \ref{fig:choro_2011} and Figure \ref{fig:choro_2001}).   

\begin{figure*}[h!]
 \centering
    {\includegraphics[width=\textwidth]{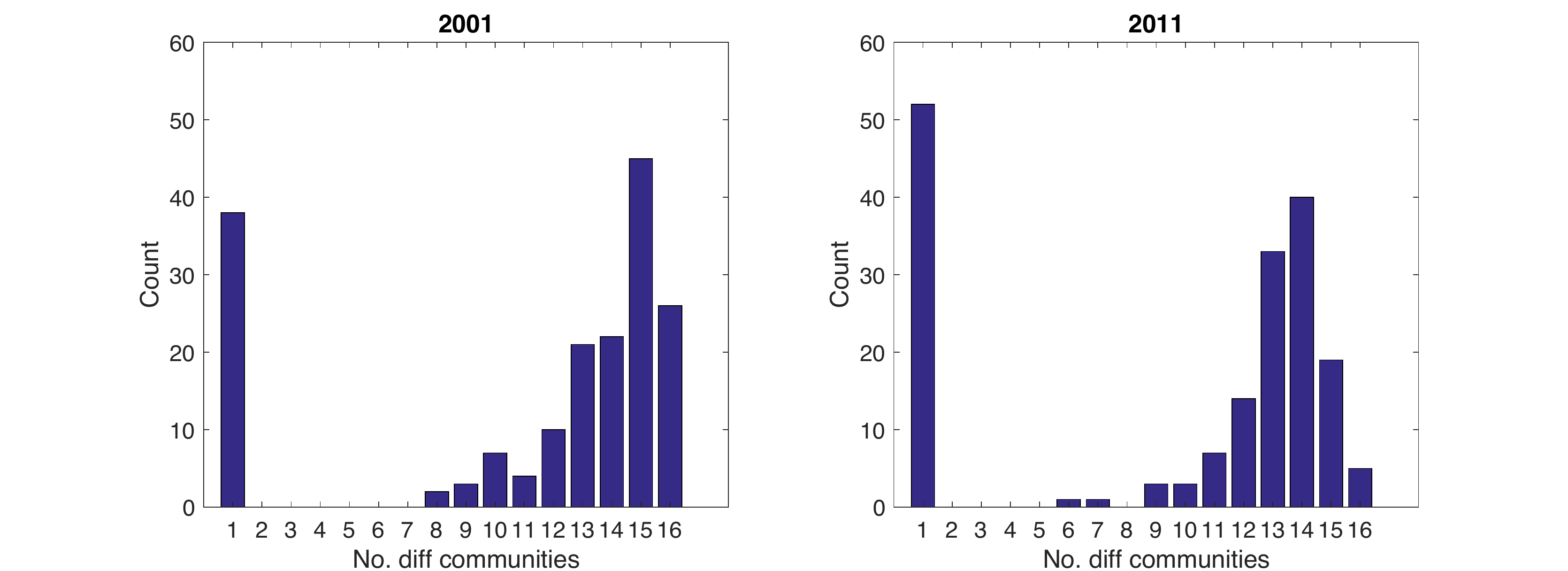}}
 \caption{\label{fig:ml_no_diff_comm_2001_2011}Multilayer community detection. Distribution of the number of different communities a country belongs to in the multi-layer. Years 2001 and 2011.}
\end{figure*}

A second interesting issue concerns exploring the shape of clusters in the multi network. To do so, we begin by studying the distribution of the number of different communities a country belongs to, which we interpret as a rough measure of country diversification in the IFTMN. The intuition is that a country belonging to a small number of different communities tends to be mostly connected with instances of ``itself'' in different commodity layers and therefore depends on the same group of other country-commodity pairs for all possible staple-food products it trades. Conversely, if a country appears in a large number of different communities in the multi-network (and thus is never isolated) then it relies on several different clusters of country-product pairs depending on the specific product it trades. As we show in Figure \ref{fig:ml_no_diff_comm_2001_2011}, the frequency distribution of this statistics are markedly bi-modal, with a peak at 1 and another peak around 14-15. This suggests that community structures in the multi-layer are polarized into two groups. The first one consists of countries that irrespective of the commodity traded always belong to the same community in the multilayer. These are countries that are poorly diversified and are the least networked in the food-trade system. Countries in the second group belong instead to several different communities depending on the commodity traded and therefore are highly diversified in the multilayer. This finding is relevant for food-security issues as it suggests that countries belonging to the first group may be more vulnerable than those in the second group to shocks that put at risk the supply of one or more food commodities. 

\begin{figure*}
 \centering
 \subfigure[2001]
   {\includegraphics[width=0.8\textwidth]{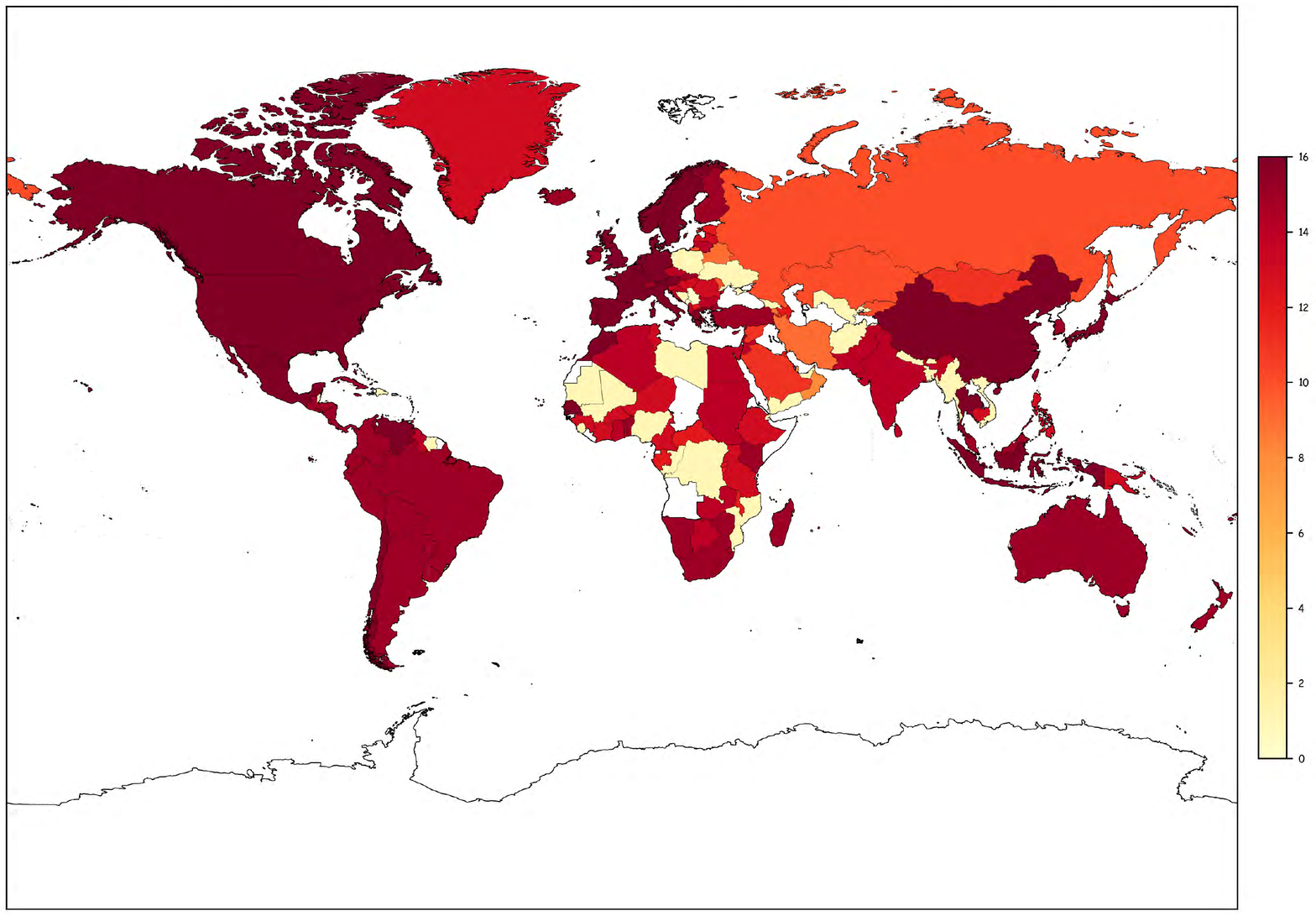}}
 \hspace{5mm}
 \subfigure[2011]
   {\includegraphics[width=0.8\textwidth]{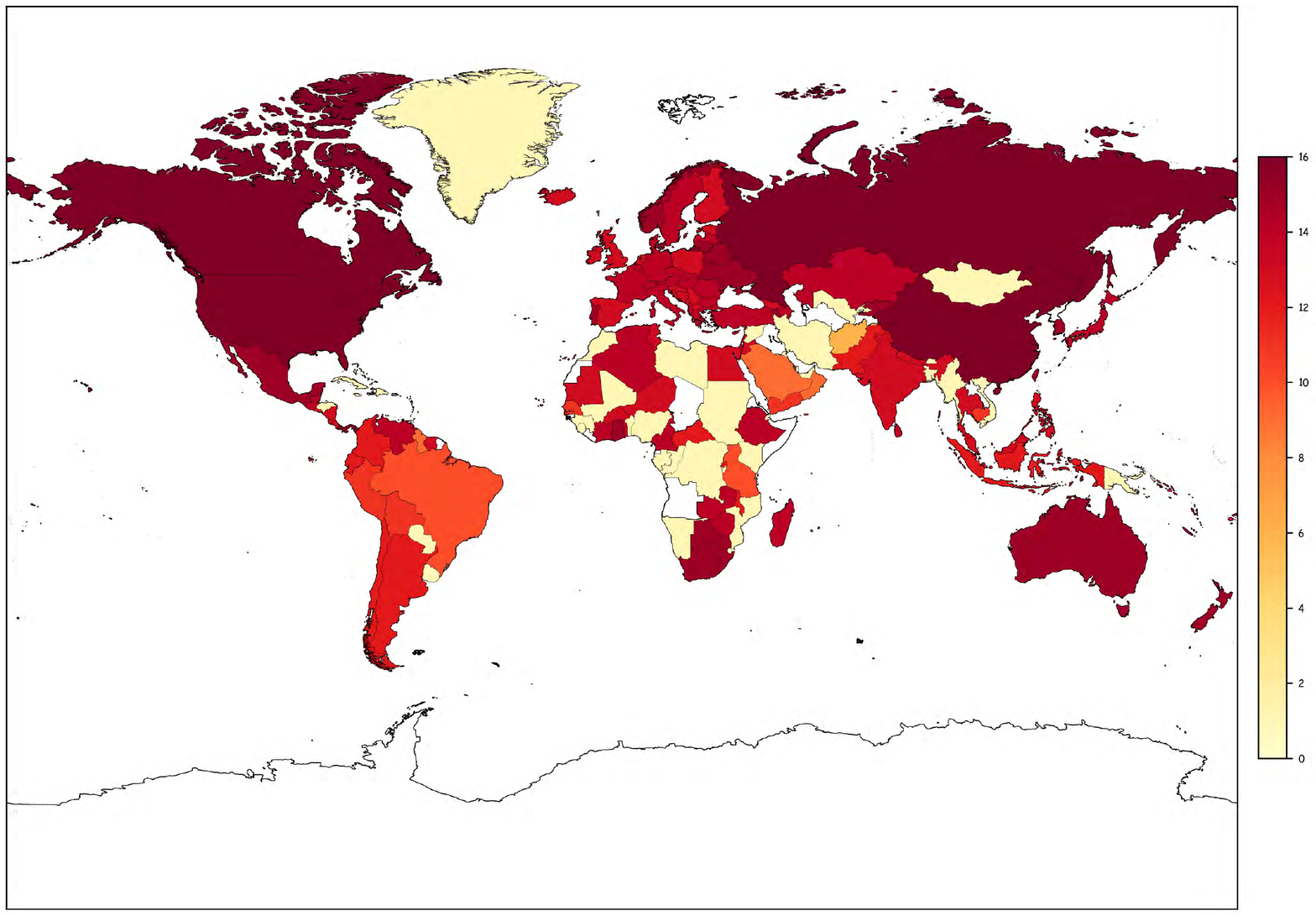}}
 \caption{\label{fig:ml_choro_diff_comm_2001_2011}Choropleth maps for the number of different communities a country belongs to in the multi-layer. Years 2001 and 2011.}
\end{figure*}

The geographical distribution of the two groups of countries is depicted in Figure \ref{fig:ml_choro_diff_comm_2001_2011} in \ref{app:add_tabs_figs}. Notice how the first group is mostly located in Africa, but also features countries in the Middle and Far East. 

\section{Discussion and Conclusions\label{sec:conclusions}}
The topology of the international food trade multi-network -- particularly its community structure -- is key to understanding how major disruptions or ``shocks'' will impact the global food system.  We find that the individual layers of this network have densely connected trading groups, a consistent characteristic over the period 1994 to 2011.  This community structure fundamentally affects how a shock would spread from country to country within the global food system.  If, for example, the epicenter of a shock is within a community, we would expect that countries in this community would face a two-fold challenge: 1) reduced supply from domestic production and/or from their usual import partners and 2) high international prices. To the extent possible, governments and companies within these countries would adjust their procurement strategies to find new sources from members of the other trading communities. Outside of the epicenter community, network characteristics like inter-community connectivity and other global dynamics like trade interventions would be critically important.  

One straightforward application of the knowledge generated from understanding commodity specific community structures is that we can improve our understanding of potential vulnerabilities to various disruption scenarios.  First let us consider a major disruption to rice production. In a scenario where China experiences a major negative production shock, how would the community structure of the rice network modulate global impacts? China would look to the international markets to make up for any shortfall that its food reserve system could not handle. Four of the top five exporters -- Thailand, Vietnam, India and Pakistan -- are co-located in Asia, where Thailand is in the same community as China, Vietnam is part of a predominantly Southeast Asian community, and India and Pakistan are both in another community. Therefore, the burden of making up for the Chinese production shortfall would fall primarily on Asian countries, with perhaps the US also contributing (considering that it is the fifth largest rice exporters). Countries like those in western Africa (e.g., Ghana and Ivory Coast) would be highly vulnerable, as they are part of the same community as China (Figure 4) and would face the task of competing with China on the global rice markets.  International rice prices would increase, assuming that rice production does not increase substantially elsewhere, there is no major release of rice reserves to the international markets (e.g., as Japan did in 2008), and that there major changes to the other global grain markets. In this situation, low- and lower-middle-income countries that are dependent on imports for their staple food supply will be at a severe disadvantage.

The community structure of the soybean network is quite different from the structure of the rice network (Figure 5), so we might expect \textit{a priori} that there are differences in shock vulnerability.  The soybean network reveals one of the most concentrated community structure, composed by only three large clusters without a clear regional scheme (Figure 4). The most important bloc -- in terms of trade volume -- includes the US and Brazil from the producing and exporting side, which together account for over 70\% of global soybean exports, and China from the importing side, which alone accounts for 56\% of global soybeans imports. If one of these main producers experiences a sharp decline in production, the global implications of the shock will largely depend on the capacity of few other major producing countries to make up for the production shortfall.  

The global wheat market has a community structure that falls in-between the structures found in the rice and soybean markets. Major producers are grouped together in three separate communities: 1) the US, Canada, and Australia, 2) Argentina and Brazil, 3) Russia and Ukraine. Interestingly, Europe belongs to yet another separate cluster, in which France is the notable producer and exporter.  One might hypothesize that this geographic diversity is advantageous for dealing with a disruption, particularly if it has as spatial component (e.g., crop disease spreading over an area, a regional conflict, or regional-scale extreme weather). Of course, community structure alone is not sufficient for understanding the impacts of shocks on these global markets.

Knowledge of community structure can be linked to the latest efforts to understand non-equilibrium conditions in the global food system.  For example, recent models of food shock propagation \cite{Puma_etal_2015,gephart2016vulnerability,marchand2016reserves} would benefit from these community-structure insights.  Improved disruption scenarios can be generated to analyze potential responses and identify vulnerabilities of the food system, at scales ranging from the individual country to the global system.  

Food reserves are increasingly seen as an essential variable that influences how shock would propagate through a trade network \cite{marchand2016reserves}. Additionally, a recent analysis showed that a simply supply-demand model with food-reserve dynamics and trade policies can explain most of the observed variations in global cereal prices over the last 40 years solely, including the most recent price peaks in 2007/08 and 2010/11 \cite{schewe2017role}. The importance of food reserves and trade policies -- particularly changes in policies when markets are out-of-equilibrium -- is connected to community structures in the markets. A natural extension is to explore the interplay among communities, food reserves, and trade policies. Market dynamics including panic buying, hoarding, and large-scale governmental intervention are poorly understood, but we should expect that community structures would play a significant role. Likewise, we might expect that country-level policy decisions on the balance between self sufficiency and import dependency in food production would be influenced by how one's country is connected to others.

More generally, the role of food price shocks in shaping the community structure of global food-trade system should be better understood \cite{Headey_2011,Anderson_Nelgen_2012}. Food price shocks can alter global trade patterns as they typically encourage countries both to rise export barriers and to lower import tariffs, which may in turn exacerbate price spikes. Such protectionist measures are often combined with other frequent responses such as panic buying, large-scale governmental intervention, hoarding and precautionary purchase. These common short-term remedies associated with price spikes are poorly understood although they may have pervasive consequences on less developed countries, generally extremely dependent on imports, thus altering the way in which they locally form their trade networks.     

Along similar lines, one may investigate more deeply the importance of other determinants of bilateral import-export flows in explaining the formation of clusters in the international web of food trade. For example, exchange rate volatility has grown significantly after the GFC. This can correlate with trade growth, as typically the more a country undergoes currency devaluation, the slower the growth in its trade \cite{Kang_2016}. Other determinants to be explored include climate-related shocks, which are especially relevant because of crop sensitivity to weather extremes \cite{Gornall_2010,Battisti_etal_2009}, regional conflicts, epidemics, agro-terrorism and crop pests \cite{McCloskey_etal_2014}. 

From a more methodological perspective, this study could be improved by additional tests aimed at checking the robustness of the main results against alternative parameterizations of (and assumptions about) the community-detection algorithms employed. For example, the well-known resolution-limit bias affecting many existing methods may be explored using the multiple-resolution community detection strategy by introduced in Ref. \cite{Arenas_etal_2008}. Furthermore, despite the fact that the foregoing analysis was focused on the identification of non-overlapping communities, this work can be extended using community-detection algorithms that look for clusters that may partly overlap \cite{Nicosia_etal_2009,Xie_etal_2013}. This is important, as knowing the degree of overlap among communities may shed more light on the way in which food crises may spread across clusters. Finally, when analyzing the IFTMN as a multi-layer network, we have implicitly assumed that any pair of layers are linked by fictional edges connecting the same country in the two layers, and that the weights of this edge are homogeneous across countries and equal to one. Such a system parameter, however, may affect the emerging community structure \cite{Mucha_etal_2010}. Therefore, experimenting with different values of such a parameter can give interesting insights on the emergence of clusters in the product-country space.

\ack Giorgio Fagiolo gratefully acknowledges support by the European Union's Horizon 2020 research and innovation program under grant agreement No. 649186 - ISIGrowth. M.J. Puma gratefully acknowledges fellowship support from the Columbia University Center for Climate and Life.


\section*{References}


\begin{thebibliography}{10}
\expandafter\ifx\csname url\endcsname\relax
  \def\url#1{{\tt #1}}\fi
\expandafter\ifx\csname urlprefix\endcsname\relax\def\urlprefix{URL }\fi
\providecommand{\eprint}[2][]{\url{#2}}

\bibitem{Porkka_etal_2013}
Porkka M, Kummu M, Siebert S and Varis O 2013 {\em PLOS ONE\/} {\bf 8} 1--12
  \urlprefix\url{https://doi.org/10.1371/journal.pone.0082714}

\bibitem{UN_2015}
{United Nations} 2015 Transforming our world: the 2030 agenda for sustainable
  development Tech. Rep. A/RES/70/1 UN General Assembly
  \urlprefix\url{https://sustainabledevelopment.un.org/resourcelibrary}

\bibitem{DOdorico_etal_2014}
{D'Odorico} P, Carr J~A, Laio F, Ridolfi L and Vandoni S 2014 {\em Earth's
  Future\/} {\bf 2} 458--469 ISSN 2328-4277 2014EF000250
  \urlprefix\url{http://dx.doi.org/10.1002/2014EF000250}

\bibitem{Godfray_etal_2010}
Godfray H~C~J, Beddington J~R, Crute I~R, Haddad L, Lawrence D, Muir J~F,
  Pretty J, Robinson S, Thomas S~M and Toulmin C 2010 {\em Science\/} {\bf 327}
  812--818 ISSN 0036-8075 (\textit{Preprint}
  \eprint{http://science.sciencemag.org/content/327/5967/812.full.pdf})
  \urlprefix\url{http://science.sciencemag.org/content/327/5967/812}

\bibitem{UN_2015_pop}
{United Nations} 2015 World population prospects: The 2015 revision, key
  findings and advance tables Tech. Rep. ESA/P/WP.241 Department of Economic
  and Social Affairs, Population Division
  \urlprefix\url{https://esa.un.org/unpd/wpp/Publications/}

\bibitem{Hazell_Wood_2008}
Hazell P and Wood S 2008 {\em Philos Trans R Soc Lond B Biol Sci\/} {\bf 363}
  495--515 ISSN 0962-8436 rstb20072166[PII]
  \urlprefix\url{http://www.ncbi.nlm.nih.gov/pmc/articles/PMC2610166/}

\bibitem{Hanjra_Qureshi_2010}
Hanjra M~A and Qureshi M~E 2010 {\em Food Policy\/} {\bf 35} 365 -- 377 ISSN
  0306-9192
  \urlprefix\url{http://www.sciencedirect.com/science/article/pii/S030691921000059X}

\bibitem{Woods_etal_2010}
Woods J, Williams A, Hughes J~K, Black M and Murphy R 2010 {\em Philosophical
  Transactions of the Royal Society of London B: Biological Sciences\/} {\bf
  365} 2991--3006 ISSN 0962-8436 (\textit{Preprint}
  \eprint{http://rstb.royalsocietypublishing.org/content/365/1554/2991.full.pdf})
  \urlprefix\url{http://rstb.royalsocietypublishing.org/content/365/1554/2991}

\bibitem{Coumou_Rahmstorf_2012}
Coumou D and Rahmstorf S 2012 {\em Nature Clim. Change\/} {\bf 2} 491--496 ISSN
  1758-678X \urlprefix\url{http://dx.doi.org/10.1038/nclimate1452}

\bibitem{Battisti_etal_2009}
Battisti D~S and Naylor R~L 2009 {\em Science\/} {\bf 323} 240--244 ISSN
  0036-8075 (\textit{Preprint}
  \eprint{http://science.sciencemag.org/content/323/5911/240.full.pdf})
  \urlprefix\url{http://science.sciencemag.org/content/323/5911/240}

\bibitem{Gornall_2010}
Gornall J, Betts R, Burke E, Clark R, Camp J, Willett K and Wiltshire A 2010
  {\em Philosophical Transactions of the Royal Society of London B: Biological
  Sciences\/} {\bf 365} 2973--2989 ISSN 0962-8436 (\textit{Preprint}
  \eprint{http://rstb.royalsocietypublishing.org/content/365/1554/2973.full.pdf})
  \urlprefix\url{http://rstb.royalsocietypublishing.org/content/365/1554/2973}

\bibitem{McCloskey_etal_2014}
McCloskey B, Dar O, Zumla A and Heymann D~L 2014 {\em The Lancet Infectious
  Diseases\/} {\bf 14} 1001--1010 ISSN 1473-3099
  \urlprefix\url{http://dx.doi.org/10.1016/S1473-3099(14)70846-1}

\bibitem{Nonhebel_Kastner_2011}
Nonhebel S and Kastner T 2011 {\em Livestock Science\/} {\bf 139} 3 -- 10 ISSN
  1871-1413 special Issue: Assessment for Sustainable Development of Animal
  Production Systems
  \urlprefix\url{http://www.sciencedirect.com/science/article/pii/S1871141311001041}

\bibitem{Tilman_etsl_2011}
Tilman D, Balzer C, Hill J and Befort B~L 2011 {\em Proceedings of the National
  Academy of Sciences\/} {\bf 108} 20260--20264 (\textit{Preprint}
  \eprint{http://www.pnas.org/content/108/50/20260.full.pdf})
  \urlprefix\url{http://www.pnas.org/content/108/50/20260.abstract}

\bibitem{Cassidy_etal_2013}
Cassidy E~S, West P~C, Gerber J~S and Foley J~A 2013 {\em Environmental
  Research Letters\/} {\bf 8} 034015
  \urlprefix\url{http://stacks.iop.org/1748-9326/8/i=3/a=034015}

\bibitem{Clapp_2009_book}
Clapp J and Cohen M~J 2009 {\em The global food crisis: Governance challenges
  and opportunities\/} (Wilfrid Laurier Univ. Press)

\bibitem{Clapp_2015}
Clapp J 2015 Food security and trade: Unpacking disputed narratives Tech. rep.
  Food and Agriculture Organization of the United Nations,Rome
  \urlprefix\url{http://www.fao.org/3/a-i5160e.pdf}

\bibitem{Puma_etal_2015}
Puma M~J, Bose S, Chon S~Y and Cook B~I 2015 {\em Environmental Research
  Letters\/} {\bf 10} 024007
  \urlprefix\url{http://stacks.iop.org/1748-9326/10/i=2/a=024007}

\bibitem{Fagiolo2009pre}
Fagiolo G, Schiavo S and Reyes J 2009 {\em Physical Review E\/} {\bf 79} 036115

\bibitem{Lee_etal_plosone_2011}
Lee K~M, Yang J~S, Kim G, Lee J, Goh K~I and Kim I~m 2011 {\em PLoS ONE\/} {\bf
  6} e18443 \urlprefix\url{http://dx.doi.org/10.1371%2Fjournal.pone.0018443}

\bibitem{Amour_etal_2016}
{d'Amour} C~B, Wenz L, Kalkuhl M, Steckel J~C and Creutzig F 2016 {\em
  Environmental Research Letters\/} {\bf 11} 035007
  \urlprefix\url{http://stacks.iop.org/1748-9326/11/i=3/a=035007}

\bibitem{Haldane_May_2011}
Haldane A~G and May R~M 2011 {\em Nature\/} {\bf 469} 351--355

\bibitem{Fagiolo_survey_2017}
Fagiolo G {\em The International Trade Network: Empirics and Modeling\/}
  chap~28

\bibitem{Barigozzi_etal_2010}
Barigozzi M, Fagiolo G and Garlaschelli D 2010 {\em Phys. Rev. E\/} {\bf 81}(4)
  046104 \urlprefix\url{https://link.aps.org/doi/10.1103/PhysRevE.81.046104}

\bibitem{Barigozzi_etal_2011}
Barigozzi M, Fagiolo G and Mangioni G 2011 {\em Physica A: Statistical
  Mechanics and its Applications\/} {\bf 390} 2051 -- 2066 ISSN 0378-4371
  \urlprefix\url{http://www.sciencedirect.com/science/article/pii/S0378437111001129}

\bibitem{Brooks_etal_2013}
Brooks D~H, Ferrarini B and Go E~C 2013 {\em Journal of International Commerce,
  Economics and Policy\/} {\bf 04} 1350015 (\textit{Preprint}
  \eprint{http://www.worldscientific.com/doi/pdf/10.1142/S1793993313500154})
  \urlprefix\url{http://www.worldscientific.com/doi/abs/10.1142/S1793993313500154}

\bibitem{Fracasso_etal_2016}
Fracasso A, Sartori M and Schiavo S 2016 {\em Science of The Total
  Environment\/} {\bf 543, Part B} 1054 -- 1062 ISSN 0048-9697 special Issue on
  Climate Change, Water and Security in the Mediterranean
  \urlprefix\url{http://www.sciencedirect.com/science/article/pii/S0048969715002028}

\bibitem{Gephart_Pace_2015}
Gephart J~A and Pace M~L 2015 {\em Environmental Research Letters\/} {\bf 10}
  125014 \urlprefix\url{http://stacks.iop.org/1748-9326/10/i=12/a=125014}

\bibitem{Wu_Guclu_2013}
Wu F and Guclu H 2013 {\em Risk Analysis\/} {\bf 33} 2168--2178 ISSN 1539-6924
  \urlprefix\url{http://dx.doi.org/10.1111/risa.12064}

\bibitem{Battiston_etal_2014}
Battiston F, Nicosia V and Latora V 2014 {\em Phys. Rev. E\/} {\bf 89}(3)
  032804 \urlprefix\url{https://link.aps.org/doi/10.1103/PhysRevE.89.032804}

\bibitem{Kivela_etal_2014}
Kivel\"{a} M, Arenas A, Barthelemy M, Gleeson J~P, Moreno Y and Porter M~A 2014
  {\em Journal of Complex Networks\/} {\bf 2} 203 (\textit{Preprint}
  \eprint{/oup/backfile/content_public/journal/comnet/2/3/10.1093_comnet_cnu016/2/cnu016.pdf})
  \urlprefix\url{+ http://dx.doi.org/10.1093/comnet/cnu016}

\bibitem{Boccaletti_etal_2014}
Boccaletti S, Bianconi G, Criado R, del Genio C, Gomez-Gardenes J, Romance M,
  Sendina-Nadal I, Wang Z and Zanin M 2014 {\em Physics Reports\/} {\bf 544} 1
  -- 122 ISSN 0370-1573 the structure and dynamics of multilayer networks
  \urlprefix\url{http://www.sciencedirect.com/science/article/pii/S0370157314002105}

\bibitem{Fortunato_2010}
Fortunato S 2010 {\em Physics Reports\/} {\bf 486} 75 -- 174 ISSN 0370-1573
  \urlprefix\url{http://www.sciencedirect.com/science/article/pii/S0370157309002841}

\bibitem{Konar_etal_2011}
Konar M, Dalin C, Suweis S, Hanasaki N, Rinaldo A and Rodriguez-Iturbe I 2011
  {\em Water Resources Research\/} {\bf 47} n/a--n/a ISSN 1944-7973 w05520
  \urlprefix\url{http://dx.doi.org/10.1029/2010WR010307}

\bibitem{Sartori_Schiavo_2014}
Sartori M and Schiavo S 2014 Virtual water trade and country vulnerability: A
  network perspective Tech. Rep.~77
  \urlprefix\url{https://ssrn.com/abstract=2518934}

\bibitem{Tamea_etal_2014}
Tamea S, Carr J~A, Laio F and Ridolfi L 2014 {\em Water Resources Research\/}
  {\bf 50} 17--28 ISSN 1944-7973
  \urlprefix\url{http://dx.doi.org/10.1002/2013WR014707}

\bibitem{billen2014biogeochemical}
Billen G, Lassaletta L and Garnier J 2014 {\em Global Food Security\/} {\bf 3}
  209--219

\bibitem{MacDonald_etal_2015}
MacDonald G~K, Brauman K~A, Sun S, Carlson K~M, Cassidy E~S, Gerber J~S and
  West P~C 2015 {\em BioScience\/} {\bf 65} 275 (\textit{Preprint}
  \eprint{/oup/backfile/content_public/journal/bioscience/65/3/10.1093_biosci_biu225/1/biu225.pdf})
  \urlprefix\url{+ http://dx.doi.org/10.1093/biosci/biu225}

\bibitem{Porter_etal_2009}
{Porter} M~A, {Onnela} J~P and {Mucha} P~J 2009 {\em ArXiv e-prints\/}
  (\textit{Preprint} \eprint{0902.3788})

\bibitem{Malliaros_etal_2013}
Malliaros F~D and Vazirgiannis M 2013 {\em Physics Reports\/} {\bf 533} 95 --
  142 ISSN 0370-1573 clustering and Community Detection in Directed Networks: A
  Survey
  \urlprefix\url{http://www.sciencedirect.com/science/article/pii/S0370157313002822}

\bibitem{Kim_Lee_2015}
Kim J and Lee J~G 2015 {\em SIGMOD Rec.\/} {\bf 44} 37--48 ISSN 0163-5808
  \urlprefix\url{http://doi.acm.org/10.1145/2854006.2854013}

\bibitem{Newman_Girvan_2004}
Newman M~E~J and Girvan M 2004 {\em Phys. Rev. E\/} {\bf 69} 026113
  \urlprefix\url{http://link.aps.org/doi/10.1103/PhysRevE.69.026113}

\bibitem{Arenas_etal_2007}
Arenas A, Duch J, Fernandez A and G\'omez S 2007 {\em CoRR\/} {\bf
  abs/physics/0702015}
  \urlprefix\url{http://dblp.uni-trier.de/db/journals/corr/corr0702.html#abs-physics-0702015}

\bibitem{Barrat_etal_2004}
Barrat A, Barth\'elemy M, Pastor-Satorras R and Vespignani A 2004 {\em
  Proceedings of the National Academy of Sciences of the United States of
  America\/} {\bf 101} 3747--3752 (\textit{Preprint}
  \eprint{http://www.pnas.org/content/101/11/3747.full.pdf})
  \urlprefix\url{http://www.pnas.org/content/101/11/3747.abstract}

\bibitem{Rotta_Noack_2011}
Rotta R and Noack A 2011 {\em J. Exp. Algorithmics\/} {\bf 16}
  2.3:2.1--2.3:2.27 ISSN 1084-6654
  \urlprefix\url{http://doi.acm.org/10.1145/1963190.1970376}

\bibitem{Blondel_etal_2008_LouvainMeth}
Blondel V~D, Guillaume J~L, Lambiotte R and Lefebvre E 2008 {\em Journal of
  Statistical Mechanics: Theory and Experiment\/} {\bf 2008} P10008
  \urlprefix\url{http://stacks.iop.org/1742-5468/2008/i=10/a=P10008}

\bibitem{Mucha_etal_2010}
Mucha P~J, Richardson T, Macon K, Porter M~A and Onnela J~P 2010 {\em
  Science\/} {\bf 328} 876--878 ISSN 0036-8075 (\textit{Preprint}
  \eprint{http://science.sciencemag.org/content/328/5980/876.full.pdf})
  \urlprefix\url{http://science.sciencemag.org/content/328/5980/876}

\bibitem{Carchiolo2011}
Carchiolo V, Longheu A, Malgeri M and Mangioni G 2011 {\em Communities
  Unfolding in Multislice Networks\/} (Berlin, Heidelberg: Springer Berlin
  Heidelberg) pp 187--195 ISBN 978-3-642-25501-4
  \urlprefix\url{https://doi.org/10.1007/978-3-642-25501-4_19}

\bibitem{muxviz}
De~Domenico M, Porter M~A and Arenas A 2015 {\em Journal of Complex Networks\/}
  {\bf 3} 159 (\textit{Preprint}
  \eprint{/oup/backfile/content_public/journal/comnet/3/2/10.1093/comnet/cnu038/2/cnu038.pdf})
  \urlprefix\url{+ http://dx.doi.org/10.1093/comnet/cnu038}

\bibitem{Winkelmann_2008}
R W 2008 {\em Econometric analysis of count data\/} (Springer, New York)

\bibitem{Baldwin_Taglioni_2006}
Baldwin R and Taglioni D 2006 Gravity for dummies and dummies for gravity
  equations Working Paper 12516 National Bureau of Economic Research
  \urlprefix\url{http://www.nber.org/papers/w12516}

\bibitem{Anderson_2011}
Anderson J~E 2011 {\em Annual Review of Economics\/} {\bf 3} 133--160

\bibitem{gephart2016vulnerability}
Gephart J~A, Rovenskaya E, Dieckmann U, Pace M~L and Br{\"a}nnstr{\"o}m {\AA}
  2016 {\em Environmental Research Letters\/} {\bf 11} 035008

\bibitem{marchand2016reserves}
Marchand P, Carr J~A, Dell?Angelo J, Fader M, Gephart J~A, Kummu M, Magliocca
  N~R, Porkka M, Puma M~J, Ratajczak Z {\em et~al.\/} 2016 {\em Environmental
  Research Letters\/} {\bf 11} 095009

\bibitem{schewe2017role}
Schewe J, Otto C and Frieler K 2017 {\em Environmental Research Letters\/} {\bf
  12} 054005

\bibitem{Headey_2011}
Headey D 2011 {\em Food Policy\/} {\bf 36} 136 -- 146 ISSN 0306-9192
  \urlprefix\url{http://www.sciencedirect.com/science/article/pii/S0306919210001065}

\bibitem{Anderson_Nelgen_2012}
Anderson K and Nelgen S 2012 {\em Oxford Review of Economic Policy\/} {\bf 28}
  235 (\textit{Preprint}
  \eprint{/oup/backfile/content_public/journal/oxrep/28/2/10.1093/oxrep/grs001/2/grs001.pdf})
  \urlprefix\url{+ http://dx.doi.org/10.1093/oxrep/grs001}

\bibitem{Kang_2016}
Kang J~W 2016 International trade and exchange rate Working Paper 498 Asian
  Development Bank
  \urlprefix\url{https://www.adb.org/sites/default/files/publication/202841/ewp-498.pdf}

\bibitem{Arenas_etal_2008}
Arenas A, Fernandez A and Gomez S 2008 {\em New Journal of Physics\/} {\bf 10}
  053039 \urlprefix\url{http://stacks.iop.org/1367-2630/10/i=5/a=053039}

\bibitem{Nicosia_etal_2009}
Nicosia V, Mangioni G, Carchiolo V and Malgeri M 2009 {\em Journal of
  Statistical Mechanics: Theory and Experiment\/} {\bf 2009} P03024
  \urlprefix\url{http://stacks.iop.org/1742-5468/2009/i=03/a=P03024}

\bibitem{Xie_etal_2013}
Xie J, Kelley S and Szymanski B~K 2013 {\em ACM Comput. Surv.\/} {\bf 45}
  43:1--43:35 ISSN 0360-0300
  \urlprefix\url{http://doi.acm.org/10.1145/2501654.2501657}

\bibitem{Danon_etal_2005}
Danon L, Diaz-Guilera A, Duch J and Arenas A 2005 {\em Journal of Statistical
  Mechanics: Theory and Experiment\/} {\bf 2005} P09008
  \urlprefix\url{http://stacks.iop.org/1742-5468/2005/i=09/a=P09008}

\bibitem{DeSousa_Lochard_2011}
{de Sousa} J and Lochard J 2011 {\em The Scandinavian Journal of Economics\/}
  {\bf 113} 553--578 ISSN 1467--9442
  \urlprefix\url{http://dx.doi.org/10.1111/j.1467-9442.2011.01656.x}

\bibitem{Fagiolo2006EcoBull}
Fagiolo G 2006 {\em Economics Bulletin\/} {\bf 3} 1--12

\bibitem{Freeman_1978}
Freeman L~C 1978 {\em Social Networks\/} {\bf 1} 215 -- 239 ISSN 0378-8733
  \urlprefix\url{http://www.sciencedirect.com/science/article/pii/0378873378900217}

\bibitem{Fagiolo2007pre}
Fagiolo G 2007 {\em Physical Review E\/} {\bf 76} 026107

\end{thebibliography}

\providecommand{\newblock}{}


%


\newpage \cleardoublepage

\noindent \textbf{Supplemental Materials}

\appendix

\setcounter{section}{0}

\section{List of Countries\label{app:countries}}
Table \ref{tab:countries} lists the countries used in our analysis with their ISO3 Code.

\begin{table*}
\caption{\label{tab:countries}List of countries used in the analysis.}
\begin{indented}
\item[]\begin{tabular}{@{}lclc}
\mr
  Country & ISO3 & Country & ISO3 \\ 
\br	
Afghanistan	&	AFG	&	Lebanon	&	LBN	\\
Albania	&	ALB	&	Libya	&	LBY	\\
Algeria	&	DZA	&	Lithuania	&	LTU	\\
Antigua and Barbuda	&	ATG	&	Luxembourg	&	LUX	\\
Argentina	&	ARG	&	Macao	&	MAC	\\
Armenia	&	ARM	&	Macedonia	&	MKD	\\
Aruba	&	ABW	&	Madagascar	&	MDG	\\
Australia	&	AUS	&	Malawi	&	MWI	\\
Austria	&	AUT	&	Malaysia	&	MYS	\\
Azerbaijan	&	AZE	&	Maldives	&	MDV	\\
Bahamas	&	BHS	&	Mali	&	MLI	\\
Bahrain	&	BHR	&	Malta	&	MLT	\\
Bangladesh	&	BGD	&	Mauritania	&	MRT	\\
Barbados	&	BRB	&	Mauritius	&	MUS	\\
Belarus	&	BLR	&	Mexico	&	MEX	\\
Belgium	&	BEL	&	Moldova	&	MDA	\\
Belize	&	BLZ	&	Mongolia	&	MNG	\\
Benin	&	BEN	&	Montenegro	&	MNE	\\
Bermuda	&	BMU	&	Morocco	&	MAR	\\
Bhutan	&	BTN	&	Mozambique	&	MOZ	\\
Bolivia	&	BOL	&	Myanmar	&	MMR	\\
Bosnia Herzegovina	&	BIH	&	Namibia	&	NAM	\\
Botswana	&	BWA	&	Nepal	&	NPL	\\
Brazil	&	BRA	&	Netherland Antilles	&	ANT	\\
Brunei	&	BRN	&	Netherlands	&	NLD	\\
Bulgaria	&	BGR	&	 New Caledonia	&	NCL	\\
Burkina-Faso	&	BFA	&	New Zealand	&	NZL	\\
Burundi	&	BDI	&	Nicaragua	&	NIC	\\
Cape Verde	&	CPV	&	Niger	&	NER	\\
Cambodia	&	KHM	&	Nigeria	&	NGA	\\
Cameroon	&	CMR	&	Norway	&	NOR	\\
Canada	&	CAN	&	Oman	&	OMN	\\
Central African Republic	&	CAF	&	Pakistan	&	PAK	\\
Chile	&	CHL	&	Panama	&	PAN	\\
China	&	CHN	&	Papua New Guinea	&	PNG	\\
Colombia	&	COL	&	Paraguay	&	PRY	\\
Congo	&	COG	&	Peru	&	PER	\\
Cook Islands	&	COK	&	Philippines	&	PHL	\\
Costa Rica	&	CRI	&	Poland	&	POL	\\
Cote d'Ivoire	&	CIV	&	Portugal	&	PRT	\\
Croatia	&	HRV	&	Qatar	&	QAT	\\
Cuba	&	CUB	&	Republic of Korea	&	KOR	\\
Cyprus	&	CYP	&	Romania	&	ROU	\\
Czech Republic	&	CZE	&	Russia	&	RUS	\\
\br
\end{tabular}
\end{indented}
\end{table*}

\setcounter{table}{0}

\begin{table*}
\caption{\label{tab:countries}\textbf{(cont'd)} List of countries used in the analysis.}
\begin{indented}
\item[]\begin{tabular}{@{}lclc}
\mr
  Country & ISO3 & Country & ISO3 \\ 
\br	
Democratic Republic of Congo	&	COD	&	Rwanda	&	RWA	\\
Denmark	&	DNK	&	Saint Kitts and Nevis	&	KNA	\\
Djibouti	&	DJI	&	Saint Lucia	&	LCA	\\
Dominica	&	DMA	&	Saint Vincent	&	VCT	\\
Dominican Republic	&	DOM	&	Sao Tome	&	STP	\\
Ecuador	&	ECU	&	Saudi Arabia	&	SAU	\\
Egypt	&	EGY	&	Senegal	&	SEN	\\
El Salvador	&	SLV	&	Serbia	&	SRB	\\
Estonia	&	EST	&	Seychelles	&	SYC	\\
Ethiopia	&	ETH	&	Sierra Leone	&	SLE	\\
Faroe Islands	&	FRO	&	Singapore	&	SGP	\\
Fiji	&	FJI	&	Slovakia	&	SVK	\\
Finland	&	FIN	&	Slovenia	&	SVN	\\
France	&	FRA	&	Solomon Islands	&	SLB	\\
French Polynesia	&	PYF	&	South Africa	&	ZAF	\\
Gabon	&	GAB	&	Spain	&	ESP	\\
Gambia	&	GMB	&	Sri Lanka	&	LKA	\\
Georgia	&	GEO	&	Sudan	&	SDN	\\
Germany	&	DEU	&	Suriname	&	SUR	\\
Ghana	&	GHA	&	Swaziland	&	SWZ	\\
Greece	&	GRC	&	Sweden	&	SWE	\\
Greenland	&	GRL	&	Switzerland	&	CHE	\\
Grenada	&	GRD	&	Syria	&	SYR	\\
Guatemala	&	GTM	&	Taiwan	&	TWN	\\
Guinea	&	GIN	&	Tanzania	&	TZA	\\
Guyana	&	GUY	&	Thailand	&	THA	\\
Honduras	&	HND	&	Togo	&	TGO	\\
Hong Kong	&	HKG	&	Tonga	&	TON	\\
Hungary	&	HUN	&	Trinidad and Tobago	&	TTO	\\
Iceland	&	ISL	&	Tunisia	&	TUN	\\
India	&	IND	&	Turkey	&	TUR	\\
Indonesia	&	IDN	&	Tuvalu	&	TUV	\\
Iran	&	IRN	&	Uganda	&	UGA	\\
Ireland	&	IRL	&	Ukraine	&	UKR	\\
Israel	&	ISR	&	United Arab Emirates	&	ARE	\\
Italy	&	ITA	&	United Kingdom	&	GBR	\\
Jamaica	&	JAM	&	Uruguay	&	URY	\\
Japan	&	JPN	&	United States of America	&	USA	\\
Jordan	&	JOR	&	Uzbekistan	&	UZB	\\
Kazakhstan	&	KAZ	&	Vanuatu	&	VUT	\\
Kenya	&	KEN	&	Venezuela	&	VEN	\\
Kiribati	&	KIR	&	Vietnam	&	VNM	\\
Kuwait	&	KWT	&	Yemen	&	YEM	\\
Kyrgyzstan	&	KGZ	&	Zambia	&	ZMB	\\
Latvia	&	LVA	&	Zimbabwe	&	ZWE	\\
\br
\end{tabular}
\end{indented}
\end{table*}

\section{Primary and secondary products employed in the analysis\label{app:products}}
Table \ref{tab:products} contains a list of the 16 commodities employed in the analysis together with secondary products considered when aggregating the kcal content (with FAOSTAT code).

\begin{table*}
\caption{\label{tab:products} List of primary and secondary products used in the analysis (with FAOSTAT code).}
\begin{indented}
\item[]\begin{tabular}{@{}lclc}
\mr
Primary	&	Code	&	Secondary	&	Code	\\
\br
Wheat	&		&		&		\\
	&		&	Wheat 	&	15	\\
	&		&	Bran 	&	17	\\
	&		&	Flour 	&	16	\\
	&		&	Macaroni 	&	18	\\
	&		&	Bread 	&	20	\\
	&		&	Bulgur 	&	21	\\
	&		&	Pastry 	&	22	\\
	&		&	Breakfast Cereals 	&	41	\\
	&		&		&		\\
Rice	&		&		&		\\
	&		&	Rice, Total 	&	30	\\
	&		&	Rice, Paddy 	&	27	\\
	&		&	Rice, Husked 	&	28	\\
	&		&	Milled Rice from Imported Husked	&		\\
	&		&	Rice 	&	29	\\
	&		&	Milled Paddy Rice 	&	31	\\
	&		&	Rice, Broken	&	32	\\
	&		&	Flour 	&	38	\\
	&		&	Bran Oil	&	36	\\
	&		&		&		\\
Maize	&		&		&		\\
	&		&	Maize 	&	56	\\
	&		&	Flour 	&	58	\\
	&		&	Germ 	&	57	\\
	&		&	Bran 	&	59	\\
	&		&	Oil 	&	60	\\
	&		&	Cake 	&	61	\\
	&		&	Maize, Green 	&	446	\\
	&		&		&		\\
Soybeans 	&		&		&		\\
	&		&	Soybeans 	&	236	\\
	&		&	Cake 	&	238	\\
	&		&	Oil 	&	237	\\
	&		&	Soya Sauce 	&	239	\\
	&		&		&		\\
Barley 	&		&		&		\\
	&		&	Barley 	&	44	\\
	&		&	Pot Barley 	&	45	\\
	&		&	Barley Pearled 	&	46	\\
	&		&	Bran 	&	47	\\
	&		&	Flour 	&	48	\\
	&		&	Malt 	&	49	\\
	&		&	Malt Extract 	&	50	\\
	&		&	Beer 	&	51	\\
	&		&		&		\\
Sorghum 	&		&		&		\\
	&		&	Sorghum 	&	83	\\
	&		&	Bran 	&	85	\\
	&		&	Beer 	&	86	\\
	&		&		&		\\
\br
\end{tabular}
\end{indented}
\end{table*}

\setcounter{table}{0}

\begin{table*}
\caption{\label{tab:products}\textbf{(cont'd)} List of primary and secondary products used in the analysis (with FAOSTAT code).}
\begin{indented}
\item[]\begin{tabular}{@{}lclc}
\mr
Primary	&	Code	&	Secondary	&	Code	\\
\br

Cassava 	&	1953	&		&		\\
	&		&	Cassava 	&	125	\\
	&		&	Starch 	&	129	\\
	&		&	Cassava, Dried 	&	128	\\
	&		&	Flour 	&	126	\\
	&		&	Tapioca 	&	127	\\
	&		&		&		\\
Sugar 	&	1955	&		&		\\
	&		&	Cane Sugar, Raw, Centrifugal 	&	158	\\
	&		&	Beet Sugar, Raw Centrifugal 	&	159	\\
	&		&	Sugar Raw, Centrifugal 	&	162	\\
	&		&	Sugar Refined 	&	164	\\
	&		&	Sugar Confectionery 	&	168	\\
	&		&	Sugar Flavoured 	&	171	\\
	&		&		&		\\
Pigmeat 	&	2073	&		&		\\
	&		&	Pig Meat 	&	1035	\\
	&		&	Pork 	&	1038	\\
	&		&	Bacon and Ham 	&	1039	\\
	&		&	Sausages of Pig Meat 	&	1041	\\
	&		&	Prep. of Pig Meat 	&	1042	\\
	&		&		&		\\
Poultry Meat 	&	2074	&		&		\\
	&		&	Chicken Meat 	&	1058	\\
	&		&	Foie Gras 	&	1060	\\
	&		&	Meat of Chicken Cannes 	&	1061	\\
	&		&	Duck Meat 	&	1069	\\
	&		&	Goose and Guinea Fowl Meat 	&	1073	\\
	&		&	Turkey Meat 	&	1080	\\
	&		&		&		\\
Milk 	&	2030	&		&		\\
	&		&	Milk, Whole Fresh Cow 	&	882	\\
	&		&	Cream Fresh 	&	885	\\
	&		&	Butter, Cow Milk 	&	886	\\
	&		&	Milk, Skimmed Cow 	&	888	\\
	&		&	Milk, Whole Condensed 	&	889	\\
	&		&	Whey, Condensed 	&	890	\\
	&		&	Yoghurt, Concentrated or Not 	&	892	\\
	&		&	Buttermilk, Curdled, Acidified Milk 	&	893	\\
	&		&	Milk, Whole Evaporated 	&	894	\\
	&		&	Milk, Whole Dried 	&	897	\\
	&		&	Milk, Skimmed Dried 	&	898	\\
	&		&	Whey, Dry 	&	900	\\
	&		&	Cheese, Whole Cow Milk 	&	901	\\
	&		&	Cheese, Processed 	&	907	\\
	&		&	Milk, Products of Natural Constituents Nes. 	&	909	\\
	&		&	Ghee, of Buffalo Milk 	&	953	\\
	&		&	Milk, Whole Fresh Sheep 	&	982	\\
	&		&	Cheese, Sheep Milk 	&	984	\\
\br
\end{tabular}
\end{indented}
\end{table*}

\setcounter{table}{0}

\begin{table*}
\caption{\label{tab:products}\textbf{(cont'd)} List of primary and secondary products used in the analysis (with FAOSTAT code).}
\begin{indented}
\item[]\begin{tabular}{@{}lclc}
\mr
Primary	&	Code	&	Secondary	&	Code	\\
\br
Cocoa 	&		&		&		\\
	&		&	Beans 	&	661	\\
	&		&	Paste 	&	662	\\
	&		&	Butter 	&	664	\\
	&		&	Powder and Cake 	&	665	\\
	&		&	Chocolate Products Nes. 	&	666	\\
	&		&		&		\\
Pulses 	&		&		&		\\
	&		&	Flour 	&	212	\\
	&		&	Pulses 	&	1954	\\
	&		& \textit{of which:}		&		\\
	&		&	Beans, Dry 	&	176	\\
	&		&	Broad Beans, Horse Beans, Dry 	&	181	\\
	&		&	Peas, Dry 	&	187	\\
	&		&	Chick Peas 	&	191	\\
	&		&	Lentils 	&	201	\\
	&		&	Bambara Beans 	&	203	\\
	&		&		&		\\
Oil, Palm 	&		&		&		\\
	&		&	Oil, Palm Fruit 	&	257	\\
	&		&	Oil, Palm Kernel 	&	258	\\
	&		&		&		\\
Oil, Sunflower 	&	268	&		&		\\
	&		&		&		\\
Nuts 	&		&		&		\\
	&		&	Walnuts, shelled 	&	232	\\
	&		&	Walnuts, with Shell 	&	222	\\
	&		&	Brazil Nuts, Shelled 	&	229	\\
	&		&	Cake, Groundnuts 	&	245	\\
	&		&	Cashew Nuts, Shelled 	&	230	\\
	&		&	Cashew Nuts, with Shell 	&	217	\\
	&		&	Groundnuts, Shelled 	&	243	\\
	&		&	Hazelnuts, Shelled 	&	233	\\
	&		&	Kola Nuts 	&	224	\\
	&		&	Nuts, Nes. 	&	234	\\
	&		&	Almonds Shelled 	&	231	\\
	&		&	Pistachios 	&	223	\\
	&		&	Chestnut 	&	220	\\
\br
\end{tabular}
\end{indented}
\end{table*}


\section{\label{app:communities}Community Detection: Methods and Algorithms} 

\noindent \textit{The IFTMN as a collection of separate layers.} 

\medskip

\noindent In this analysis, we employ a new heuristic for modularity clustering, inspired to the fast modularity optimization algorithm originally introduced by \cite{Blondel_etal_2008_LouvainMeth}. The well-known Louvain algorithm is a multi-level coarsening procedure by iterated vertex moving based on a local optimization of Newman-Girvan modularity in the neighborhood of each node. More specifically, it follows a two-stage procedure that is iterated, until the gain in modularity is below a given threshold. The first step is represented by community reassignments. We define a network with $N$ nodes, each of which is initially assigned to a separate community, thus obtaining $N$ single-vertex clusters. For each node $i$ we consider its neighboring nodes $j$ and we evaluate the gain, in terms of increased modularity, which would be obtained by removing $i$ from his community and assigning it to that of $j$. Node $i$ at this point is moved to the communities to which this gain is maximum. If no increase in modularity is possible, the node is not moved. This process is applied repetitively and sequentially for all nodes, until modularity falls below a given tolerance threshold. The second step follows a coarse-graining scheme. We use the clusters discovered at the end of the community reassignment stage previously mentioned, in order to define a new, coarse-grained network. The formerly identified communities constitute the nodes of this second-stage graph. The edge weight between the nodes representing two communities is solely the sum of the edge weights between the lower-level nodes of each community. The links within each community generate self-loops in the new, coarse-grained network. It is now possible to apply again the first step, using as input the network obtained at the end of the second phase and to repeat the method. The algorithm stops when results impossible to get any further improvement in terms of modularity. 

In this work, the optimization of $Q$ is performed by using an extension of the Louvain algorithm described above. More specifically, we adopt the multilevel local search algorithm for modularity clustering introduced by \cite{Rotta_Noack_2011} and implemented in Pajek (\texttt{mrvar.fdv.uni-lj.si/pajek/}), a popular software for analysis and visualization of large networks. Starting from the intrinsic logic behind the Louvain algorithm, Rotta and Noack \cite{Rotta_Noack_2011} define a new heuristic proceeding in two phases: a coarsening stage and a refinement stage. The coarsening phase produces a sequence of graphs called coarsening levels: the first coarsening level is the input graph. On each coarsening level, a clustering is computed by means of a coarser, which in this particular case is a Clustering Joining heuristic (CJ henceforth). In this first phase, indeed, a multi-level CJ algorithm iteratively joins the cluster pair, starting from single-vertex clusters, until this join would not increase the modularity. The cluster pair of each join is chosen according to a parameter of the algorithm which represents a certain priority criterion. This prioritizer assigns to each cluster pair $(C, D)$ a real number called merge priority and thereby determines the order in which the CJ algorithm selects cluster pairs. The Modularity Increase (MI), $\Delta Q_{C,D}$ resulting from joining the clusters $C$ and $D$ is an obvious and widely used prioritizer.

The subsequent refinement phase, further improves the clustering computed in the first stage. It visits the coarsening levels in reverse order, that is, from the coarsest graph to the original graph and computes a clustering for each level of the refinement phase. This multi-level refinement is significantly more effective than the conventional single- level refinement or no refinement at all. The multi-level version applies the refinement on all coarsening levels, while the conventional single-level form moves just vertices of the original graph. More specifically, Rotta and Noack \cite{Rotta_Noack_2011} stress that the multi-level refinement, by local vertex moving (VM henceforth), at reduction factor 50\% (i.e., the one used during the coarsening phase) clearly outperforms other methods. Finally, the priority criterion of the refiner is again the MI. Note that, typically, the number of coarsening levels increases with decreasing reduction factor. On the one hand, this means that the refiner has more opportunities to improve the clustering, but on the other hand, the more frequent contraction in coarsening and the more thorough refinement tend to increase the runtime. For instance, with a reduction factor of 100\%, coarsening by CJ produces exactly two coarsening levels: thus, the refiner works on only one level, namely the original graph (as in the conventional single-level refinement). 

We use a resolution parameter equal to 1, which represents the standard Louvain method's resolution. Furthermore, we run the algorithm with 10 restarts: the heuristic, indeed, usually returns different results in each execution, therefore it is recommended to repeat the proceedings several times in order to ascertain the stability of the final outcome and to select the best partition. Finally, we leave unchanged the standard maximum number of iterations in each restart (i.e., equal to 20), the maximum number of levels in each iteration (i.e., equal to 20) and the maximum number of repetitions in each level (i.e., equal to 50). In general, these default values work fine in the most cases.

\bigskip

\noindent \textit{The IFTMN as a multi-layer network.}

\medskip

\noindent As described above, we perform a multilayer community detection by analyzing how communities span across the different layers. To do that we employ the generalization $Q_{\ast}^t$ of the modularity function as introduced in Ref. \cite{Mucha_etal_2010}. $Q_{\ast}^t$ is derived by considering a generalization of the null model for multilayer networks and introducing a set of parameters to control for the coupling between different layers. More specifically, each layer $x$ is represented by its adjacency matrix $A_{ij,x}^t$, while inter-layer coupling (connection) between a generic node $j$ in layer $x$ and itself in layer $y$ is represented by $C_{j,xy}^t$. By exploiting the continuous-time Laplacian dynamics the authors derive the following definition of $Q_{\ast}^t$:

\scriptsize
\begin{equation*}
  Q_{\ast}^t=\frac{1}{2\mu}\sum_{ij,xy}\left\{ \left( A_{ij,x}^t -\gamma_x\frac{k_{i,x}^t k_{j,x}^t}{2m_s}\right) \delta_{xy} + \delta_{ij}C_{j,xy}^t \right\} \delta \left(\xi_{ix},\xi_{jy}\right)
\label{eq:ms}
\end{equation*}
\normalsize
where $k_{ix} = \sum_j A_{ij,x}^t$ is the degree of node $i$ in layer $x$, $\gamma_x$ is a resolution parameter in each layer and $\mu$ is the total degree of the multilayer network by considering intra- and inter-layers connections. The definition for $Q_{\ast}$ can be easily generalized to the case of weighted directed layers. 

We performed multilayer community detection via a directed optimization of $Q_{\ast}$ by using a generalization of the Louvain algorithm previously described. As in the Louvain method, we employ a two-phase iterative procedure: community reassignments and coarsening. These two phases are applied iteratively until the gain in $Q_{\ast}$ is below a given threshold.        

\medskip

\section{\label{app:nmi}Assessing dissimilarity between CSs} 

The issue of comparing CSs across commodities (and time periods) is addressed in this paper using the Normalized Information Index (NMI), see Ref. \cite{Danon_etal_2005}. To define the NMI measure, we first introduce the ``confusion matrix''. Given two community partitions $P_A$ and $P_B$ of the same set of units (i.e., nodes), the confusion matrix $F$ is defined as a matrix whose generic entry $f_{ij}$ records the number of nodes in the cluster $i$ of the partition $P_A$ that appear in the cluster $j$ of the partition $P_B$. The NMI is defined as: 

\begin{equation*}
	NMI=\frac{-2\sum_{i=1}^{C_A}\sum_{j=1}^{C_B}{f_{ij}\log (\frac{f_{ij}F}{f_{i\cdot}f_{\cdot j}})}}{\sum_{i=1}^{C_A}{f_{i\cdot}\log(\frac{f_{i\cdot}}{F})}+\sum_{j=1}^{C_B}{f_{\cdot j}\log(\frac{f_{\cdot j}}{F})}}
\end{equation*}
where $C_A$ and $C_B$ are the number of communities in partitions $A$ and $B$; $(f_{i\cdot},f_{\cdot j})$ are the row and column sums of the confusion matrix; and $F=\sum_i \sum_j f_{ij}$. The NMI index ranges between 0 and 1: it is equal to 0 if the two partitions are independent, and takes a value of 1 if the two partitions are identical. 

Therefore, the NMI index measures similarity between non-overlapping CSs of a same set of units.

\section{\label{app:covariates}Covariates employed in regression analyses}

We consider the following factors traditionally employed in the empirical trade literature, provided by CEPII gravity dataset (\texttt{cepii.fr}) and the WTO RTA dataset (\texttt{rtais.wto.org}).

\begin{itemize}
	\item  \textbf{Economic variables}: \textit{Combined economic size}, defined as the product of the economic sizes (GDPs) of the two countries; and \textit{Combined economic development}, defined as  the product of per-capita GDPs of the two countries (i.e. a measure of combined country incomes).
	\item  \textbf{Trade policy variables}: \textit{Free trade agreements}, which is 1 when each pair of countries has a free trade agreement and 0 otherwise or six specific dummy variables, namely \textit{AFTA}, \textit{EFTA}, \textit{NAFTA}, \textit{European Union}, \textit{NAFTA}, \textit{MERCOSUR} and \textit{COMESA}, representing relevant regional free trade agreements, which are 1 if pairs of countries belong to each specific RTA and 0 otherwise\footnote{See \texttt{https://en.wikipedia.org/wiki/List\_of \_multilateral\_free-trade\_agreements} for a complete list.}.
	\item  \textbf{Geographical variables}: \textit{Contiguity}, which is 1 if the two countries share common borders; \textit{Distance}, which is the simple distance, in terms of kilometers between the most representative cities in the pairs; \textit{Region}, which is 1 only if the two countries belong to the same regional bloc (namely East Asia and Pacific, Europe and Central Asia, Latin America and Caribbean, Middle East and North Africa, North America, Sub-Saharan Africa, South Asia).
	\item  \textbf{Historical and political variables}: \textit{Colonial Relationship} is 1 for pairs that were ever in colonial relationship; \textit{Common Colonizer} is 1 when the two countries have had a common colonizer after 1945; \textit{Same Country} is 1 if the two country were part of the same country.
	\item  \textbf{Cultural variables}: \textit{Common Language}, which is 1 when the country pair speak the same official language; \textit{Common Ethnicity}, that is 1 when a language is spoken by at least the 91\% of the population in both countries.
\end{itemize}

\noindent See Table \ref{tab:covars} for a detailed list and description of variables and their data sources. 

Several theoretical and empirical considerations suggest the expected sign that these variables should have in our regression analyses explaining the probability of country-pair co-presence in a cluster. In general, the combined level of GDP of country pairs is expected to have a positive effect on trade intensity: this reflects the fact that countries with larger economic size also have relevant production capacity and market size. The impact of combined income level, being a proxy for the purchasing power of country pairs, in instead ambiguous and may be product specific, as it is not necessary true that richer countries trade more intensively in all agricultural products. Free and regional trade agreements are expected, in general, to strengthen trade relationships between country pairs. Geographical proximity, being a proxy of trade frictions, is expected to have a positive impact on the probability of countries to belong to the same community: thus, we expect positive signs both for \textit{contiguity} and \textit{region} variables, whereas distance should enter negatively. In general, cultural, historical and political proximity is expected to facilitate trade relationship, but these variables may have a more nuanced effect on commodity-specific trade relationship.

\begin{table*}
\caption{\label{tab:covars}Covariates employed in the econometric analyses. Definitions and Data Sources.}
\begin{indented}
\item[]\begin{tabular}{@{}p{5cm}p{9cm}p{2.5cm}}
\mr
Covariate Definition & Description & Source \\
\br
Combined economic size & Product of GDP$^a$ of country $i$ and
GDP of country $j$, in year $t$ & CEPII$^b$ \\
Combined economic
development &
Product of GDP per capita of
country $i$ and GDP per capita of
country $j$, in year $t$ &
CEPII \\

Free trade agreements & =1 if country i and country j have a free trade agreement in year $t$. & See Ref. \cite{DeSousa_Lochard_2011}  \\

NAFTA \par AFTA \par COMESA \par EFTA \par EU \par MERCOSUR \par & =1 if country $i$ and country $j$ belong to a specific regional trade agreement in year $t$. & RTA Database$^c$ \\

Contiguity & =1 if country $i$ and country $j$ share a
border. & CEPII \\

Distance & Distance, in km, between country $i$
and country $j$. & CEPII \\

Region & =1 if country $i$ and country $j$ belong
to the same geographical region. & CEPII\\

Colonial relationship & =1 if country $i$ and country $j$ ever shared a colonial relationship. & CEPII \\

Common colonizer & =1 if country $i$ and country $j$ shared common colonizer after 1945. & CEPII \\

Same country & =1 if country $i$ and country $j$ were part of the same
country. & CEPII \\

Common language & =1 if country $i$ and country $j$ share common official language. & CEPII \\

Common ethnicity & =1 if a language is spoken by at least the 9\% of the populations in both country $i$ and $j$. & CEPII \\

\br
\end{tabular}
\item[] $^a$ Gross Domestic Product (in nominal US dollars) 
\item[] $^b$ See the Gravity Dataset maintained by CEPII, available at \texttt{cepii.fr}
\item[] $^c$ See the RTA Database maintained by WTO, available at \texttt{rtais.wto.org/UI/PublicMaintainRTAHome.asp}
\end{indented}
\end{table*}


\section{Properties of the IFTMN\label{app:netprop}}
We characterize IFTMN topological properties using the following network statistics, computed over weight $\mathbf{W}^t_c$ and adjacency $\mathbf{A}^t_c$ matrix of each layer $(c,t)$: (i) \textit{Density}, defined as the existence number of links over all possible $N(N-1)$ directed edges; (ii) \textit{Bilateral Density}, defined as the ratio of reciprocated links; (iii) \textit{Weighted Asymmetry} as defined in \cite{Fagiolo2006EcoBull}; (iv) \textit{Size of Largest Connected Component} (LCC), i.e. the number of nodes in the largest connected subgraph, where connectivity is defined in a weak form (i.e., disregarding directionality); (v) \textit{Centralization}, see \cite{Freeman_1978}, which measures how much the binary structure is centralized; (vi) \textit{Binary/Weighted Assortativity}, that is the correlation coefficient between node average nearest-neighbor degree/strength (ANND/S) and total node degree/strength, see \cite{Fagiolo_survey_2017}; (v) \textit{binary/weighted Average Clustering}, that is the average across nodes of node total binary/weighted clustering coefficients as defined in \cite{Fagiolo2007pre}; (vi) \textit{Average and Standard Deviation of Link Weights}, that is arithmetic average and standard deviation of (logs) of export flows in a single layer. 

Note that: (a) whereas bilateral density measures symmetry at a binary level, the weighted-asymmetry index employs link weights to assess how much reciprocity is present in the weighted directed graph; (b) if the assortativity indexes are positive (resp. negative) the graph is assortative (resp. disassortative); (c) the average of link weights equals total volume per link, i.e. the average intensity of export flows. 

\begin{figure*}
 \centering
 \subfigure[Year=2001]
   {\includegraphics[width=8cm]{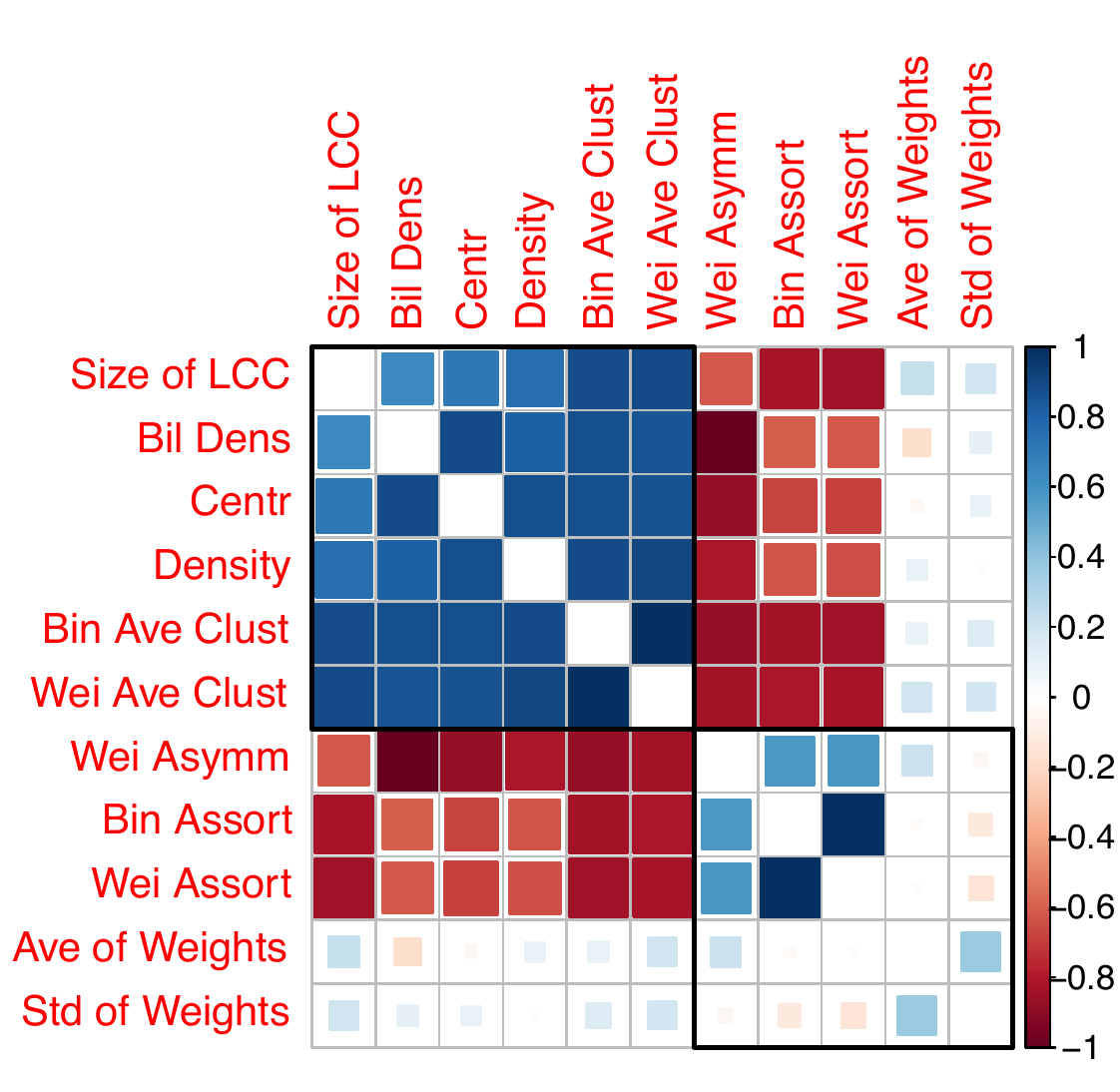}}
 \hspace{5mm}
 \subfigure[Year=2011]
   {\includegraphics[width=8cm]{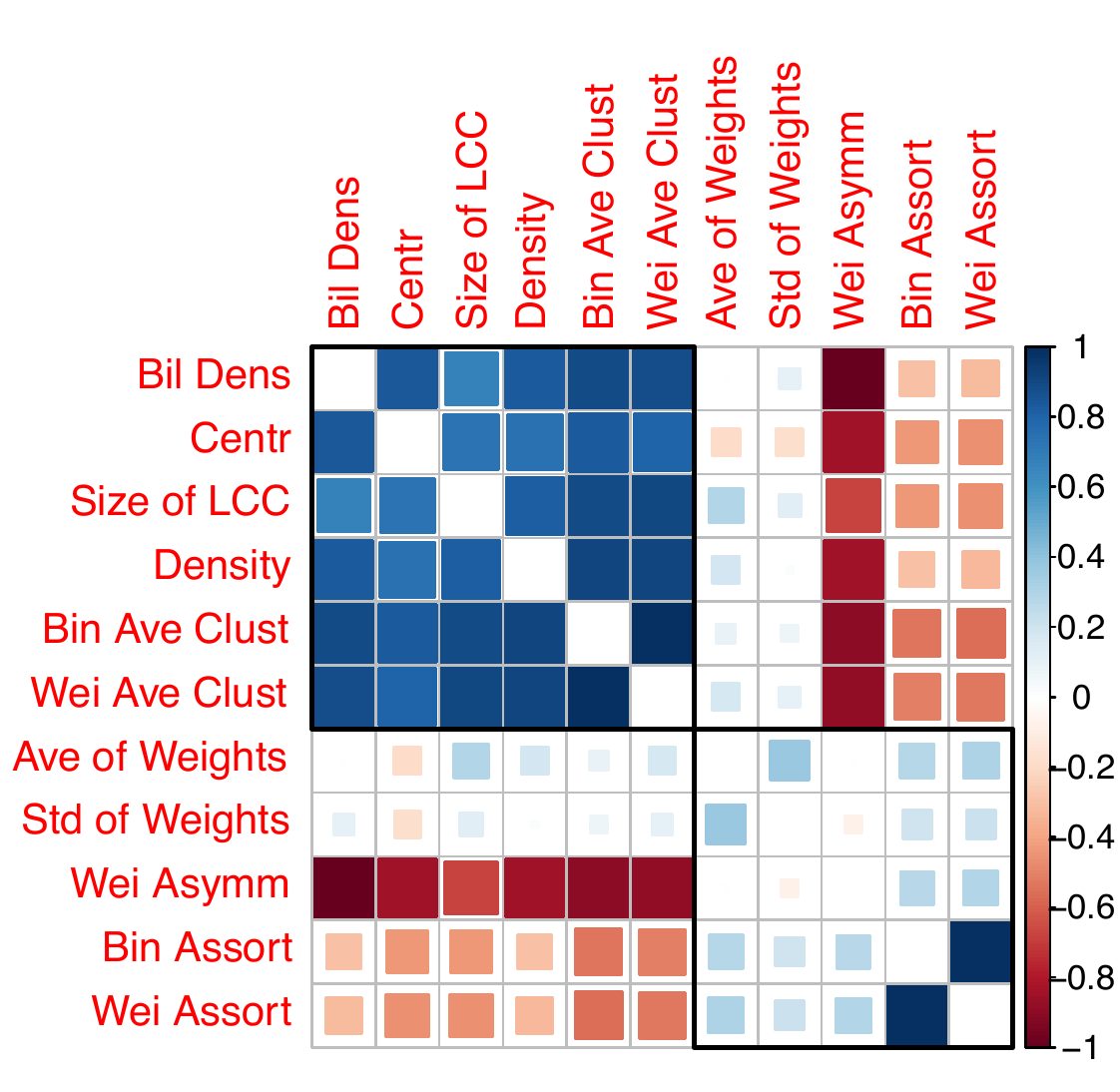}}
 \caption{Correlation coefficients between network statistics across commodity layers. Commodities have been ordered using a (Ward) hierarchical clustering.\label{fig:corr_stats}}
\end{figure*}

As an illustrative example, we report for two selected years (2001 and 2011), the values of network statistics in Tables \ref{tab:stats_2001}-\ref{tab:stats_2001} and (Pearson) correlation matrices in Fig. \ref{fig:corr_stats}, plotted after performing a (Ward) hierarchical clustering. All correlation coefficients turn out tuo be statistically different from zero. Before performing a principal-component analysis using 2001 and 2011 network statistics, we notice that some pairs of statistics are trivially very-highly correlated (i.e., more than 0.9 in absolute level) in both years. For example, weighted asymmetry is strongly negatively correlated with bilateral density. This means that weighted asymmetry does not pick up additional information as compared to bilateral density. The same happens for both binary and weighted clustering, which are strongly positively correlated with density. Finally, weighted assortativity is almost perfectly positively correlated with its binary counterpart. Therefore, in both years, we remove from the analysis the following variables: weighted asymmetry, binary and weighted clustering, weighted assortativity. This leaves us with a space of 7 variables. 

We then perform a PCA over the remaining 7 variables, weighting by the inverse of the variance. We choose the first two principal components, which explain together respectively 83\% and 85\% of total variance (with the first PC explaining about 50-55\%). In both years, the first PC is positively related to density, size of LCC, centralization and bilateral density, and negatively related, especially in year 2001, with binary assortativity. The second PC is instead related to both average and standard deviation of link weights.   

\begin{table*}
\caption{\label{tab:stats_2001}Network Statistics. Year: 2001.}
\begin{indented}
\item[]\begin{tabular}{@{}lccccccccccc}
\mr
	&	Density	&	Bil	&	Wei	&	Size	&	Centr	&	Bin	&	Wei	&	Bin Ave	&	Wei Ave	&	Ave of	&	Std	of\\
	&		&	Dens	&	Asymm	&	of LCC	&		&	Assort	&	Assort	&	Clust	&	Clust	&	Weights	&	Weights	\\
	\br	
Wheat	&	0.13	&	0.45	&	0.52	&	171	&	0.43	&	0.03	&	0.00	&	0.52	&	10.67	&	20.28	&	3.28	\\
Soybeans	&	0.06	&	0.30	&	0.70	&	160	&	0.16	&	0.07	&	0.04	&	0.40	&	8.18	&	20.36	&	3.99	\\
Maize	&	0.06	&	0.35	&	0.65	&	159	&	0.31	&	0.07	&	0.05	&	0.39	&	8.09	&	20.49	&	3.41	\\
Sugar	&	0.05	&	0.24	&	0.77	&	159	&	0.26	&	0.05	&	0.04	&	0.36	&	7.53	&	20.92	&	3.49	\\
Rice	&	0.05	&	0.17	&	0.83	&	161	&	0.14	&	-0.05	&	-0.07	&	0.32	&	6.59	&	20.45	&	3.22	\\
Barley	&	0.07	&	0.42	&	0.57	&	165	&	0.38	&	-0.18	&	-0.19	&	0.51	&	9.82	&	19.17	&	3.68	\\
Oil, palm	&	0.03	&	0.19	&	0.82	&	152	&	0.12	&	0.06	&	0.04	&	0.27	&	5.45	&	19.99	&	3.06	\\
Oil, sunflower	&	0.03	&	0.20	&	0.79	&	139	&	0.08	&	0.07	&	0.06	&	0.29	&	6.01	&	20.55	&	2.83	\\
Milk	&	0.10	&	0.29	&	0.69	&	162	&	0.28	&	0.03	&	0.00	&	0.44	&	8.75	&	19.95	&	2.78	\\
Cassava	&	0.01	&	0.19	&	0.80	&	111	&	0.10	&	0.25	&	0.21	&	0.13	&	2.38	&	18.74	&	3.14	\\
Pulses	&	0.07	&	0.35	&	0.64	&	159	&	0.28	&	0.01	&	-0.01	&	0.41	&	8.09	&	19.68	&	2.75	\\
Cocoa	&	0.10	&	0.41	&	0.57	&	164	&	0.35	&	-0.09	&	-0.10	&	0.51	&	9.99	&	19.34	&	2.63	\\
Pig meat	&	0.04	&	0.32	&	0.65	&	145	&	0.15	&	0.09	&	0.08	&	0.36	&	6.93	&	19.49	&	2.90	\\
Poultry meat	&	0.05	&	0.26	&	0.72	&	153	&	0.17	&	0.08	&	0.06	&	0.32	&	6.04	&	19.06	&	2.80	\\
Nuts	&	0.07	&	0.33	&	0.66	&	158	&	0.33	&	0.00	&	-0.02	&	0.39	&	7.62	&	19.39	&	2.68	\\
Sorghum	&	0.01	&	0.16	&	0.84	&	87	&	0.05	&	0.28	&	0.27	&	0.11	&	2.25	&	19.88	&	2.91	\\
\br
\end{tabular}
\end{indented}
\end{table*}

\begin{table*}
\caption{\label{tab:stats_2011}Network Statistics. Year: 2011.}
\begin{indented}
\item[]\begin{tabular}{@{}lccccccccccc}
\mr
	&	Density	&	Bil	&	Wei	&	Size	&	Centr	&	Bin	&	Wei	&	Bin Ave	&	Wei Ave	&	Ave of	&	Std	of\\
	&		&	Dens	&	Asymm	&	of LCC	&		&	Assort	&	Assort	&	Clust	&	Clust	&	Weights	&	Weights	\\
	\br	
Wheat	&	0.16	&	0.51	&	0.46	&	169	&	0.43	&	0.21	&	0.17	&	0.53	&	10.97	&	20.47	&	3.31	\\
Soybeans	&	0.07	&	0.33	&	0.66	&	152	&	0.20	&	0.26	&	0.23	&	0.36	&	7.31	&	20.27	&	4.18	\\
Maize	&	0.07	&	0.41	&	0.59	&	157	&	0.31	&	0.16	&	0.15	&	0.40	&	8.10	&	20.46	&	3.67	\\
Sugar	&	0.07	&	0.30	&	0.70	&	155	&	0.24	&	0.12	&	0.12	&	0.36	&	7.49	&	20.83	&	3.42	\\
Rice	&	0.07	&	0.26	&	0.74	&	155	&	0.21	&	0.13	&	0.11	&	0.36	&	7.38	&	20.38	&	3.18	\\
Barley	&	0.09	&	0.47	&	0.51	&	160	&	0.38	&	0.05	&	0.03	&	0.50	&	9.59	&	19.08	&	3.59	\\
Oil, palm	&	0.04	&	0.23	&	0.78	&	150	&	0.18	&	0.08	&	0.05	&	0.30	&	6.24	&	20.06	&	3.30	\\
Oil, sunflower	&	0.04	&	0.27	&	0.72	&	137	&	0.12	&	0.32	&	0.31	&	0.27	&	5.63	&	20.73	&	2.95	\\
Milk	&	0.11	&	0.32	&	0.66	&	164	&	0.26	&	0.25	&	0.22	&	0.39	&	7.77	&	20.10	&	2.88	\\
Cassava	&	0.02	&	0.19	&	0.82	&	123	&	0.16	&	0.12	&	0.10	&	0.20	&	3.84	&	18.85	&	2.77	\\
Pulses	&	0.08	&	0.39	&	0.60	&	158	&	0.38	&	-0.01	&	-0.03	&	0.45	&	8.94	&	19.71	&	2.73	\\
Cocoa	&	0.13	&	0.48	&	0.49	&	161	&	0.39	&	-0.01	&	-0.03	&	0.53	&	10.32	&	19.51	&	2.74	\\
Pig meat	&	0.06	&	0.39	&	0.58	&	141	&	0.30	&	0.27	&	0.25	&	0.34	&	6.84	&	19.97	&	2.99	\\
Poultry meat	&	0.06	&	0.33	&	0.66	&	151	&	0.41	&	0.32	&	0.29	&	0.32	&	6.18	&	19.44	&	2.83	\\
Nuts	&	0.08	&	0.36	&	0.62	&	157	&	0.38	&	-0.06	&	-0.08	&	0.42	&	8.26	&	19.66	&	2.74	\\
Sorghum	&	0.01	&	0.22	&	0.79	&	101	&	0.08	&	0.33	&	0.31	&	0.13	&	2.57	&	19.56	&	3.17	\\
	\br
\end{tabular}
\end{indented}
\end{table*}

\begin{figure*}[h!]
 \centering
    {\includegraphics[width=10cm]{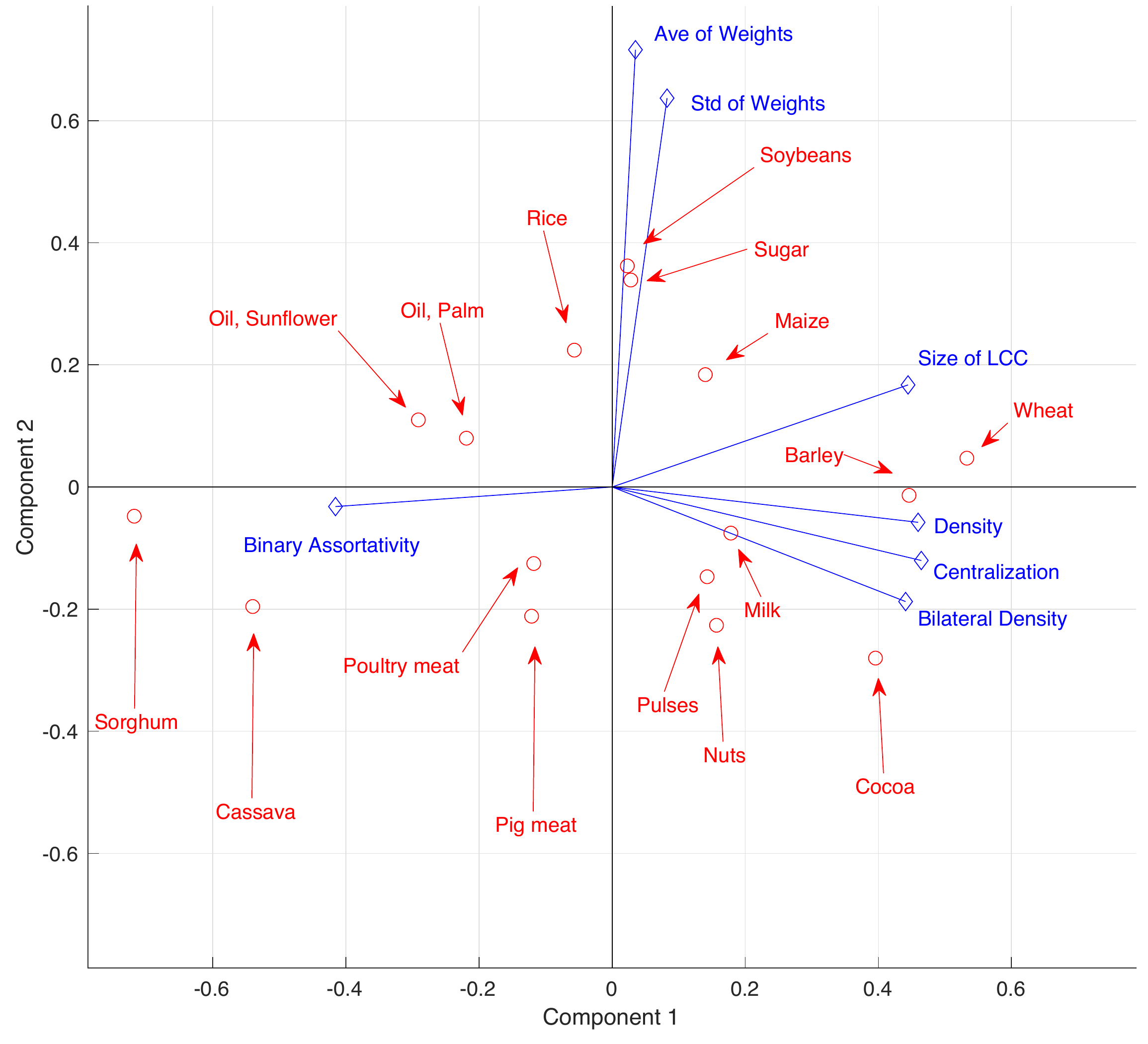}}
 \caption{The IFTMN in year 2001. Principal component (PC) analysis in the space of network statistics. First two PCs explain 85\% of total variance.\label{fig:pca_2001}}
\end{figure*}

\begin{figure*}[h!]
 \centering
    {\includegraphics[width=10cm]{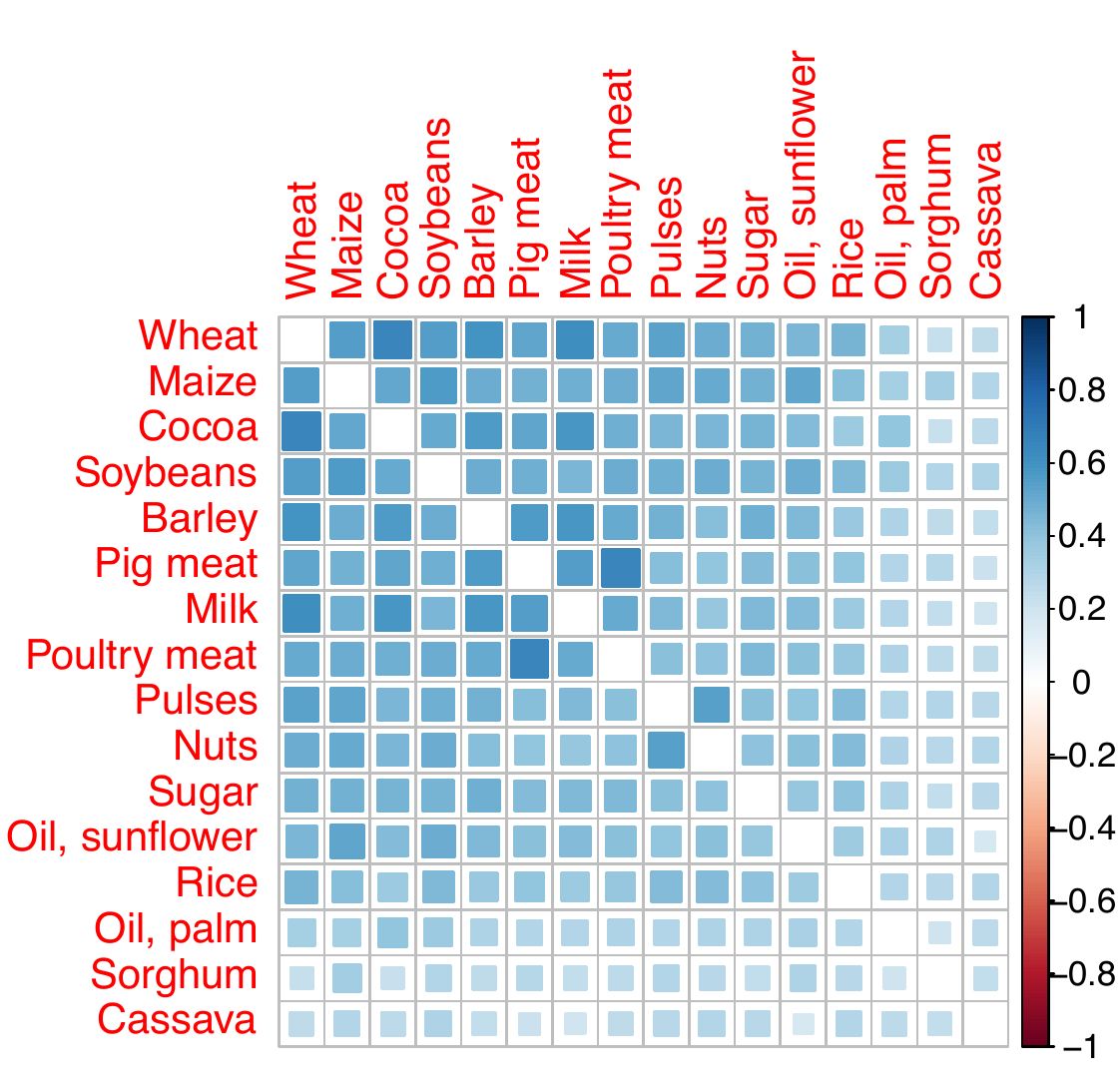}}
 \caption{\label{fig:lw_corr_2001}Correlation between logged link weights of commodity layers. Year=2001. Commodities have been ordered using a (Ward) hierarchical clustering.}
\end{figure*}


\section{Single-Layer Community Detection in 2011: Some Remarks\label{app:remarks_2011}}

In what follows we highlight some economically-relevant features of aggregate and commodity-specific community structures in 2011. We focus on 10 commodity classes, those exhibiting the most relevant geopolitical and economic patterns.\footnote{The specific choices made in the course of the process of aggregation of secondary products may have strongly influenced communities detected in at least three cases, namely milk, pulses and nuts networks (i.e., we aggregate several derivative products, not always of less importance if compared with the main primary commodity).}

\setcounter{footnote}{0}

\begin{itemize}
	
\item \textbf{Wheat}: major producers belong in pairs to separate communities: i) North America and Australia, ii) Argentina and Brazil, iii) Russia and Ukraine. Interestingly, Europe belongs to yet another separate cluster, characterized by the presence of a relevant producer and exporter such as France. Despite being not a big producer, Europe is not an open market for agricultural products and this finding may be linked to protectionist agricultural policies of the European Union, at least for coarse grains. Balkans countries not belonging to the EU\footnote{Serbia, Bosnia Herzegovina, Montenegro, Macedonia, Albania and Croatia which joined the European
Union only in 2013.} set up a small independent community inside the European cluster (except for Albania which stays in the Russian orbit). South East Asia and Far East belong to the North American-Australian cluster,\footnote{60\% of total Japanese imports of wheat comes from the United States, while Australia supplies
48\% and 60\% of total wheat internationally demanded respectively by China and Indonesia.} together with Central America, Chile and Caribbean countries. The African continent is split among the aforementioned four major clusters: North Eastern countries such as Egypt,\footnote{Egypt is the top importer of wheat in 2011 and Russia supplies 40\% of its total imports.} Ethiopia together with Tunisia and Morocco stay in the Russian orbit, South Eastern countries belong to the South American cluster, Western Africa is included in the European community, while some isolated cases such as Nigeria, Ghana, Congo and the Democratic Republic of Congo belong to the North American one. Middle East is split between Russian and North American clusters.\footnote{Ukrainian and Russian droughts during summer 2010 may have contributed to wheat shortages in
several countries belonging to their community such as Tunisia, Morocco, Egypt, Syria, among others,
where food price spikes have a crucial role in the Arab spring.}

\setcounter{footnote}{0}

\item \textbf{Soybeans}: the soybean network reveals one of the most concentrated community structure, composed by only three large clusters without a clear regional scheme. The most important bloc --in terms of trade volume-- gathers a handful of countries extremely relevant for global trade in soybeans: US and Brazil from the producing and exporting side (which together account for more than 70\% of global soybeans exports) and especially China from the importing side which alone accounts for 56\% of global soybeans imports.\footnote{China imports soybeans almost to the same extent from Brazil (47\%) and from the US (42\%). 60\% of soybeans exports of the US and 66\% of Brazil's are intended for China market.} A second giant community is characterized by less relevant exporters such as Argentina, Paraguay and Ukraine but embraces almost whole Africa (with relevant exporters represented by North African countries), Eastern and Central Asia as well as Russia and several countries in South East Asia, including India, Indonesia and Australia. Interestingly, South European countries such as Portugal, Spain, Italy, Croatia and Greece belong to this community and not to the European bloc. An important trade corridor is the one between Canada and the Netherlands, which then serves the West and Central Europe cluster. Besides these three communities there are three other small independent blocs, composed by just few neighboring countries.
\item \textbf{Maize}: world's maize network is divided in three major communities each one dominated by a couple of relevant producers and exporters: i) North America, ii) South America, iii) Europe together with Russia. Few relevant players form the first community: US, Canada from the exporting side and Mexico, Japan,\footnote{The most intense maize trade corridor is the one characterized by imports of Japan from United States,
that accounts for almost 14\% of total maize traded in 2011.} South Korea, and China from the importing side. The South American cluster, dominated by Argentina and Brazil, includes North Africa, the Arabian peninsula, India and South East Asia. The giant community formed by Russia and Europe contains important maize producing countries such as Ukraine,\footnote{Maize is one of the top export for Ukraine, mainly directed to West European countries (Spain, Italy,
the Netherlands), North African and Middle East ones (Egypt, Tunisia, Algeria, Iran). An interesting
extension to this study could be an in-depth investigation of Ukrainian maize trade profile after 2011,
since in the last years the country started to trade intensively also with China, South Korea and Japan.} Hungary and France. The limited importance of France in the maize network with respect to the wheat one probably determines the Eastern enlargement of the community and the circumscribed influence of the European Union in Africa. The lion's share of the West and North African countries belong to the South American cluster, while other countries set up small independent communities, such as a couple of countries belonging to Central Asia region.
\item \textbf{Sugar}: one distinctive feature of international sugar trade is the role played by preferential trade arrangements. Notably, this scheme seems to be largely mirrored in the community structure detected. The sugar network is divided into five communities: the largest cluster is the one formed around Brazil,\footnote{In 2011, Brazil it accounted for more than 50\% of total world exports and in 2001 for just over 20\%
(i.e., evidence of the significant expansion of the sugar-ethanol complex in Brazil over the past 15 years).} the major global producer and exporter of sugar, and Cuba, which exports 66\% of its traded sugar to China under the Cuba-China Protocol. Besides China, from the importing side Russia, Canada and North and West Africa are the most important importers of sugar joining this cluster. The US belong to the same community of Central and Southwest America: Guatemala, El Salvador, Nicaragua, Mexico and Colombia are the most important producers and exporters of this cluster, trading intensively with Chile, Peru and especially with the US under the Tariff Rate Import Quota (TRQ) and the North American Free Trade Agreement (NAFTA).\footnote{Mexico exports 98\% of its total traded sugar exclusively to the US.} Europe is another relevant importer of sugar and it belongs to a third community served mainly by Fiji, Belize, Guyana, Jamaica and some countries in Southeast Africa under the African, Caribbean, Pacific (ACP) Sugar Protocol, the Special Preferential Sugar (SPS) and the Everything But Arms (EBA) arrangements.\footnote{The Sugar Protocol has been a feature of EU policy to ACP countries since 1975. It officially expired
in 2009 and, following a six year transition period, the Protocol --which provides a group of 19 ACP
countries with guaranteed access to the EU market for fixed quantities of sugar at preferential prices-- has
been replaced by a non-reciprocal duty and quota-free preferential trade system in 2015. Although it
expired at the end of 2009, the transition period provision allows us to assume that its effects are valid
also for 2011. European Union is one of the principal importer of raw sugar for refining. 99\% of
Fiji sugar exports and 74\% of Belize sugar exports in 2011 are directed to the refineries of the UK.}
India, which has emerged as an important player in the sugar market starting from early
2000s, mainly trades with Southeast Africa and Middle Eastern countries whose
refineries, locate in the Persian Gulf, are increasingly important. Finally, Australian and
Thai intense export relations with Japan and South Korea principally define the South
East Asian Bloc.

\item \textbf{Rice}: in this case we identify six communities in total. The rice network
undoubtedly has a Southeast Asian focal point: of the five top exporters, four, namely
Thailand, Vietnam, India and Pakistan, are from Asia and they set up three different
communities. The dominance of Asian countries in rice production dwarfs the
contribution from countries in other regions, but Brazil in South America and Egypt in
North Africa are also relevant rice producers. Furthermore, the US are the only non-
Asian country among the top five exporters: it belongs to an independent community
together with Canada and Mexico, linked to Russia, Central Asia and few Middle
Eastern countries. Rice is mostly consumed in the same country where it is produced, so
trade in rice is thin both in absolute terms and as a proportion of global production, if
compared with the other two major cereals, representing only 1/4 and 1/3 of wheat
market and maize market, respectively. It is a critical staple food in South East Asia,
Middle East and Africa and rice trade flows are often controlled by preferential trade
agreements and government-to-government contracts. This feature of the rice trade
network may be reflected in the detected community structure which indeed displays a
highly fragmented Asian scheme: i) Thailand, Myanmar, China, Australia together with
several African countries, ii) Vietnam, Indonesia, Malaysia and Philippines, iii) India,
Pakistan, Bangladesh together with Middle East and several Sub-Saharan countries.
Interestingly Cambodia belongs to the European cluster: in fact Cambodia, unlike
Thailand and Vietnam, benefits from the Everything But Arms (EBA) trade scheme,
enjoying a privilege position as the largest rice external supplier to the EU together with
Egypt.\footnote{It is worth to mention that alongside Cambodia, the EU's preferential access list includes also another
rice exporter such as Myanmar but, from 1997 to 2013, this agreement was suspended due to serious and
systematic violations of principles of core international labour conventions. The community scheme
detected for rice network seems to reflect the withdrawal of Myanmar preferences by the EU. Since
2011, the EU has progressively re-engaged with Myanmar. Updated bilateral trade data might reveal a
completely different community picture for 2015: Myanmar rice exports in 2011 are mainly directed to
Africa and Asia (86\% of total rice exports), while in 2014, after the formal reinstatement of the country
into EBA, more than 30\% are intended for EU markets and only 28\% for African ones (OEC, 2016).}

\item \textbf{Barley}: we detect six large communities, beside some small clusters formed by
only few neighbouring countries. Europe is the global largest exporter of barley and,
interestingly unlike the case of the other coarse grains, it is divided in three different
communities: i) a West-Central European cluster formed around the most relevant
European producer and exporter which is France, ii) a Eastern European bloc orbiting
around Romania, Hungary and to a less extent Bulgaria, iii) an Ukrainian-Russian cluster, which embraces Central Asia and Middle East (especially Iran and Saudi Arabia),
besides, interestingly Scandinavian countries. Australia is the most important exporter in
the Southeast Asian cluster, which is linked with the North American one (China is the
top importer in this community). South America sets up an independent cluster
dominated by Argentina. Africa is split among the West-European community (as in the
wheat trade network) and the American cluster.

\item \textbf{Oil, Palm}: its community structure is extremely fragmented and does not display
any clear regional scheme or geographically defined trade bloc. There are three large and
scattered communities rotating around a handful of producing and exporting countries,
namely Indonesia and Malaysia accounting together for by 87\% of global production,
but also Thailand, Papua New Guinea and to a lesser extent Colombia and Ecuador.
Top importers such as India, China, the Netherlands, Pakistan and Italy also belong to
different clusters. Interestingly Italy, which trade intensively with Indonesia, is separated
from the European community, which instead is linked to Papua New Guinea through
the Netherlands. The latter is the first European importer and an important trade and
refining hub of palm oil. Indeed, a large part of Dutch palm oil imports are refined and
re-exported, mostly to other member states.\footnote{Important international firms such as Cargill, Sime Darby, Wilmar and IOI Loders all have palm
refineries in Rotterdam. Palm oil enters the country through the port of Rotterdam to be partly transshipped
directly to other European countries.} In the near future, increasing global
commitments to sustainability and health issues might influence the palm oil trade
geography. Furthermore, the growing prominence of Eastern European countries as a
destination for European exports will probably lead to a reorganizations of traditional
trade channels: trade hubs such as the Netherlands will increasingly target Eastern
European countries and gradually shift away from Western Europe, thus accompanying
the shift in European consumption patterns. At the same time, direct imports by
Eastern Europe are expected to increase.
Besides these three large and splintered communities, there are only few small
regional clusters, formed by a couple of neighboring countries.

\item \textbf{Oil, Sunflower}: European Union is, beyond Ukraine, the largest producer and
exporter of sunflower oil but interestingly member states are not gathered under a
unified community: i) France sets up a Northern-Central European cluster, ii) Southern
Eastern countries form a community around Romania, Bulgaria and Hungary, iii) the
Baltic republics belong to a small independent trade bloc, iv) Spain and Portugal are
intensively linked with Ukraine. The latter forms a giant cluster gathering Russia, the
lion's share of Central and Far East Asian countries as well as Middle East. Japan,
Thailand, Malaysia and Australia are few meagre exceptions belonging to the American
cluster.

\item \textbf{Cocoa}: world cocoa market is divided into five communities. Top African
exporters are in three different communities: i) Cote D'Ivoire belongs to the American
cluster, which also gathers important producers such as Dominican Republic, Ecuador
and Brazil, ii) Ghana belongs to the South East community, iii) Togo, Nigeria, Uganda
and Cameroon mainly serve the European cluster. Two more clusters are visible but
less relevant since they do not gather important players of the cocoa market, neither
importers or exporters: the first one formed by some Eastern European Countries,
Russia and Central Asia and the second one which embraces some African countries,
Middle East, and the Balkans.
Europe and the United States are the main importers of post-processing cocoa
products, even if China, starting from 2008 is becoming an increasingly relevant
importer of cocoa powder, paste and cake.

\item \textbf{Poultry Meat}: community structure detected for poultry meat products is
extremely concentrated. We identified only three balanced communities: i) Latina
American cluster around one of the leading world exporter, Brazil, serving important
importers such as Saudi Arabia, UAE, several Sub-Saharan and South East Asian
countries, ii) North American cluster mainly linked to Russia, Middle East and selected
relevant importers in South East Asia such as Vietnam, iii) European cluster linked to
some North African and Sub-Saharan countries. An interesting extension of the present
work could be an in-depth investigation of interdependencies between maize, soybeans
and poultry meat networks within an input-output perspective.

\end{itemize}

\newpage


\section{Additional Tables and Figures\label{app:add_tabs_figs}}

\begin{table*}
\caption{\label{tab:comm_stats_2001_2011}Number of communities identified and maximum modularity attained.}
\begin{indented}
\item[]\begin{tabular}{@{}lcccc}
\mr
	&	\multicolumn{2}{c}{2001}			&	\multicolumn{2}{c}{2011}			\\
Commodity	&	\# Communities 	&	Modularity	&	\# Communities	&	Modularity	\\
\mr
Wheat	&	6	&	0.46	&	7	&	0.48	\\
Soybeans	&	4	&	0.32	&	6	&	0.25	\\
Maize	&	5	&	0.39	&	8	&	0.47	\\
Sugar	&	7	&	0.54	&	5	&	0.5	\\
Rice	&	9	&	0.46	&	6	&	0.43	\\
Barley	&	6	&	0.39	&	8	&	0.45	\\
Oil, palm	&	6	&	0.26	&	6	&	0.22	\\
Oil, sunflower	&	7	&	0.49	&	5	&	0.47	\\
Milk	&	5	&	0.36	&	7	&	0.39	\\
Cassava	&	6	&	0.09	&	8	&	0.05	\\
Pulses	&	8	&	0.32	&	6	&	0.25	\\
Cocoa	&	6	&	0.33	&	5	&	0.31	\\
Pigmeat	&	5	&	0.4	&	7	&	0.36	\\
Poultry meat	&	5	&	0.4	&	3	&	0.43	\\
Nuts	&	5	&	0.36	&	6	&	0.35	\\
Sorghum	&	10	&	0.29	&	9	&	0.48	\\	
	\br
\end{tabular}
\end{indented}
\end{table*}

\begin{figure*}[ht!]
 \centering
    {\includegraphics[width=15cm]{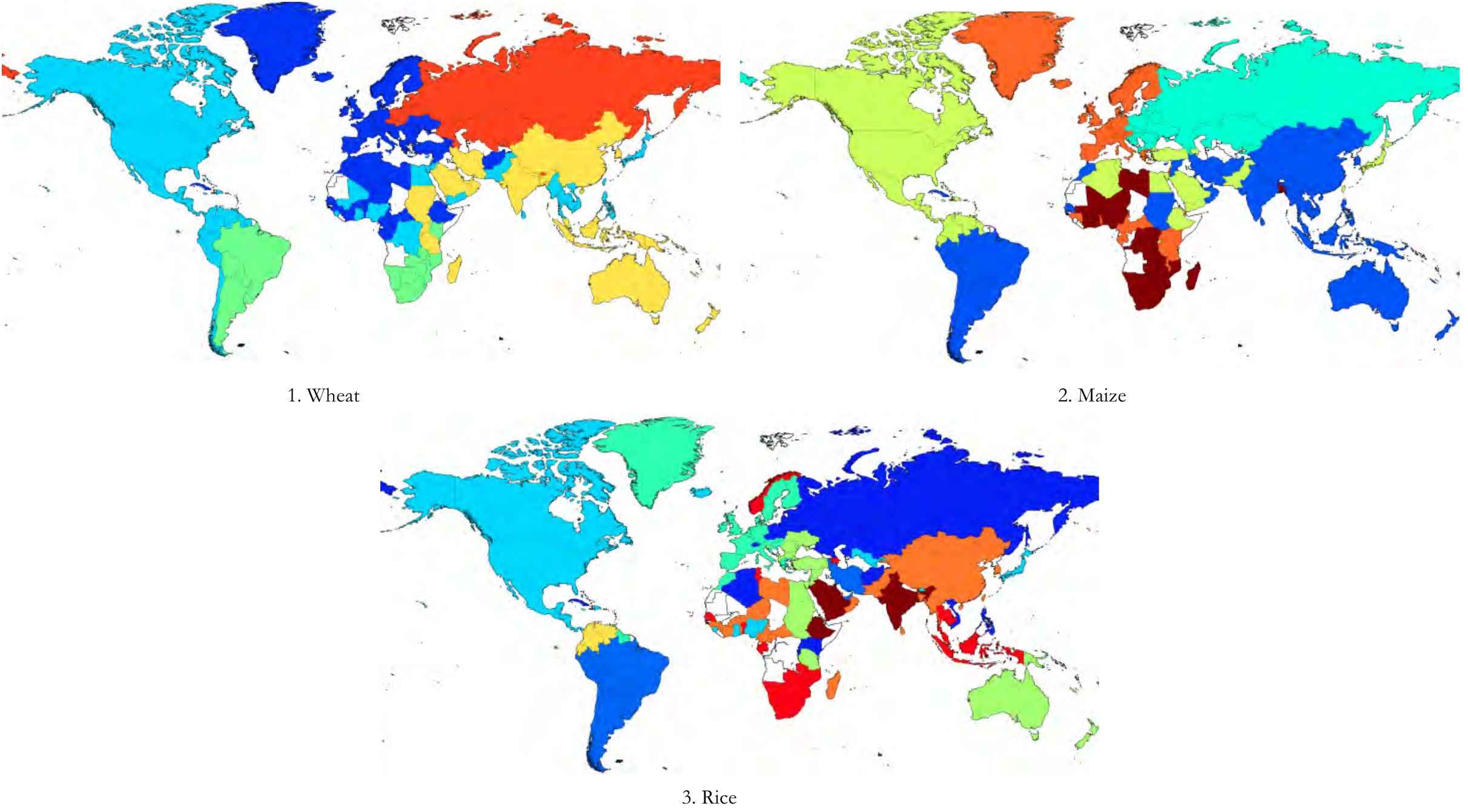}}
 \caption{Community detection in year 2001. Choropleth maps display country membership to communities for selected commodities. In white, countries not belonging to any community or for which no data are available.\label{fig:choro_2001}}
\end{figure*}

\begin{figure*}[ht!]
 \centering
    {\includegraphics[width=15cm]{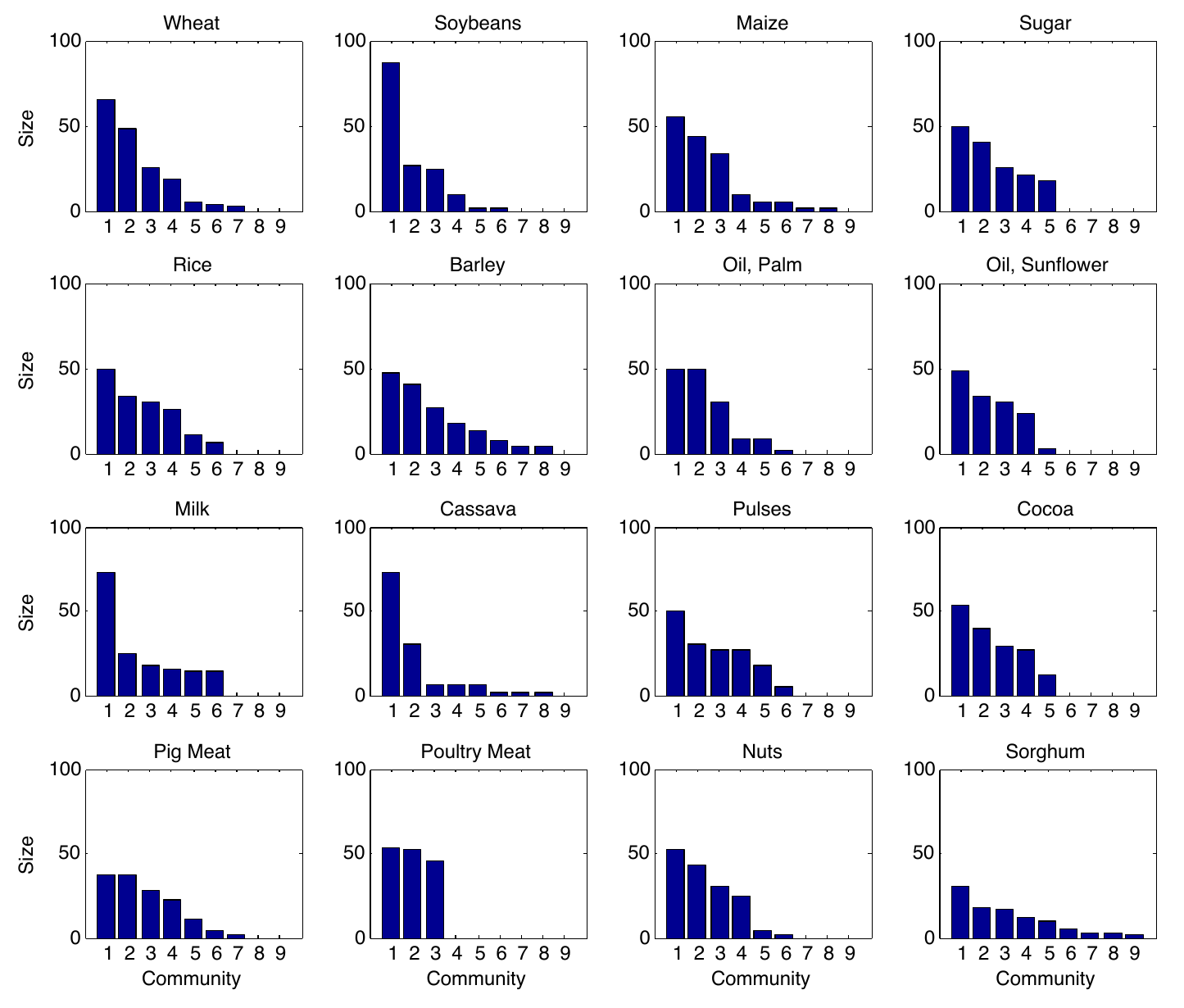}}
 \caption{Community detection in year 2011. Distributions of the size of communities across commodities.\label{fig:sizedistr_2011}}
\end{figure*}

\begin{figure*}[ht!]
 \centering
    {\includegraphics[width=15cm]{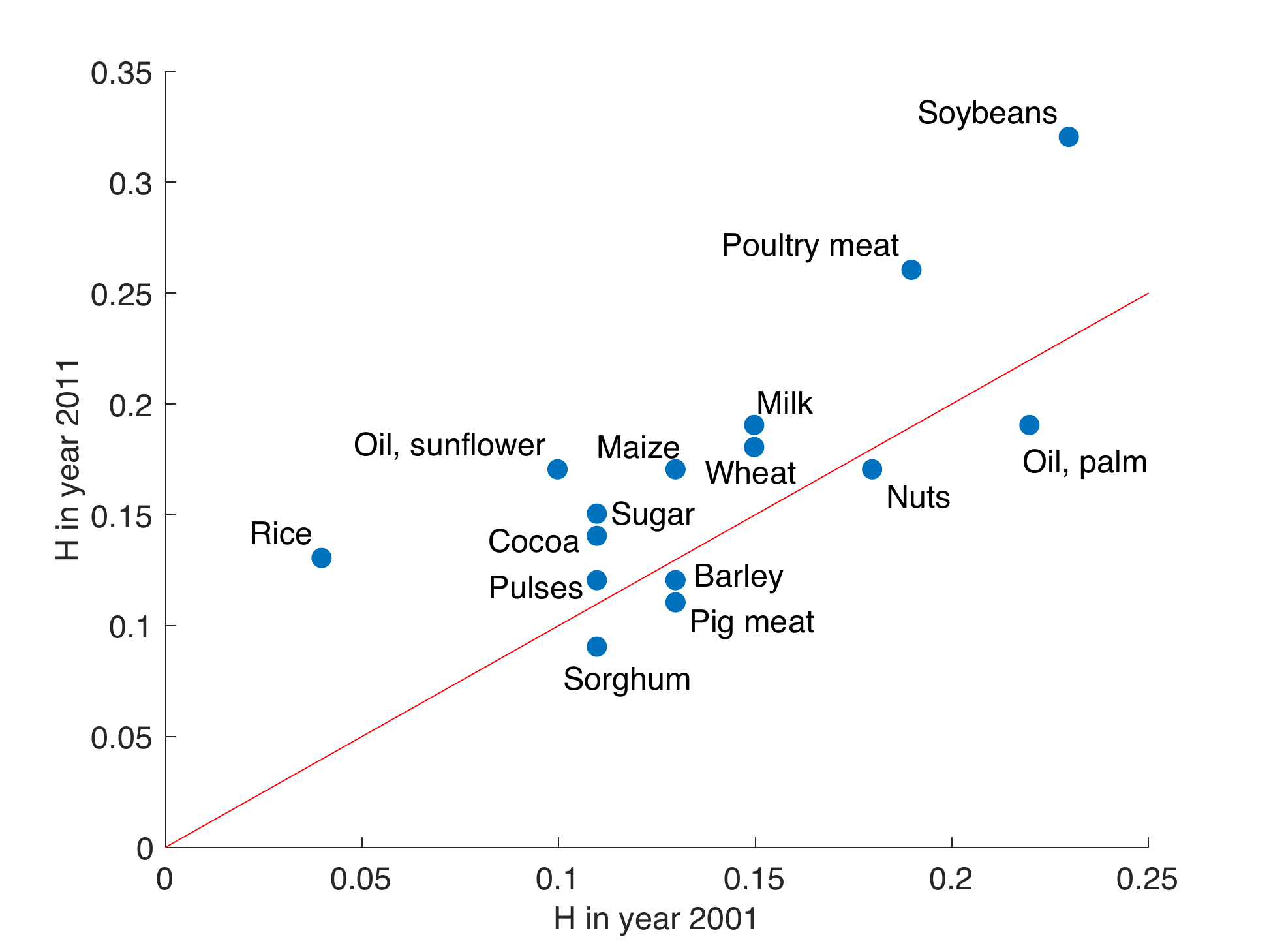}}
 \caption{Community detection in years 2001 and 2011. Herfindahl concentration index computed on community size distributions.\label{fig:h_index_2001_2011}}
\end{figure*}

\begin{figure*}[ht!]
 \centering
    {\includegraphics[width=17cm]{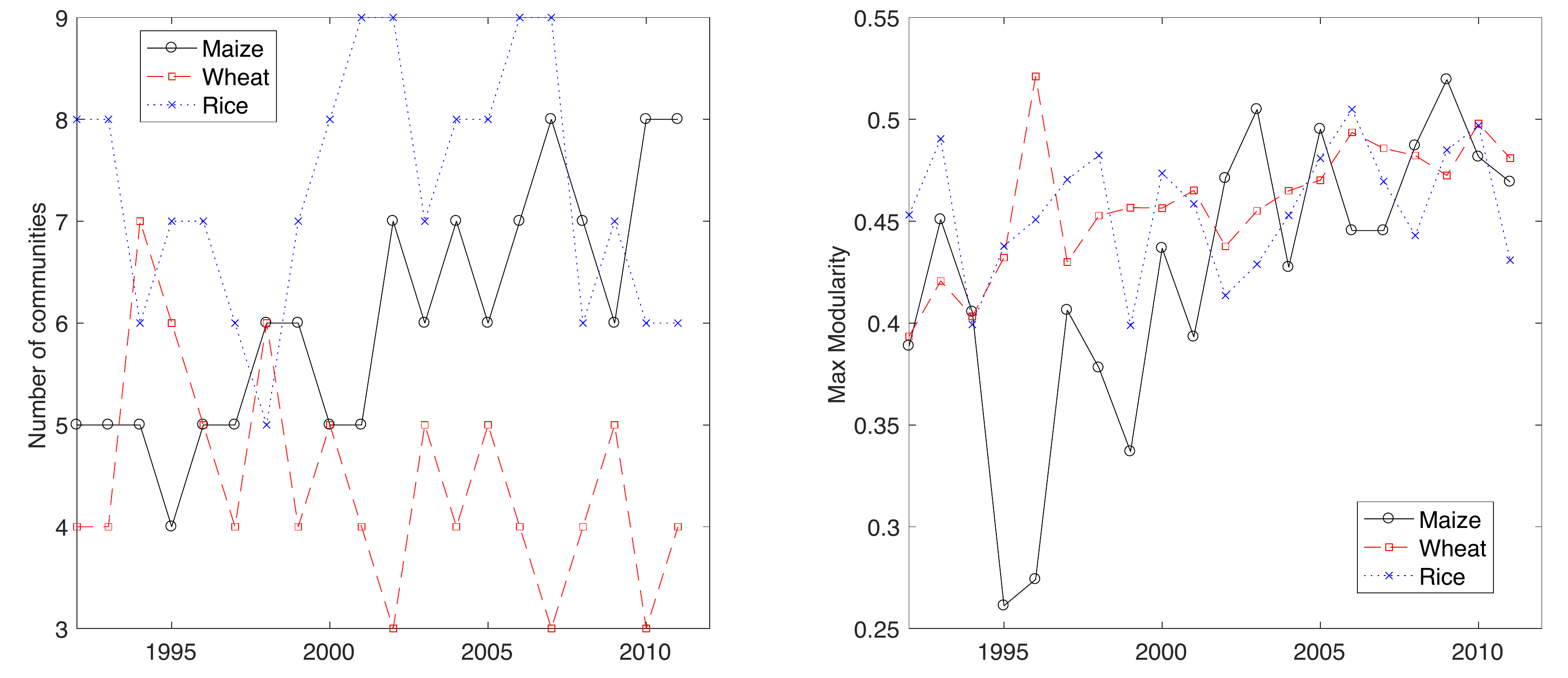}}
 \caption{Community detection for wheat, maize and rice in the sample 1992-2011. Number of communities detected (left) and maximum modularity attained (right).\label{fig:maize_wheat_rice_1992_2011}}
\end{figure*}

\begin{figure*}[h!]
 \centering
    {\includegraphics[width=\textwidth]{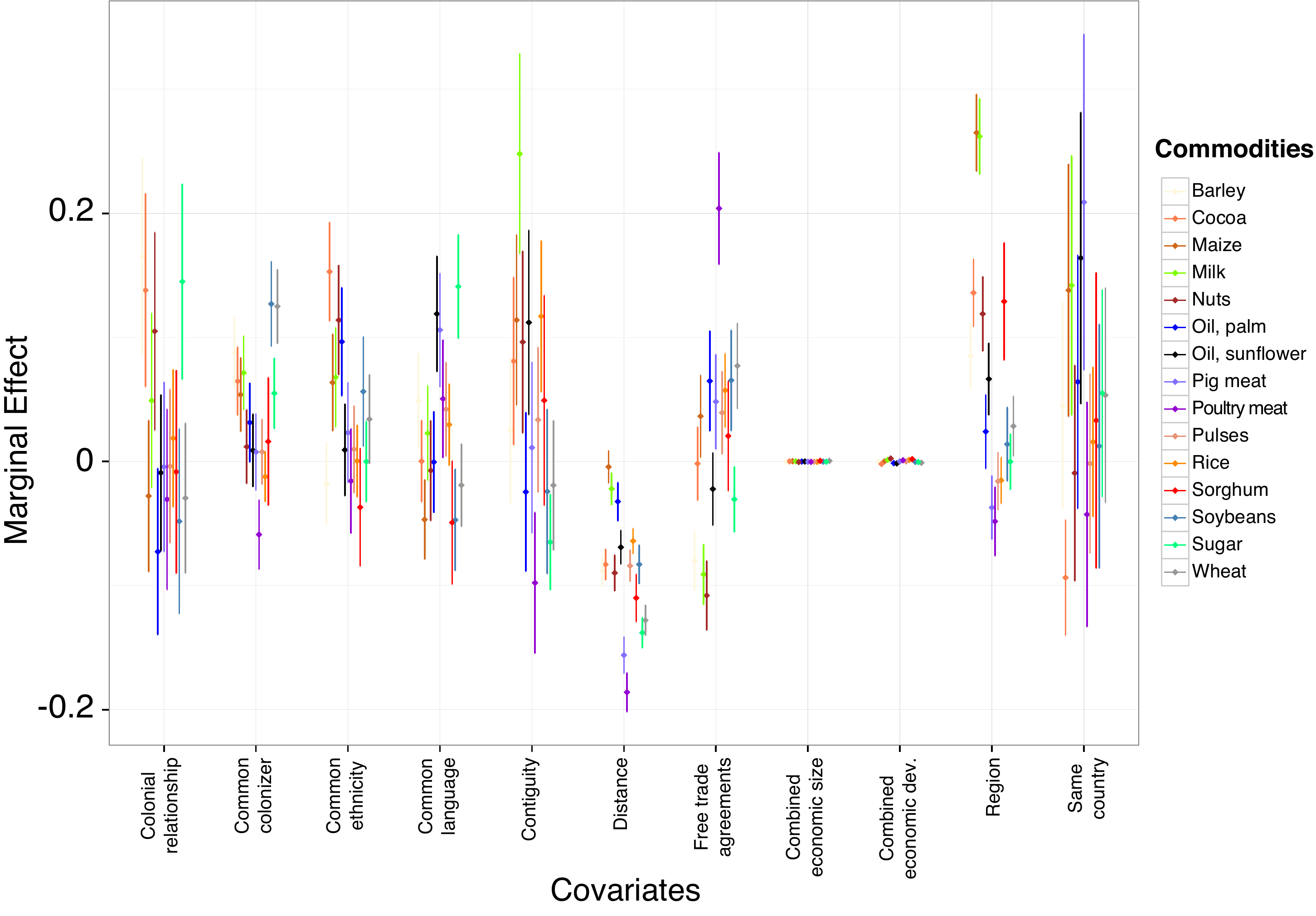}}
 \caption{\label{fig:cs_regs_2001} Probit estimation for year 2001. Marginal effects obtained fitting Eq. (\ref{eq:probit}) to each commodity layer separately using maximum-likelihood. X-axis: covariates used in the model. Y-axis: marginal effect of the covariate on the probability that two countries belong to the same community. Dots represent the point estimate of marginal effects and bars are 95\% confidence intervals.}
\end{figure*}

\begin{figure*}
 \centering
 \subfigure[Specification \#1: Free-trade agreement variable]
   {\includegraphics[width=0.8\textwidth]{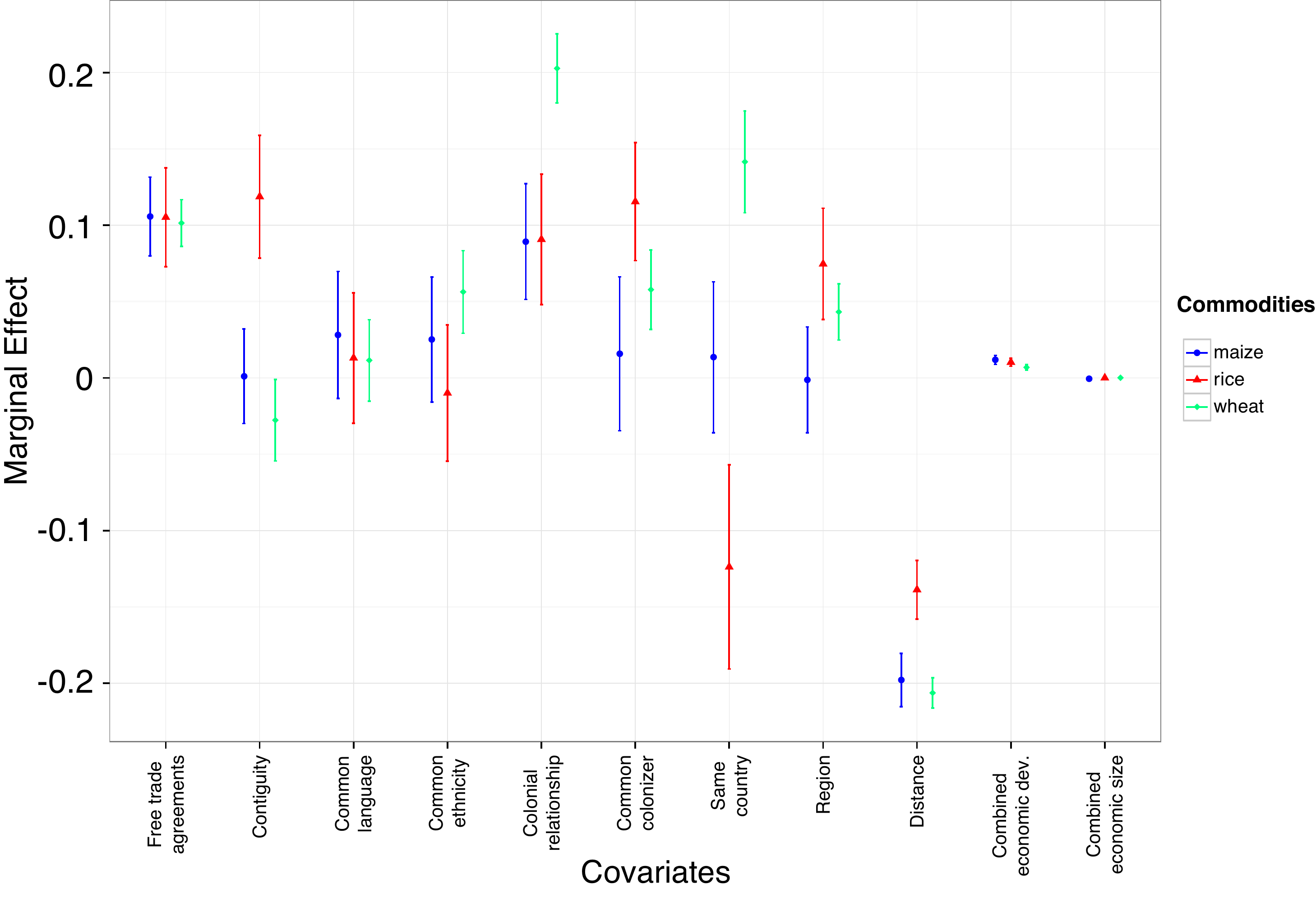}}
 \hspace{5mm}
 \subfigure[Specification \#2: Dummies for specific free-trade agreements]
   {\includegraphics[width=0.8\textwidth]{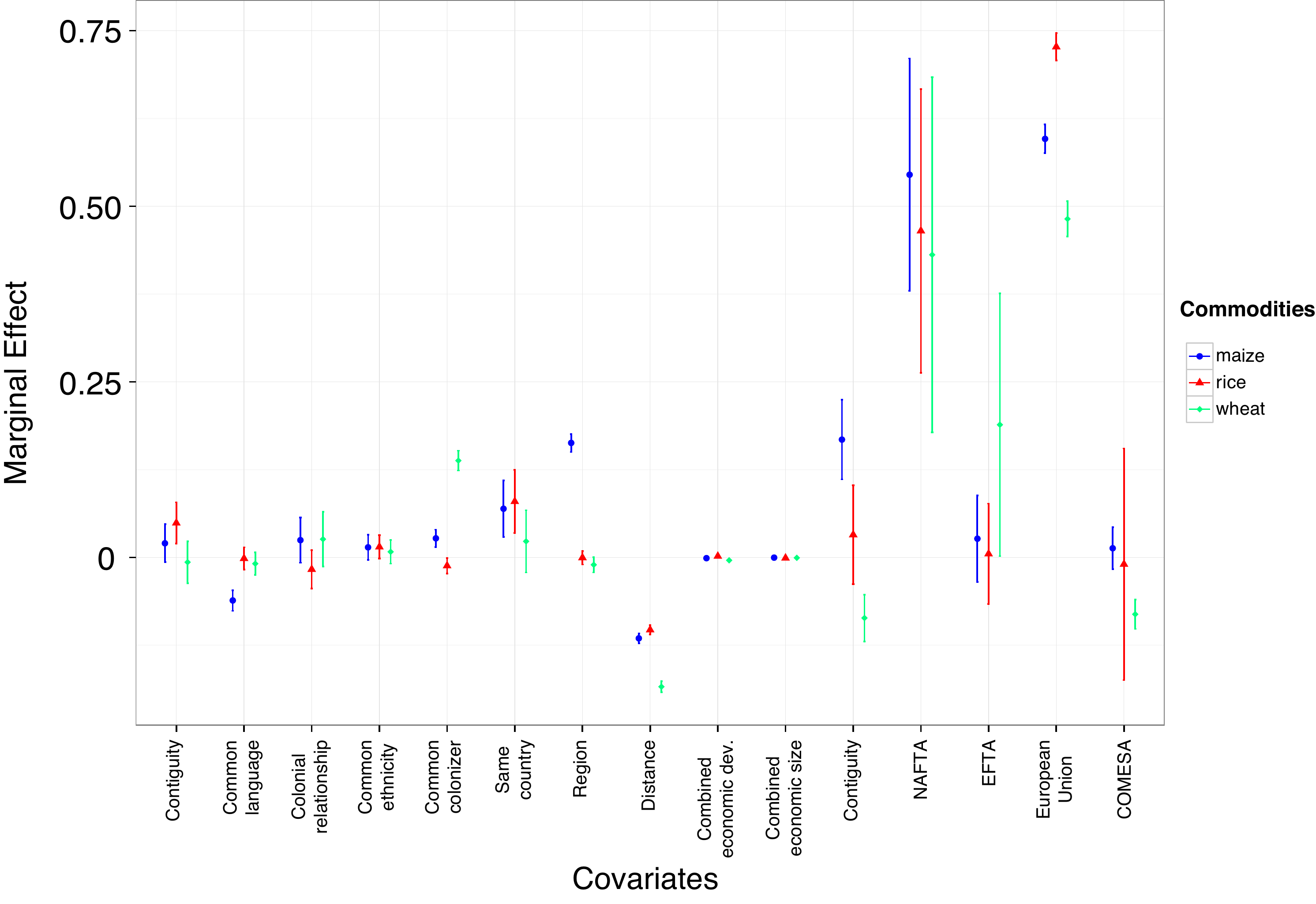}}
 \caption{\label{fig:panel_regs} Panel data estimation of probit models.  Marginal effects obtained fitting Eq. (\ref{eq:probit}) to each commodity layer separately using maximum-likelihood. X-axis: covariates used in the model. Y-axis: marginal effect of the covariate on the probability that two countries belong to the same community. Dots represent the point estimate of marginal effects and bars are 95\% confidence intervals.}
\end{figure*}

\begin{figure*}[h!]
 \centering
    {\includegraphics[width=\textwidth]{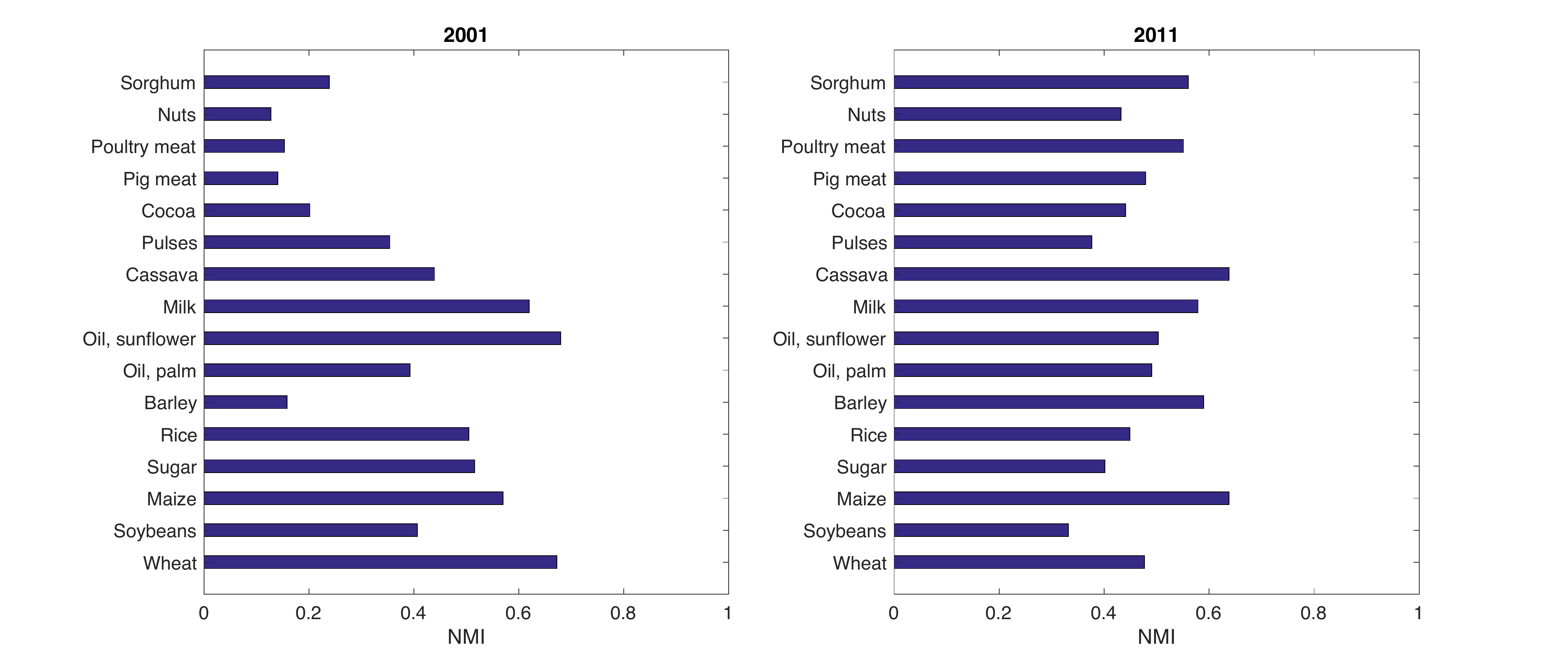}}
 \caption{\label{fig:ml_nmi_2001_2011_vs_sl} Multi-layer community detection. Normalized mutual information (NMI) index comparing communities obtained when the IFTMN is considered as a multi-layer network and when commodity layers are taken as independent. Higher values of NMI means more similar community structures. Years 2001 and 2011.}
\end{figure*}

\begin{figure*}
 \centering

 \subfigure[Wheat, 2001]
   {\includegraphics[width=8cm]{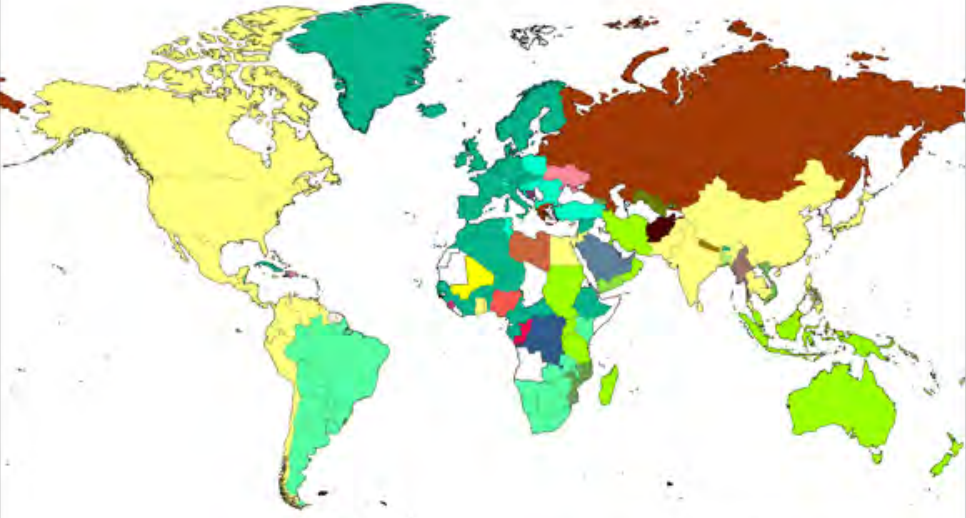}}
 \hspace{5mm}
 \subfigure[Wheat, 2011]
   {\includegraphics[width=8cm]{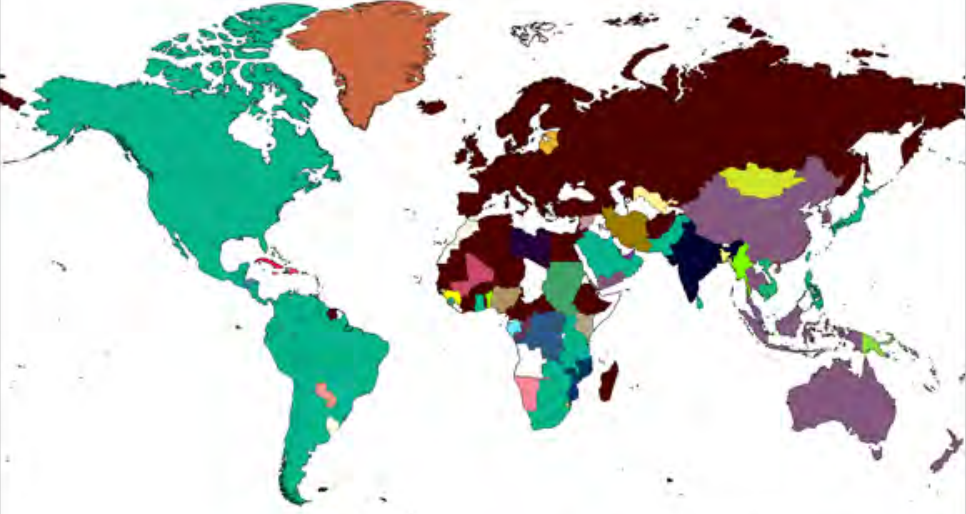}}

 \subfigure[Rice, 2001]
   {\includegraphics[width=8cm]{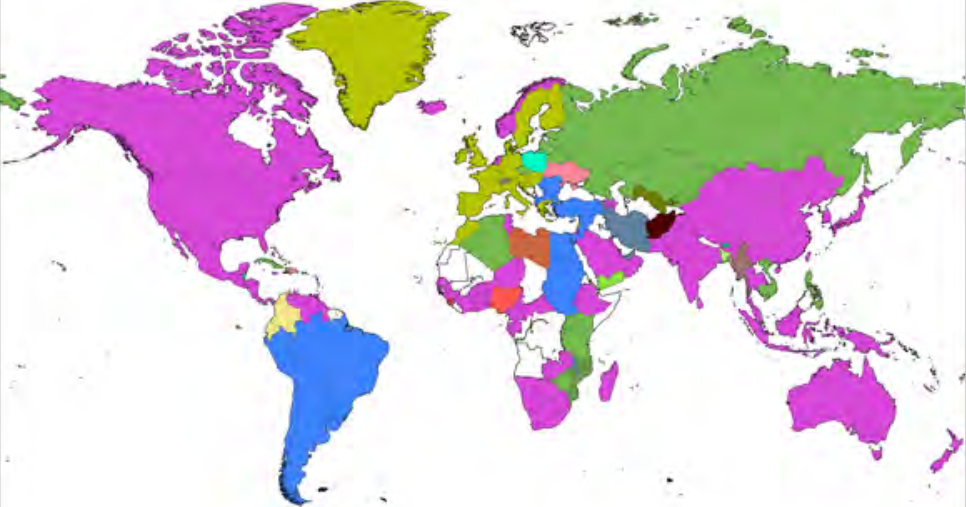}}
 \hspace{5mm}
 \subfigure[Rice, 2011]
   {\includegraphics[width=8cm]{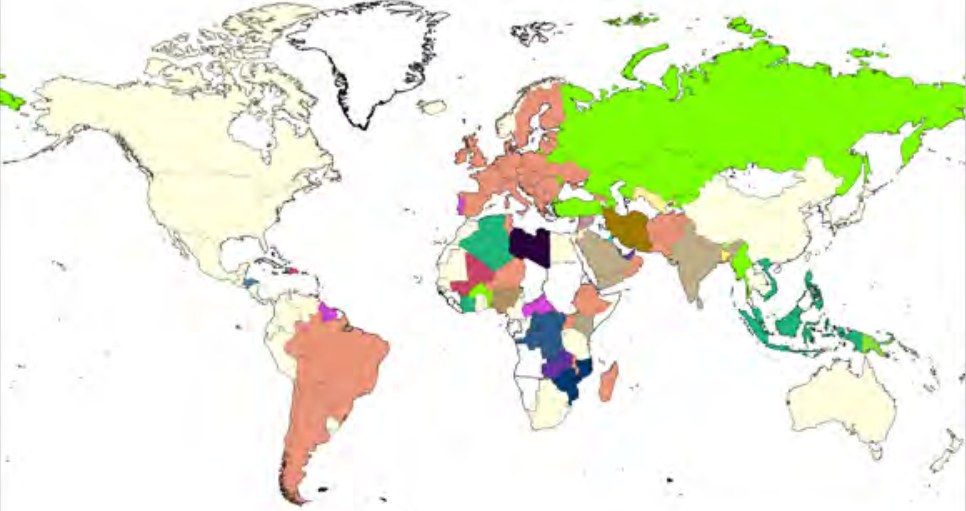}}

 \subfigure[Maize, 2001]
   {\includegraphics[width=8cm]{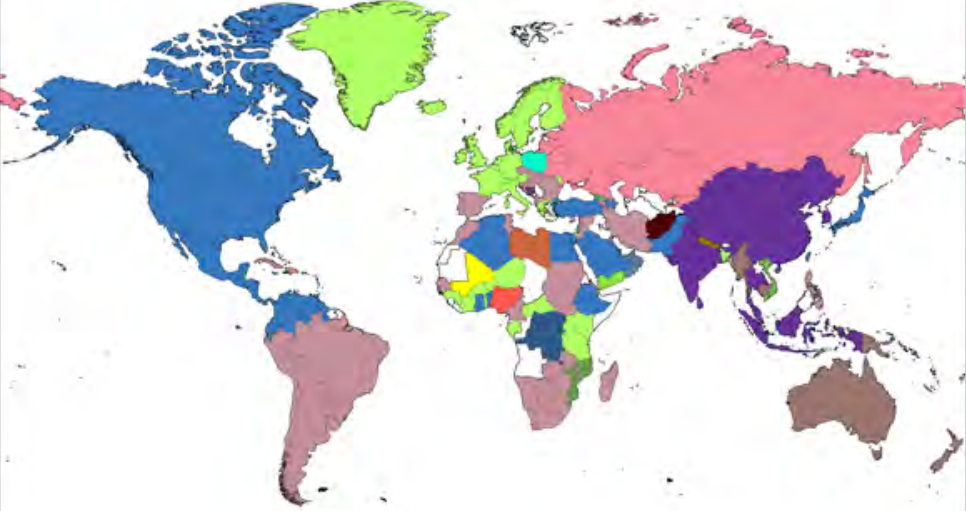}}
 \hspace{5mm}
 \subfigure[Maize, 2011]
   {\includegraphics[width=8cm]{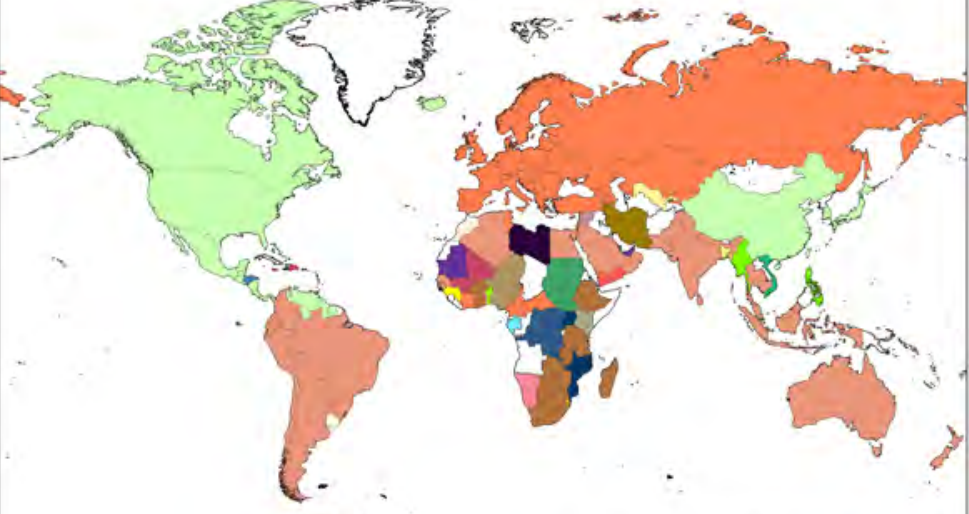}}

 \caption{\label{fig:choro_ml}Multi-layer community detection. Choropleth maps of community structures for wheat, rice and maize in 2001 and 2011. Maps are obtained projecting multi-layer communities into the space of commodities. Colors are consistent across multi-layer communities (i.e., if two countries are filled with the same color across different maps it means that they belong to the same country-product cluster in the multi-layer).}
\end{figure*}

%
%
%
%

\end{document}